\documentclass[]{article}
\usepackage[pdftex]{graphicx}
\usepackage{amsmath}
\usepackage{blindtext}
\usepackage{color}
\usepackage{amsfonts}
\usepackage{amssymb}
\usepackage{cite}
\usepackage{color}
\usepackage{verbatim}
\usepackage{epstopdf}
\usepackage{wrapfig}

\begin{document}

\title{\textbf{Notes on photonic topological insulators and scattering-protected edge states - a brief introduction}}
\author{George W. Hanson, S. Ali Hassani Gangaraj, Andrei Nemilentsau \\ \\ Department of Electrical Engineering, University of Wisconsin-Milwaukee \\ Milwaukee Wisconsin 53211, USA \\ \\ Email: George@uwm.edu }

\maketitle

\begin{quotation}
Disclaimer: The topic of photonic topological insulators and scattering-protected edge states bridges concepts from condensed matter physics and electromagnetics, and necessitates understanding the Berry phase, potential, and curvature, and
related concepts. These notes are an attempt at a moderately self-contained
introduction to the topic, including two detailed photonic examples drawn
from the literature. We made these notes in the process of trying to
understand this topic ourselves, and we are posting this material in the
spirit of helping other researchers start to understand this material. We claim
no novelty in the material or its presentation, nor is this work intended as a comprehensive review.
\end{quotation}

\section{Berry Phase Concepts}

In this section we introduce the main idea of Berry phase, potential, and curvature, and summarize
some related concepts.

\subsection{Motivation - backscattering-immune one-way SPP propagation}

\begin{figure}[ht]
\begin{center}
\noindent  \includegraphics[width=6in]{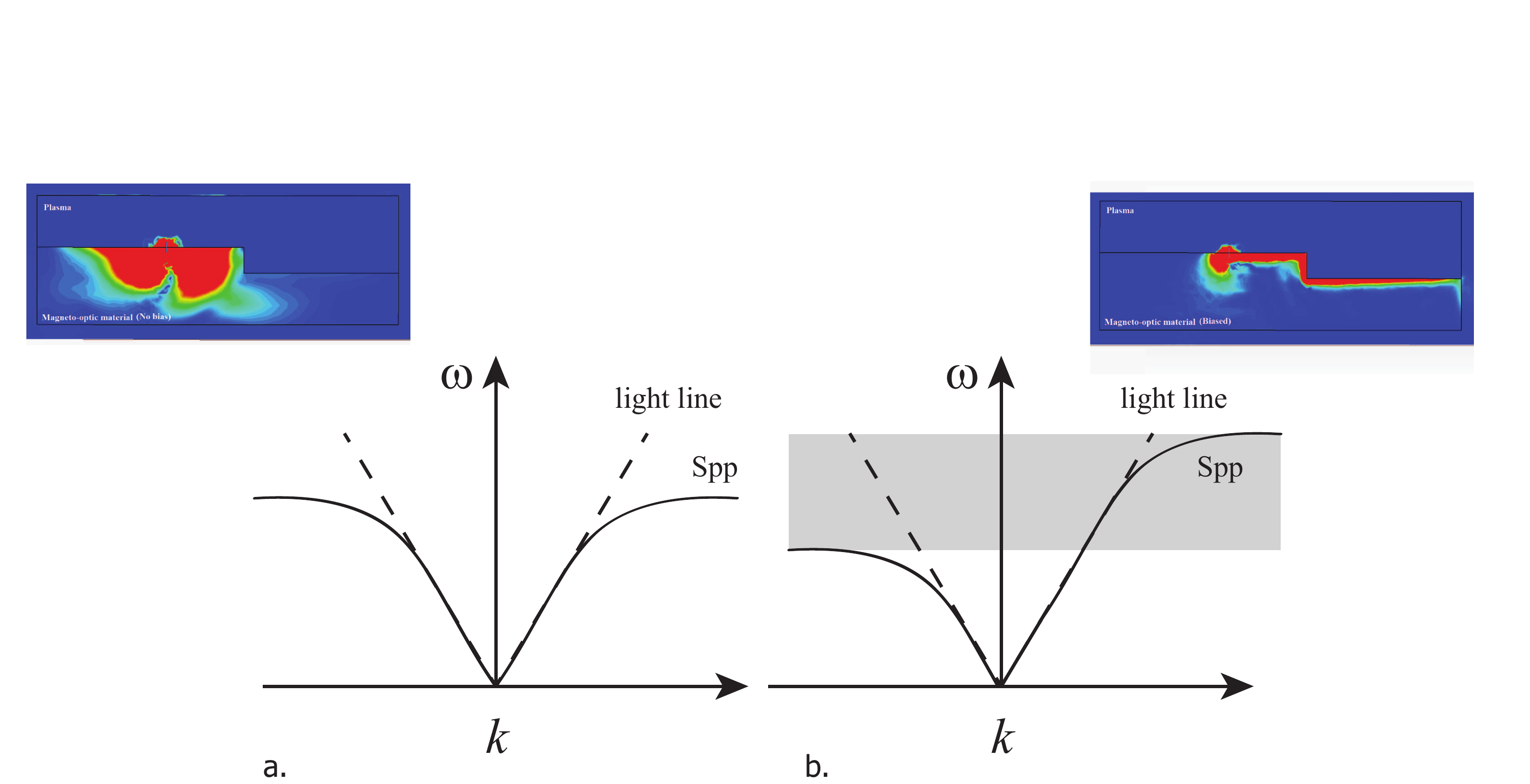}  
\end{center}
\caption{a. Dispersion of reciprocal SPP. Upper-left insert shows SPP power flow excited by a vertical dipole source near a step change in height at the interface between a reciprocal medium (below) and a different reciprocal medium (above), b. nonreciprocal SPP; shaded region depicts frequency range of uni-directional propagation. Upper-right insert is the same as upper-left insert, except that the lower medium is now non-reciprocal and we operate in the gap, ensuring one-way propagation. }
\label{SPP}
\end{figure}

Surface plasmons polaritons (SPPs) are well-known and long-studied waves
that can be guided at the interface between two materials (nominally, for an
SPP to exist in an isotropic environment one material has relative permittivity $\varepsilon
=\varepsilon _{1}<0$ and the other has $\varepsilon =\varepsilon _{2}>0$,
such as an air-plasma (metal) interface). For a wave travelling as $e^{\pm ikz}$ ($z$ parallel to the interface), the SPP dispersion relation is 
\begin{equation}
k=\frac{\omega }{c}\sqrt{\frac{\varepsilon _{1}\varepsilon _{2}}{\varepsilon
_{1}+\varepsilon _{2}}},
\end{equation}%
where $\varepsilon _{\alpha }=\varepsilon _{\alpha }\left( \omega \right) $. For example, for a simple lossless plasma $\varepsilon \left( \omega \right) =1-\omega
_{p}^{2}/\omega ^{2}$ with $\omega _{p}$ being the plasma frequency.
Plotting the dispersion equation (Fig. \ref{SPP}.a) 
we see that propagation is
reciprocal, $\omega \left( -k\right) =\omega \left( k\right) $, so that
forward-propagating ($k$) and backward-travelling ($-k$) waves exist at the same
frequency. A source near the surface will excite SPPs
travelling in both directions ($\pm z$), and upon encountering a
discontinuity an SPP travelling in, say, the $+z$ direction will undergo
both reflection and transmission, again resulting in both forward and
backward travelling waves. This is shown in the upper-left insert of the figure (see also Fig. \ref{Power} and associated example later in the text). 

Waves can be excited in a single direction using a directive source (e.g., planar Yagi-Uda antenna, or by a circularly-polarized source that couples to the SPPs spin polarization), but upon encountering a discontinuity, partial reflection of the wave will occur since the material itself allows propagation in both directions.

However, if the medium only supports modes that can travel in one direction, say, via non-reciprocity as depicted in Fig. \ref{SPP}.b (e.g., via a magnetic-field biased plasma having a tensor permittivity with non-zero off-diagonal elements), then upon encountering a discontinuity an SPP cannot be reflected (back-scattered), as shown in the upper-right insert of the figure. This is a rather remarkable occurrence, and has important applications in
waveguiding (e.g., defect- immune waveguides). In general, there will be a range of energies where only
propagation in one direction is possible (e.g., in Fig. \ref{SPP}.b in the indicated
frequency band only forward propagating modes can exist, there are no states
with $-k$). Fig. \ref{Power} at the end of these notes shows power density excited by a source over a magnetoplasma material, as discussed in \cite{Arthur}-\cite{Mario2}. 

However, the idea of one-way (backscattering-immune) surface-wave
propagation is more general then indicated above. In particular, one does
not necessarily need a non-reciprocal material. A broad class of materials
exist known as photonic topological insulators (PTIs) which have this
characteristic, generally supporting Hall/chiral edge states. This class of materials includes biased non-reciprocal
magneto-plasmas (more generally, materials with broken time-reversal
symmetry), but it also includes time-reversal-invariant materials with
broken inversion symmetry. In the latter case, photon states are separated
in two `spin' sub-spaces (usually through geometry such as via a hexagonal
lattice), and `spin-orbit' coupling is introduced through inversion
symmetry-breaking. Here focus on the simplest subclass formed
by photonic topological media with a broken time reversal
symmetry, sometimes also designated as Chern-type insulators
(the analogs of quantum Hall insulators). 

On the electronic side, topological insulators (TIs) and quantum Hall edge state materials (which utilize many of the same concepts described here) came first, and, noting the analogy between electronic and optical systems, the first work on PTIs was described in \cite{Haldane2} and \cite{Haldane}. The first experimental demonstration of an optical TI was shown in \cite{Rechtsman}, and in various material systems \cite{Rechtsman1}, \cite{Rechtsman2}, \cite{Poo}, \cite{Dong}, and \cite{Skirlo}, among others.
Understanding the broad field of PTIs necessitates
understanding the Berry phase, potential, curvature, and the concept of Chern invariants, which is
the subject of these notes. After an introduction to these concepts, we examine two previous PTI results from the literature, and provide details of the various computations necessary to characterize the materials.

\subsection{Origin of the Berry Phase}\label{sec1}

Here we derive the Berry phase, following the usual procedure for electronic systems. For a derivation that only considers classical electromagnetics, as well as a more thorough introduction, see \cite{MGA}. 

We consider a system described by a Hamiltonian dependent
on parameters that vary in time, $H=H(\mathbf{R})$, such that $%
\mathbf{R}=\mathbf{R}\left( t\right) =\left( R_{1}\left( t\right)
,R_{2}\left( t\right) ,R_{3}\left( t\right) ,....\right) $ We will consider
a path in parameter space $C$ along which $\mathbf{R}$ changes. For example, 
$\mathbf{R}$ could describe the position of a particle $\left( x\left(
t\right) ,y\left( t\right) ,z\left( t\right) \right) $ and $C$ could be a
path in physical space. However, here we are primarily interested in the
case when $\mathbf{R}\left( t\right) $ lives in momentum (reciprocal) space.

The evolution of the system is assumed to be adiabatic, such that the
parameters $\mathbf{R}(t)$ of the Hamiltonian change slowly along path $C
$ in parameter space. The adiabatic theorem states that if a system is
initially in the nth eigenstate of the initial Hamiltonian $H\left( \mathbf{R%
}\left( 0\right) \right) $, and the system is moved slowly-enough as $%
\mathbf{R}\left( t\right) $ changes, it will arrive at the nth eigenstate of
the final Hamiltonian $H\left( \mathbf{R}\left( T\right) \right) $.
Development of the adiabatic solution below will show how the Berry phase
comes about.

We will assume that the time dependent states evolves through an evolution
equation%
\begin{equation}
i\hbar \partial _{t}\left\vert \Psi (t)\right\rangle =H(\mathbf{R}%
(t))\left\vert \Psi (t)\right\rangle \label{evol}
\end{equation}%
which is typically taken to be the Schr\"{o}dinger equation, where $%
\left\vert \Psi (t)\right\rangle $ is a scalar, but it could also represent
the Dirac equation where $\left\vert \Psi (t)\right\rangle $ is a spinor,
and classical Maxwell's equations ($\hslash =1$), where $\left\vert \Psi
(t)\right\rangle $ is the six-vector of EM fields. 

Because of the slow variation of the Hamiltonian parameters we can assume that at every time the instantaneous eigenstates of the Hamiltonian satisfy%
\begin{equation}
H(\mathbf{R})\left\vert n(\mathbf{R})\right\rangle =E_{n}\left\vert
n(\mathbf{R})\right\rangle .  \label{IH}
\end{equation}%
However, (\ref{IH}) does not uniquely determine the function $\left\vert n(%
\mathbf{R})\right\rangle $, since we could include an arbitrary phase
factor (gauge choice) that depends on $\mathbf{R}\left( t\right) $.

To motivate the following derivation, note that if the Hamiltonian is
independent of time, then a system that starts out in the $n$th
eigenstate $\left\vert n\right\rangle $, remains in $n$th eigenstate but
simply pick up a phase factor,%
\begin{equation}
\left\vert \Psi _{n}(t)\right\rangle =\left\vert n\right\rangle e^{-\frac{i}{%
\hbar }E_{n}t}.
\end{equation}%
So, to represent the evolution of the system with slowly varying
Hamiltonian we use a superposition of these instantaneous eigenvectors,
adjusting the phase factor to account for the time variation,%
\begin{equation}
\left\vert \Psi (t)\right\rangle =\sum_{n}a_{n}(t)e^{-\frac{i}{\hbar }%
\int_{0}^{t}E_{n}(\mathbf{R}(t^{\prime }))dt^{\prime }}\left\vert n(%
\mathbf{R}(t))\right\rangle =\sum_{n}a_{n}(t)e^{i\alpha _{n}}\left\vert
n(\mathbf{R}(t))\right\rangle 
\end{equation}%
where $\alpha _{n}(t)=-\frac{1}{\hbar }\int_{0}^{t}E_{n}(\mathbf{R}%
(t^{\prime }))dt^{\prime }$ is called the dynamical phase. If we substitute
this general form of solution in the evolution equation (\ref{evol})
we obtain%
\begin{align}
& i\hbar \sum_{n}(\partial _{t}a_{n}+ia_{n}\partial _{t}\alpha
_{n})e^{i\alpha _{n}}\left\vert n\right\rangle +i\hbar
\sum_{n}a_{n}e^{i\alpha _{n}}\left\vert \partial _{t}n\right\rangle
=H\left\vert \Psi \right\rangle   \nonumber \\
& i\hbar \sum_{n}(\partial _{t}a_{n})e^{i\alpha _{n}}\left\vert
n\right\rangle +\sum_{n}E_{n}a_{n}e^{i\alpha _{n}}\left\vert n\right\rangle
+i\hbar \sum_{n}a_{n}e^{i\alpha _{n}}\left\vert \partial _{t}n\right\rangle
=H\left\vert \Psi \right\rangle ,
\end{align}%
and taking the inner product of both sides by $\left\langle m\right\vert $,
yields%
\begin{equation}
\partial _{t}a_{m}=-\sum_{n}a_{n}e^{i(\alpha _{n}-\alpha _{m})}\left\langle
m|\partial _{t}n\right\rangle .  \label{M1}
\end{equation}%
In the adiabatic limit, where excitation to other instantaneous eigenvectors
is negligible \footnote{%
To prove the statement that the excitation probability of states $n\neq m$
is small, the time derivative of the energy state equation is $\partial
_{t}H\left\vert n\right\rangle +H\left\vert \partial _{t}n\right\rangle
=\partial _{t}E_{n}\left\vert n\right\rangle +E_{n}\left\vert \partial
_{t}n\right\rangle $, where we can set $\partial _{t}E_{n}=0$ due to slow
variation. The inner product with $\left\langle m\right\vert $ yields $%
\left\langle m|\partial _{t}n\right\rangle =\left\langle m|\partial
_{t}H|n\right\rangle /(E_{m}-E_{n})~~(n\neq m)$ so we have from (\ref{M1}) $%
\partial _{t}a_{m}=-\sum_{n}a_{n}e^{i(\alpha _{n}-\alpha _{m})}\left\langle
m|\partial _{t}H|n\right\rangle /(E_{m}-E_{n})~~(n\neq m).$ Choose the
initial state to be one of the instantaneous eigenstates $\left\vert \Psi
(t=0)\right\rangle =\left\vert n(\mathbf{R}(0))\right\rangle $, so $%
a_{n}(t=0)=1$ and $a_{m}(t=0)=0$ for $n\neq m$. Then, for $n\neq m$ we have $%
\partial _{t}a_{m}\approx -e^{i(\alpha _{n}-\alpha _{m})}\left\langle
m|\partial _{t}H|n\right\rangle /(E_{m}-E_{n}).$ Since the time dependencies
of $\left\langle m|\partial _{t}H|n\right\rangle $ and $E_{n}-E_{m}$ are
slow, the most important time dependence will be in the exponential, which
can be approximated by $e^{i(\alpha _{n}-\alpha
_{m})}=e^{i(E_{m}-E_{n})t/\hbar }$. Neglecting the other slow time
dependencies then yields%
\begin{align}
& \partial _{t}a_{m}\approx -e^{i(\alpha _{n}-\alpha _{m})}\left\langle
m|\partial _{t}H|n\right\rangle /(E_{m}-E_{n})  \notag \\
& a_{m}(t)=-\int_{0}^{t}e^{i(E_{m}-E_{n})t/\hbar }\frac{\left\langle
m|\partial _{t}H|n\right\rangle }{(E_{m}-E_{n})}\cdot dt=\frac{i}{\hbar }%
\frac{\left\langle m|\partial _{t}H|n\right\rangle }{\omega _{mn}^{2}}%
\{e^{i\omega _{mn}t/\hbar }-1\},
\end{align}%
$\omega _{mn}=(E_{m}-E_{n})/\hbar ,~~(n\neq m)$. Due to adiabatic
approximation we have adopted, $\left\langle m|\partial _{t}H|n\right\rangle 
$ is slow compared to the transition frequency $\omega
_{mn}=(E_{m}-E_{n})/\hbar $. Therefore, the magnitude of the excitation
probability to other states $|a_{m}(t)|^{2}$ is small for $n\neq m$. For
further reading see \cite{ballentine}.} , the choice of initial state $%
\left\vert \Psi (t)\right\rangle =\left\vert n(\mathbf{R}%
(t=0))\right\rangle $ will imply that $|a_{n}(t)|=1$, $a_{m}(t)=0$ for $%
m\neq n$. We then have%
\begin{align}
& \partial _{t}a_{m}=-\sum_{n}a_{n}e^{i(\alpha _{n}-\alpha
_{m})}\left\langle m|\partial _{t}n\right\rangle   \notag \\
& \partial _{t}a_{n}=-a_{n}\left\langle n|\partial _{t}n\right\rangle
\rightarrow a_{n}=e^{i\gamma _{n}};~~\partial _{t}\gamma _{n}=i\left\langle
n|\partial _{t}n\right\rangle .
\end{align}%
Therefore, the adiabatic evolution of the state vector becomes%
\begin{equation}
\left\vert \Psi (t)\right\rangle =e^{i\gamma _{n}}e^{i\alpha _{n}}\left\vert
n(\mathbf{R}(t))\right\rangle .  \label{Eq:2}
\end{equation}%
We have
\begin{align}
\gamma _{n}& =i\int_{0}^{t}\left\langle n(\mathbf{R}(t^{\prime
}))\right\vert \frac{\partial }{\partial t}\left\vert n(\mathbf{R}(t^{\prime
}))\right\rangle dt^{\prime }  \nonumber \\
& =i\int_{0}^{t}\left\langle n(\mathbf{R}(t^{\prime }))\right\vert \frac{%
\partial }{\partial \mathbf{R}}\left\vert n(\mathbf{R}(t^{\prime
}))\right\rangle \cdot \frac{\partial \mathbf{R}}{\partial t^{\prime }}%
dt^{\prime }  \nonumber \\
& =\int_{R_{i}}^{R_{f}}d\mathbf{R}\cdot i\left\langle n(\mathbf{R}%
)\right\vert \nabla _{\mathbf{R}}\left\vert n(\mathbf{R})\right\rangle
=\int_{R_{i}}^{R_{f}}d\mathbf{R}\cdot \mathbf{A}_{n}(\mathbf{R})  \label{e4}
\end{align}%
(setting $\nabla _{\mathbf{R}}=\partial /\partial \mathbf{R)}$, where $R_{i}$
and $R_{f}$ are the initial and final values of $\mathbf{R}(t)$ in parameter
space, and where
\begin{equation}
\mathbf{A}_{n}(\mathbf{R})=i\left\langle n(\mathbf{R})\right\vert
\nabla _{\mathbf{R}}\left\vert n(\mathbf{R})\right\rangle =-\mathrm{Im}\left\langle
n(\mathbf{R})\right\vert \nabla _{\mathbf{R}}\left\vert n(\mathbf{R}%
)\right\rangle   \label{AEM}
\end{equation}
is called the \textit{Berry vector potential} (also called \textit{the Berry
connection} since it connects the state at $\mathbf{R}$ and the state at $\mathbf{R}+d\mathbf{R}$) and $\gamma _{n}$ is called the \textit{Berry phase}.\footnote{%
One can also obtain this result by assuming the existence of this extra
phase \cite{Berry}, $\left\vert \Psi (t)\right\rangle =e^{i\gamma
_{n}}e^{i\alpha _{n}}\left\vert n(\mathbf{R}(t))\right\rangle $, and
inserting into Schr\"{o}dinger's equation. Taking the inner product with $%
\left\langle n(\mathbf{R}(t))\right\vert $ and using $\left\langle n(%
\mathbf{R}(t))|H(\mathbf{R})|n(\mathbf{R}(t))\right\rangle =E_{n}
$ leads to the same result as above.}.

Eq. (\ref{e4}) shows that, in addition to the dynamical phase, the state will acquire an additional phase $\gamma _{n}$ during the adiabatic
evolution (note that $\gamma _{n}$ is real-valued; $e^{i\gamma _{n}\left(
t\right) }$ is a phase, not a decay term\footnote{$\left\langle n(%
\mathbf{R})\right\vert \nabla _{\mathbf{R}}\left\vert n(\mathbf{R}%
)\right\rangle $ can easily seen to be itself imaginary since $\left\langle
n(\mathbf{R})\right\vert \left. n(\mathbf{R})\right\rangle =1$, and
so taking a derivative on both sides yields $\left\langle n(\mathbf{R}%
)\right\vert \nabla _{\mathbf{R}}\left\vert n(\mathbf{R})\right\rangle
=-\left\langle n(\mathbf{R})\right\vert \nabla _{\mathbf{R}}\left\vert n(%
\mathbf{R})\right\rangle ^{\ast }$.}). The existence of this phase has been known since the early days of quantum mechanics, but it was thought to be non-observable since a gauge-transformation could remove it. It was Berry who, in 1984, showed that for cyclic variation ($\mathbf{R_f}=\mathbf{R_i}$) the phase is not removable under a gauge transformation\cite{Berry} (discussed below), and was also observable\footnote{Physically observable quantities must be gauge-independent}. This net phase change depends only on the path $C$ in parameter
space that is traversed by $\mathbf{R}(t)$, but not on the rate at which it
is traversed (assuming the adiabatic hypothesis still holds). It is
therefore called a \textit{geometrical phase}, in distinction to the dynamical phase
which depends on the elapsed time. This geometric phase has been generalized
for non-adiabatic evolution \cite{Anandan}.

\subsubsection{Geometric Phase}
\begin{figure}[ht]
\begin{center}
\noindent  \includegraphics[width=3in]{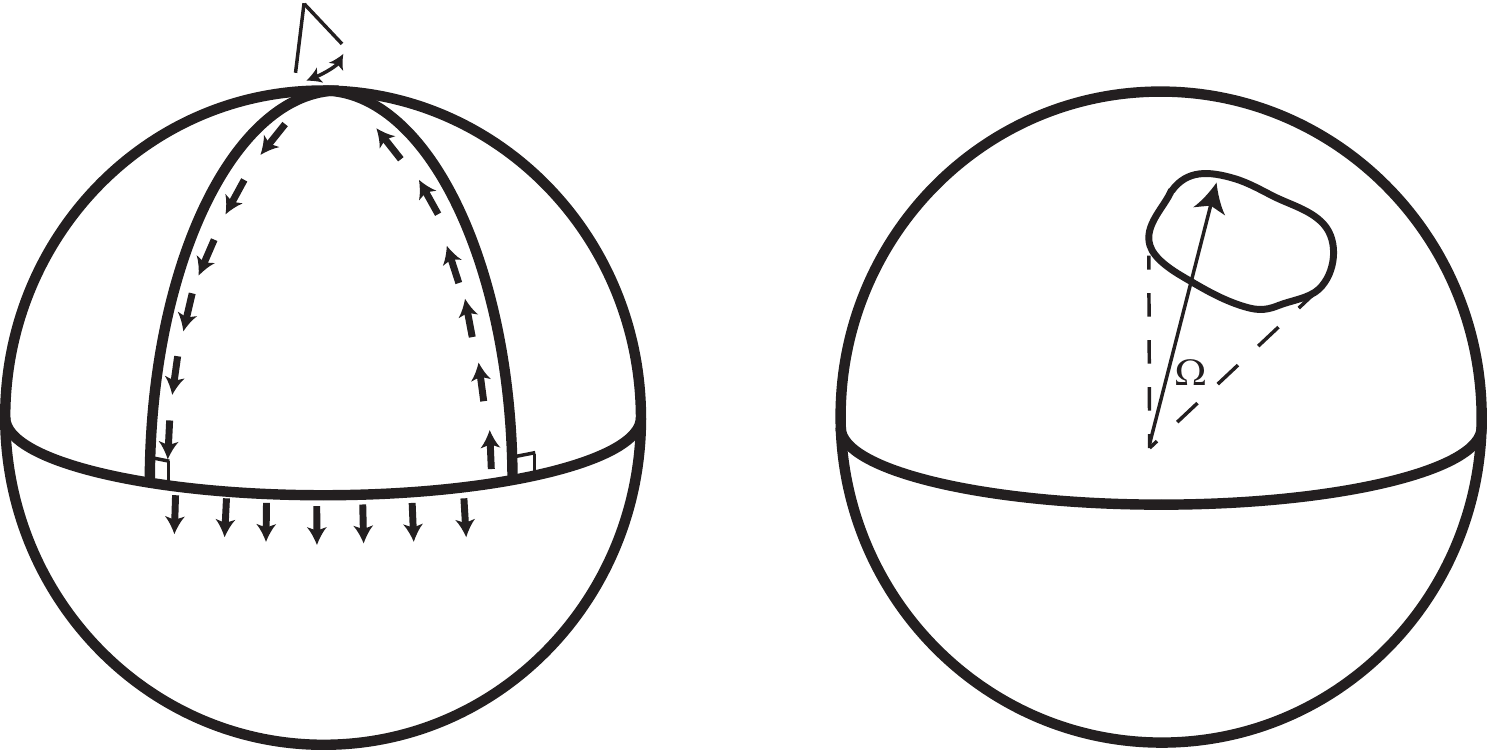}  
\end{center}
\caption{a. Parallel transport around a sphere. b. Parallel transport about a closed contour on a sphere and solid angle subtended.}
\label{sphere}
\end{figure}

Geometric phases have a long history, and arise in many branches of physics \cite{RL1}. They are well-illustrated by considering parallel transport of a vector along a curved surface. To consider an intuitive example, as widely discussed (see, e.g., \cite{griffiths}) and depicted in Fig. \ref{sphere}, consider at $t=0$ a pendulum at the north pole of a sphere, swinging along a longitude line. If the pendulum is moved along the longitude line to the equator, across the equator some distance, and at $t=T$ arriving back at the north pole via a different longitude line (and assuming the movement is sufficiently slow, in keeping with the adiabatic assumption), the angle of the pendulum swing with some fixed reference is obviously different from it's initial angle (this difference is called the defect angle, which is a mechanical analogue of phase). The defect angle is given by the solid angle $\Omega$ subtended by the path of movement. For example, $\Omega_\text{sphere}=4 \pi$ for a sphere, and so if the longitude lines are 180 degrees apart the subtended angle is $\Omega_\text{sphere}/4=\pi$. For the electronic case, moving along a contour in parameter space, the Berry phase is equal to $s\Omega$, where $s$ is the particle spin. Parallel transport along an non-curved surface does not lead to a defect angle, and so we see that a non-zero Berry phase has it's origins in the curvature of parameter space. 
In optics, an optical fiber wound into a helix has been used to demonstrate Berry phase \cite{TC1986}, among other results (see, e.g., \cite{xu2014}). In these cases the momentum is 
$\mathbf{p}=\widehat{\mathbf{x}}p_{x}+\widehat{\mathbf{y}}p_{y}+\widehat{\mathbf{z}}p_{z}=\hslash \mathbf{k},$
where $\mathbf{k}$ is the propagation vector of the optical wave, $\left\vert \mathbf{k}\right\vert =k=2\pi /\lambda $. Therefore, $ \left\vert \mathbf{p}\right\vert
^{2}=p_{x}^{2}+p_{y}^{2}+p_{z}^{2}=(\hslash k)^{2} $, which is a sphere, and so rotation of momentum is equivalent to movement on the surface of a sphere. 

When $\mathbf{R}$ is a real-space parameter, consider an electron in the ground state of an atom. As the atom is slowly moved through a static magnetic field the electron stays in the ground state (adiabatic)
but picks up a Berry phase, which is the Aharonov-Bohm phase. An example of $\mathbf{R}$ as a parameter space is given in Section \ref{example} for an electron fixed in space but exposed to a time-varying magnetic field $\mathbf{B}\left( t\right) $. As detailed in the electromagnetic examples below,
we will be more interested in the case when the parameter space $\mathbf{R}$ is
momentum space, $\mathbf{R}=\mathbf{k}$. In this case, we can simply
consider moving through $\mathbf{k}$-space without necessitating the time
variable, and simply consider $\gamma _{n}\left( \mathbf{k}\right) $, which
will depend on the path taken in $\mathbf{k}$-space.

\subsubsection{Gauge}
Obviously, $A_{n}(\mathbf{R})$ is a gauge dependent quantity. If we make a
gauge transformation $\left\vert n(\mathbf{R})\right\rangle \rightarrow e^{i\xi (\mathbf{R}%
)}\left\vert n(\mathbf{R})\right\rangle $
with $\xi (\mathbf{R})$ an arbitrary smooth function (this is equivalent to
the EM gauge transformation\footnote{
In EM, the gauge transform is%
\begin{equation}
\Phi ^{\prime }\left( \mathbf{r},t\right) =\Phi \left( \mathbf{r},t\right) -%
\frac{\partial \chi \left( \mathbf{r},t\right) }{\partial t},\ \ \mathbf{A}%
^{\prime }\left( \mathbf{r},t\right) =\mathbf{A}\left( \mathbf{r},t\right)
+\nabla \chi \left( \mathbf{r},t\right) ,
\end{equation}%
which leaves the fields 
\begin{equation}
\mathbf{E}\left( \mathbf{r},t\right) =-\nabla \Phi \left( \mathbf{r}%
,t\right) -\frac{\partial \mathbf{A}\left( \mathbf{r},t\right) }{\partial t}%
,\ \ \mathbf{B}\left( \mathbf{r},t\right) =\nabla \times \mathbf{A}\left( 
\mathbf{r},t\right) 
\end{equation}%
unchanged. Then,%
\begin{equation}
i\hslash \frac{d}{dt}\left\vert \psi \right\rangle =\widehat{H}\left\vert
\psi \right\rangle 
\end{equation}%
with the Hamiltonian%
\begin{equation*}
\widehat{H}\left( \mathbf{r},t\right) =\frac{1}{2m}\left[ \widehat{\mathbf{p}%
}+e\mathbf{A}\left( \mathbf{r},t\right) \right] ^{2}-e\Phi \left( \mathbf{r}%
,t\right) +V\left( r\right) ,
\end{equation*}%
becomes 
\begin{equation}
i\hslash \frac{d}{dt}\left\vert \psi ^{\prime }\right\rangle =\widehat{H}%
^{\prime }\left\vert \psi ^{\prime }\right\rangle 
\end{equation}%
where $\left\vert \psi ^{\prime }\right\rangle =e^{-ie\chi \left( \mathbf{r}%
,t\right) /\hslash }\left\vert \psi \right\rangle $ and 
\begin{equation*}
\widehat{H}^{\prime }=\frac{1}{2m}\left[ \widehat{\mathbf{p}}+e\mathbf{A}%
^{\prime }\left( \mathbf{r},t\right) \right] ^{2}-e\Phi ^{\prime }\left( 
\mathbf{r},t\right) +V\left( r\right) .
\end{equation*}%
Therefore, Schr\"{o}dinger's equation is invariant under the gauge
transformation, and the EM change of gauge is equivalent to a phase change
in the wavefunction, $\left\vert \psi ^{\prime }\right\rangle =e^{-ie\chi
\left( \mathbf{r},t\right) /\hslash }\left\vert \psi \right\rangle $.})
the Berry potential transforms to 
$A_{n}(\mathbf{R})\rightarrow A_{n}(\mathbf{R})-\nabla _{\mathbf{R}}\xi (%
\mathbf{R})$. Consequently, the additional phase $\gamma _{n}$ will be
changed by $\xi (\mathbf{R}_{i})-\xi (\mathbf{R}_{f})$ after the gauge
transformation, where $\mathbf{R}_{i}$ and $\mathbf{R}_{f}$ are the initial
and final points of the path $C$. For an arbitrary path one can choose a
suitable $\xi (\mathbf{R})$ such that accumulation of that extra phase term
vanishes, and we left only with the dynamical phase. However, by considering
a closed path (cyclic evolution of the system) $C$ where $\mathbf{R}_{f}=%
\mathbf{R}_{i}$ and noting that the eigenbasis should be single-valued, $%
\left\vert n(\mathbf{R}_{i}\right\rangle =\left\vert n(\mathbf{R}%
_{f}\right\rangle $, then $e^{i\xi (\mathbf{R}_{i})}\left\vert n(\mathbf{R}%
_{i}\right\rangle =e^{i\xi (\mathbf{R}_{f})}\left\vert n(\mathbf{R}%
_{f}\right\rangle =e^{i\xi (\mathbf{R}_{f})}\left\vert n(\mathbf{R}%
_{i})\right\rangle $, and so $e^{i\xi (\mathbf{R}_{i})}=e^{i\xi (\mathbf{R}%
_{f})}$. That is, single-valuedness of the eigenbasis means that  $e^{i\xi (\mathbf{R})}$ (but not necessarily $\xi (\mathbf{%
R})$) must be single-valued, and therefore we must have 
\begin{equation}
\xi (\mathbf{R}_{i})-\xi (\mathbf{R}_{f})=2\pi m
\end{equation}%
where $m$ is an integer. This shows that $\gamma _{n}$ can be only changed
by an integer multiple of $2\pi $ under a gauge transformation using a
smooth gauge function; this phase cannot be removed. Note that this holds
for $\mathrm{Dim}\mathbf{(}R)>1$, so that we have a path integral in (\ref%
{e4}). For a one-parameter space $R$, (\ref{e4}) becomes a simple integral
over a vanishing path; for $R_{f}=R_{i}=R$ $\int_{R}^{R}i\left\langle
n(R)\right\vert \frac{\partial }{\partial R}\left\vert n(R)\right\rangle dR=0
$. However, when applied to periodic solids (for which the Berry phase
is also called the \textit{Zac phase}, electrons crossing the Brillouin zone can indeed pick up a
Berry phase, which persists in 1D because of the periodicity of the Brillouin zone; assuming period $a$, when $k$
sweeps across the BZ due to, say, an applied field, a phase can be acquired since $\int_{-\pi /a}^{\pi /a}\left( \cdot \right) dk=\oint\nolimits_{-\pi /a}^{\pi
/a}\left( \cdot \right) dk$. 

As described below, we will only be interested in paths $C$ that are closed
in parameter space, and so we write 
\begin{equation}
\gamma _{n}=\gamma _{n}\left( \mathbf{R}\right) =\oint\nolimits_{C}d\mathbf{R%
}\cdot \mathbf{A}_{n}(\mathbf{R}).  \label{SL2}
\end{equation}%
In the space of gauge functions $\xi $ where $e^{i\xi (\mathbf{R})}$ is
single-valued, (\ref{SL2}) is gauge-dependent (one could say it is
gauge-invarient up to factors of $2\pi $). If we restrain the class of gauge
functions $\xi $ to be themselves single-valued, then (\ref{SL2}) is
gauge-independent\footnote{%
This is easily seen since $\oint\nolimits_{C}d\mathbf{R}\cdot \nabla _{%
\mathbf{R}}\xi (\mathbf{R})=0$ for $\xi $ a smooth, single-valued function.}.

For a two-dimensional periodic material (such as graphene as an electronic
example, or a hexagonal array of infinite cylinders as an electromagnetic
example), $C$ is typically the boundary of the first Brillouin zone and $%
\mathbf{S}$ is its surface in $\mathbf{k}$-space. In this case, the
\textquotedblleft cyclic\textquotedblright\ variation forming the closed
path $C$ in $\mathbf{k}$-space is the perimeter of the first Brillouin zone.

\subsection{Berry curvature, flux, and tensor, and Chern number} \label{sec2}

Eq. (\ref{SL2}) is an analogy to the equation for magnetic flux $\Phi _{\text{%
mag}}$, in terms of the real-space magnetic field and magnetic vector potential $\mathbf{A}%
_{\text{mag}}$ in electromagnetics,%
\begin{equation}
\Phi _{\text{mag}}=\int_{S}d\mathbf{S}\cdot \mathbf{B}\left( \mathbf{r}%
\right) =\oint\nolimits_{C}d\mathbf{l}\cdot \mathbf{A}_{\text{mag}}\left( 
\mathbf{r}\right) .  \label{MF1}
\end{equation}%
where $\oint\nolimits_{C}d\mathbf{l}\cdot \mathbf{A}_{\text{mag}%
}\left( \mathbf{r}\right) $ is also related to the Aharonov-Bohm phase in
quantum mechanics. For the magnetic flux density in electromagnetics we have 
\begin{equation}
\mathbf{B}\left( \mathbf{r}\right) =\nabla _{\mathbf{r}}\times \mathbf{A}_{%
\text{mag}}(\mathbf{r}).
\end{equation}%
By analogy to electromagnetics, when $2\leq \dim \left( \mathbf{R}\right)
\leq 3$ (other dimensional are considered below) a vector wave can be
obtained from the Berry vector potential $\mathbf{A}_{n}(\mathbf{R})$ as 
\begin{align}
\mathbf{F}_{n}(\mathbf{R})& =\nabla _{\mathbf{R}}\times \mathbf{A}_{n}(%
\mathbf{R}) \\
& =i\nabla _{\mathbf{R}}\times \left\langle n(\mathbf{R})\right\vert \nabla
_{\mathbf{R}}\left\vert n(\mathbf{R})\right\rangle =i\left\langle \nabla _{%
\mathbf{R}}n(\mathbf{R})\right\vert \times \left\vert \nabla _{\mathbf{R}}n(%
\mathbf{R})\right\rangle \label{BCV}.
\end{align}%
This field is called the \textit{Berry curvature}, and is obviously gauge-independent. It is a geometrical property of the parameter space, and can be viewed as an
effective magnetic field in parameter space; just as the motion of a moving charge is perpendicular to the magnetic field ($\mathbf{F}_B=\mathbf{v}\times \mathbf{B}$), i.e., the curvature of the magnetic vector potential, the Berry curvature will induce transverse particle motion (an electronic or optical Hall effect). To continue the analogy, first
note that the magnetic flux can also be written as%
\begin{equation}
\Phi _{\text{mag}}=\int_{S}d\mathbf{S}\cdot \mathbf{B}\left( \mathbf{r}%
\right) ,  \label{MF2}
\end{equation}%
where (\ref{MF1}) and (\ref{MF2}) are equal via Stokes' theorem, i.e., 
\begin{equation}
\oint\nolimits_{C}d\mathbf{l}\cdot \mathbf{A}_{\text{mag}}\left( 
\mathbf{r}\right) =\int_{S}d\mathbf{S}\cdot \nabla _{\mathbf{r}}\times 
\mathbf{A}_{\text{mag}}(\mathbf{r})=\int_{S}d\mathbf{S}\cdot \mathbf{B}%
\left( \mathbf{r}\right) .
\end{equation}%
For Stokes' theorem to hold the fields must be nonsingular on and within the
contour $C$. Given that magnetic monopoles, which would serve as
singularities of the field, do not seem to exist, Stokes' theorem is valid
to apply in this case. One can similarly use Stokes' theorem to connect the
Berry phase and the Berry curvature,%
\begin{equation}
\gamma _{n}=\oint\nolimits_{C}d\mathbf{R}\cdot \mathbf{A}_{n}(\mathbf{R})%
\overset{\mathbf{?}}{\mathbf{=}}\int_{S}d\mathbf{S}\cdot \mathbf{F}_{n}(%
\mathbf{R})  \label{SL}
\end{equation}%
where $C$ and $S$ are a contour and surface in parameter space, and where
the right side could be called the \textit{Berry flux}. However, it must be
kept in mind that the relation (\ref{SL}) is not always valid, since for the
parameter-space fields $\mathbf{A}_{n}\left( \mathbf{R}\right) $, $\mathbf{F}%
_{n}\left( \mathbf{R}\right) $ singularities can occur, such that Stokes'
theorem does not generally hold\footnote{%
The obstruction to Stokes' theorem 
\begin{equation}
D=\frac{1}{\pi }\left[ \oint\nolimits_{C}d\mathbf{R}\cdot \mathbf{A}_{n}(%
\mathbf{R})\mathbf{-}\int_{S}d\mathbf{S}\cdot \mathbf{F}_{n}(\mathbf{R})%
\right] \neq 0
\end{equation}%
can lead to a $\mathbb{Z}_{2}$ invariant that characterizes the system \cite{FK}, \cite{M2}.}%
. Nevertheless, a gauge-independent Berry phase $\gamma _{n}$ can be
computed from the Berry flux integral,%
\begin{equation}
\gamma _{n}=\int_{S}d\mathbf{S}\cdot \mathbf{F}_{n}(\mathbf{R}),
\end{equation}
Stokes' theorem holding modulo $2 \pi$.

The above form of $\mathbf{F}_n$ (and of $\mathbf{A}_n$) can be inconvenient for numerical work since it involves derivatives of the state function (the problem this engenders is described below). In the following an alternative tensor formulation is shown, applicable for any dimension parameter space, and which also leads to a more convenient form for numerical computations. 

For $R^{\mu }$ and $R^{\nu }$ elements of $\mathbf{R}$, with $\mu, \nu$ covering all of $\mathbf{R}$, then the Berry curvature tensor can be defined as
\begin{equation}
F_{\mu \nu }^{n}=\frac{\partial }{\partial R^{\mu }}A_{n}^{\nu }-\frac{%
\partial }{\partial R^{\nu }}A_{n}^{\mu }=i\left[ \left\langle \frac{%
\partial }{\partial R^{\mu }}n(\mathbf{R})\right\vert \left. \frac{\partial 
}{\partial R^{\nu }}n(\mathbf{R})\right\rangle -\left\langle \frac{\partial 
}{\partial R^{\nu }}n(\mathbf{R})\right\vert \left. \frac{\partial }{%
\partial R^{\mu }}n(\mathbf{R})\right\rangle \right]   \label{Eq:19}
\end{equation}%
which serves as a generalization of the vector Berry curvature. We can also
write the Berry curvature tensor in terms of the Berry curvature vector; for 
$\dim \left( \mathbf{R}\right) =3$ 
\begin{equation}
F=-\mathbf{F}\times \mathbf{I}_{3\times 3}=\left[ 
\begin{array}{ccc}
0 & F_{z} & -F_{y} \\ 
-F_{z} & 0 & F_{x} \\ 
F_{y} & -F_{x} & 0%
\end{array}%
\right] 
\end{equation}%
where $\mathbf{I}_{3\times 3}$ is the identity. More generally, the Berry
curvature tensor $F_{\mu \nu }^{n}$ and vector $\mathbf{F}_{n}$ are related
by $F_{\mu \nu }^{n}=\epsilon _{\mu \nu \xi }(\mathbf{F}_{n})_{\xi }$ with $%
\epsilon _{\mu \nu \xi }$ the Levi-Civita anti-symmetry tensor.

Importantly, the Berry curvature tensor (\ref{Eq:19}) can be
also written as a summation over the eigenstates,%
\begin{equation}
F_{\mu \nu }^{n}=i\sum_{n^{\prime },~n^{\prime }\neq n}\frac{\left\langle
n\right\vert \partial H/\partial R^{\mu }\left\vert n^{\prime }\right\rangle
\left\langle n^{\prime }\right\vert \partial H/\partial R^{\nu }\left\vert
n\right\rangle -\left\langle n\right\vert \partial H/\partial R^{\nu
}\left\vert n^{\prime }\right\rangle \left\langle n^{\prime }\right\vert
\partial H/\partial R^{\mu }\left\vert n\right\rangle }{(E_{n}-E_{n^{\prime
}})^{2}}.  \label{Eq:15}
\end{equation}%

To obtain this result, note that
\begin{equation}
A_n^{\nu,\mu} = i \left< n(\mathbf{R})\right|\frac{\partial}{\partial R^{\nu,\mu}}\left|n(\mathbf{R})\right>,\\	\frac{\partial}{\partial R^{\alpha}} A_n^{\nu,\mu} = i  \left< \frac{\partial}{\partial R^{\alpha}} n(\mathbf{R}) | \frac{\partial}{\partial R^{\nu,\mu}} n(\mathbf{R})\right> + i  \left<  n(\mathbf{R}) | \frac{\partial}{\partial R^{\alpha} \partial R^{\nu,\mu}} n(\mathbf{R})\right>. \notag 
\end{equation}
Inserting into (\ref{Eq:19}),
\begin{equation}\label{Eq:10}
	F_{\mu \nu}^n = \frac{\partial}{\partial R^{\mu}}A_n^{\nu} - \frac{\partial}{\partial R^{\nu}}A_n^{\mu} = i \left[ < \frac{\partial}{\partial R^{\mu}} n(\mathbf{R})  | \frac{\partial}{\partial R^{\nu}} n(\mathbf{R}) > - < \frac{\partial}{\partial R^{\nu}} n(\mathbf{R}) | \frac{\partial}{\partial R^{\mu}} n(\mathbf{R})> \right].
	\end{equation}
Then,
\[
H(\mathbf{R})\left\vert n(\mathbf{R})\right\rangle =E_{n}(%
\mathbf{R})\left\vert n(\mathbf{R})\right\rangle \rightarrow
\partial H/\partial \mathbf{R}\left\vert n\right\rangle +H\left\vert
\partial n/\partial \mathbf{R}\right\rangle =\partial E_{n}/\partial 
\mathbf{R}\left\vert n\right\rangle +E_{n}\left\vert \partial n/\partial 
\mathbf{R}\right\rangle 
\]%
and because of the adiabatic assumption we can drop the first term on the right side. By changing the kets to bras and multiplying
by $\left\vert n^{\prime }\right\rangle $ we obtain
\begin{align}
& \left\langle n\right\vert \partial H/\partial \mathbf{R}\left\vert
n^{\prime }\right\rangle +\left\langle \partial n/\partial \mathbf{R}%
\right\vert H\left\vert n^{\prime }\right\rangle =E_{n}\left\langle \partial
n/\partial \mathbf{R}|n^{\prime }\right\rangle   \nonumber \\
& \left\langle n\right\vert \partial H/\partial \mathbf{R}\left\vert
n^{\prime }\right\rangle =(E_{n}-E_{n^{\prime }})\left\langle \partial
n/\partial \mathbf{R}|n^{\prime }\right\rangle ,~~n\neq
n^{\prime }  
\end{align}%
from which the result (\ref{Eq:15}) follows.

In general, the fact that the wavefunction
itself is gauge-dependent, creates an issue in computing $\mathbf{A}_{n}$
via (\ref{SL2}), and therefore $\mathbf{F}_{n}$ via (\ref{BCV}), since for slightly
different $\mathbf{R}$ values a numerical algorithm will generally output
eigenstates with unrelated phases, thus prohibiting one from numerically
taking the required derivative of the eigenvector unless care is taken to
make sure the phases are smooth. However, (\ref{Eq:15}) only requires the
derivative of the Hamiltonian, and so any numerical phase will disappear in
taking the inner product.  

Similar manipulations lead to the Berry phase%
\begin{equation}
\gamma _{n}=i\int d\mathbf{S\cdot }\sum_{n^{\prime },~n^{\prime }\neq n}%
\frac{\left\langle n\right\vert \nabla _{\mathbf{R}}H\left\vert n^{\prime
}\right\rangle \times \left\langle n^{\prime }\right\vert \nabla _{\mathbf{R}%
}H\left\vert n\right\rangle }{(E_{n}-E_{n^{\prime }})^{2}}. \label{Eq:15a}
\end{equation}%
Although the previous forms depend only on a certain state and it's
derivative, the forms (\ref{Eq:15}) and (\ref{Eq:15a}) involving summation over $n^{\prime }\neq n$
show that the Berry properties can be thought of as resulting from
interactions between the $n$th state and all other states -- it is a global
property of the bandstructure.  

Equations (\ref{Eq:15}) and (\ref{Eq:15a}) show that the Berry curvature becomes singular if
two energy levels $E_{n}$ and $E_{n^{\prime }}$ are brought together at a
certain value of $\mathbf{R}$, resulting in the \textquotedblleft Berry
monopole.\textquotedblright\ In fact, the adiabatic approximation assumes no
degeneracies on the path $C$, but degeneracies can occur within the space
enclosed by the path.

It is easy to show the conservation law\footnote{%
When we also do a summation over $n$ then in fact we are taking all of the
non-diagonal elements of the operators $\partial H/\partial R^{\mu ,v}$ into
account. Then, for any states like $\left\vert n\right\rangle \equiv
\left\vert a\right\rangle ,~\left\vert n^{\prime }\right\rangle \equiv
\left\vert b\right\rangle ;~a\neq b$ there are another set of states like $%
\left\vert n\right\rangle \equiv \left\vert b\right\rangle ,~\left\vert
n^{\prime }\right\rangle \equiv \left\vert a\right\rangle ;~a\neq b$ such
that $\left\langle n\right\vert \partial H/\partial R^{\mu }\left\vert
n^{\prime }\right\rangle \left\langle n^{\prime }\right\vert \partial
H/\partial R^{\nu }\left\vert n\right\rangle \mid _{(n=a,n^{\prime
}=b)}=\left\langle a\right\vert \partial H/\partial R^{\mu }\left\vert
b\right\rangle \left\langle b\right\vert \partial H/\partial R^{\nu
}\left\vert a\right\rangle $ and $\left\langle n\right\vert \partial
H/\partial R^{\nu }\left\vert n^{\prime }\right\rangle \left\langle
n^{\prime }\right\vert \partial H/\partial R^{\mu }\left\vert n\right\rangle
\mid _{(n=b,n^{\prime }=a)}=\left\langle b\right\vert \partial H/\partial
R^{\nu }\left\vert a\right\rangle \left\langle a\right\vert \partial
H/\partial R^{\mu }\left\vert b\right\rangle =\left\langle a\right\vert
\partial H/\partial R^{\mu }\left\vert b\right\rangle \left\langle
b\right\vert \partial H/\partial R^{\nu }\left\vert a\right\rangle $ cancel
out each other at the numerator.}%
\begin{equation}
\sum_{n}F_{\mu \nu }^{n}=0, \label{cons}
\end{equation}%
which demonstrates, among other things, that the sum over all bands of the
Berry curvature is zero.

As discussed later for the photonic case, under time-reversal (TR) and inversion (I) symmetries, 
\begin{align}
\text{TR}& \text{: \ }\mathbf{F}\left( -\mathbf{k}\right) =-\mathbf{F}\left( 
\mathbf{k}\right)  \\
\text{I}& \text{: \ }\mathbf{F}\left( -\mathbf{k}\right) =\mathbf{F}\left( 
\mathbf{k}\right)  \\
\text{TR+I}& \text{: \ }\mathbf{F}\left( \mathbf{k}\right) =\mathbf{0}.
\end{align}%
Therefore, a non-zero Berry curvature will exist when either TR or I are broken. 

\subsection{Chern number, bulk-edge correspondence, and topologically
protected edge states}

From elementary electromagnetics, Gauss's law relates the total flux over a
closed surface $S$ to the total charge within the surface, 
\begin{equation}
\oint_{S}\varepsilon _{0}\mathbf{E}\left( \mathbf{r}\right) \cdot d\mathbf{S}%
=Q^{T}=mq,
\end{equation}%
where, assuming identical charged particles, $m$ is the number of particles
and $q$ the charge of each particle (although often approximated as a
continuum, $Q^{T}$ is quantized). To keep things simple we'll assume a
monopole charge of strength $mq$ located at the origin. The electric field
is given by Coulombs law, 
\begin{equation}
\mathbf{E}=\left( \frac{mq}{4\pi \varepsilon _{0}}\right) \frac{\mathbf{r}}{%
r^{3}}.
\end{equation}%
The analogous magnetic form 
\begin{equation}
\oint_{S}\mathbf{B}\left( \mathbf{r},t\right) \cdot d\mathbf{S}=0
\end{equation}%
indicates that there are no magnetic monopoles. However, in parameter space
the flux integral over a closed manifold of the Berry curvature is quantized
in units of $2\pi $, indicating the number of Berry monopoles (degeneracies)
within the surface, 
\begin{equation}
\oint_{S}d\mathbf{S}\cdot \mathbf{F}_{n}(\mathbf{R})=2\pi m_{n}=2\pi C_{n}
\end{equation}%
where $m_{n}=C_{n}$ is an integer for the $n$th band known as the \textit{%
Chern number}. The Chern number can be seen to be the flux associated with a
Berry monopole of strength $2\pi C_{n}$, leading to the Berry curvature  
\begin{equation}
\mathbf{F}_n=\left( \frac{C_n}{2}\right) \frac{\mathbf{k}}{k^{3}}.
\end{equation}%
The Berry monopole plays the role of source/sink of Berry curvature $\mathbf{%
F}$, just the electric charge monopole $mq$ servers as a source/sink of
electric field $\mathbf{E\propto r/}r^{3}$. The Chern number can also be
written in terms of the gauge form. For two dimensions, e.g., $\mathbf{R}=\mathbf{k}=\left( k_{x},k_{y}\right) ,$%
\begin{equation}
C_{n}=\frac{1}{2\pi }\int_{S}dk_{x}dk_{y}F_{xy}^{n}=\frac{1}{2\pi }%
\int_{S}dk_{x}dk_{y}\left( \frac{\partial }{\partial k_{x}}A_{n}^{y}-\frac{%
\partial }{\partial k_{y}}A_{n}^{x}\right) ,\label{CNNF}
\end{equation}%
or, from (\ref{Eq:15}),%
\begin{equation}
C_{n}=\frac{i}{2\pi }\int_{S}dk_{x}dk_{y}\sum_{n^{\prime },~n^{\prime }\neq
n}\frac{\left\langle n\right\vert \partial H/\partial k_{x}\left\vert
n^{\prime }\right\rangle \left\langle n^{\prime }\right\vert \partial
H/\partial k_{y}\left\vert n\right\rangle -\left\langle n\right\vert
\partial H/\partial k_{y}\left\vert n^{\prime }\right\rangle \left\langle
n^{\prime }\right\vert \partial H/\partial k_{x}\left\vert n\right\rangle }{%
(E_{n}-E_{n^{\prime }})^{2}},  \label{CNF}
\end{equation}%
which is a form used later for computation. 

Importantly, the Chern number is topologically invariant, meaning it is
unaffected by smooth deformations in the surface that preserve topology
(e.g., for a real-space surface, a teacup deforming into a torus). Moreover,
the sum $\sum_{n}C_{n}$ over all energies or bands $n$ is zero (this comes
from the curvature conservation equation (\ref{cons})), which plays a role
in what is know as bulk-edge correspondence. This is an extremely important
point in understanding the most significant aspect of Topological Insulators
(TIs), which is backscattering-protected edge propagation. Note that in the presence of TR but with I broken, integration over the entire BZ yields zero Chern number for each band, whereas in the presence of I but with TR broken, the band Chern number will generally be nonzero.

In periodic media (e.g., for electrons, in a crystalline solid, and for
photons, EM waves in a photonic crystal), the Berry phase $\gamma _{n}$ is a
geometric (in parameter space) attribute of the $n$th band. The Brillouin
zone is equivalent to a torus, forming the closed surface over which the
Berry curvature of any non-degenerate band is integrated to compute the
Chern number $C_{n}$ for that band.

\subsubsection{Example - Two-level systems in parameter space} \label{example}
A common example that demonstrates Berry phase, curvature, and Chern number
concepts is cyclic evolution of a two level system \cite{griffiths}, such as electronic spin or valley pseudospin. Consider the evolution of spin for an electron at the
origin immersed in a magnetic field. Let the tip of the magnetic field
vector trace out a closed curve on a sphere of radius $r=B_{0}$, $\mathbf{B}=B_{0}\widehat{\mathbf{r}}\left( t\right) $ -- in this case the magnetic
field itself plays the role of parameter space, $\mathbf{R}=\left(
B_{x},B_{y},B_{z}\right) $. The Hamiltonian is the projection of spin onto the magnetic field coordinates,
\begin{equation}
H=-\mathbf{\mu }\cdot \mathbf{B}=-\gamma \mathbf{B}\cdot \mathbf{S},
\end{equation}%
where $\mathbf{\mu }$ is the magnetic moment ($\mathbf{\mu }=\gamma \mathbf{S%
}$), $\gamma $ is the gyromagnetic ratio ($\gamma =q_{e}/2mc$ for orbital
electrons, $\gamma =gq_{e}/2mc$ where $g\sim 2$), $\mathbf{S}=\left( \hslash
/2\right) \mathbf{\sigma }$, and $\mathbf{\sigma }=(\sigma_x ,\sigma_y ,\sigma_z)$ are the Pauli matrices%
\begin{equation}
\sigma _{x}=\left( 
\begin{array}{cc}
0 & 1 \\ 
1 & 0%
\end{array}%
\right) ,\ \ \sigma _{y}=\left( 
\begin{array}{cc}
0 & -i \\ 
i & 0%
\end{array}%
\right) ,\ \ \sigma _{z}=\left( 
\begin{array}{cc}
1 & 0 \\ 
0 & -1%
\end{array}%
\right) .
\end{equation}%
Whereas for magnetic moment due to a current loop the torque $\mathbf{T}=%
\mathbf{\mu }\times \mathbf{B}$ acts to align $\mathbf{\mu }$ and $\mathbf{B}
$, for angular momentum and spin the torque causes a precession of $\mathbf{\mu }$
around $\mathbf{B}$, with the precession frequency $\mathbf{\omega }%
_{0}=-\gamma \mathbf{B}$.

Writing the Hamiltonian as $H=\mathbf{h} \cdot \sigma$, where $\mathbf{h}=h \widehat{\mathbf{r}}$ with $h=-\gamma \hslash B_0 /2$, 
\begin{equation}
\mathbf{\sigma }_{r}=h \widehat{\mathbf{r}} \cdot \mathbf{\sigma }=h\left( 
\begin{array}{cc}
\cos \theta  & e^{-i\phi }\sin \theta  \\ 
e^{i\phi }\sin \theta  & -\cos \theta 
\end{array}%
\right) .
\end{equation}%
Eigenvalues and eigenvectors satisfy $\left( h\mathbf{\sigma }\cdot \widehat{%
\mathbf{r}}\right) \left\vert u\right\rangle =\lambda \left\vert
u\right\rangle $ and are 
\begin{align}
\lambda ^{+}& =+h\text{, \ }\left\vert u^{+}\right\rangle =\left( 
\begin{array}{c}
e^{-i\phi }\cos \left( \theta /2\right)  \\ 
\sin \left( \theta /2\right) 
\end{array}%
\right) , \\
\lambda ^{-}& =-h,\ \ \left\vert u^{-}\right\rangle =\left( 
\begin{array}{c}
e^{-i\phi }\sin \left( \theta /2\right)  \\ 
-\cos \left( \theta /2\right) 
\end{array}%
\right) .
\end{align}%

The Berry potential is $\mathbf{A}^{\pm }(\mathbf{R})=i\left\langle u^{\pm
}\right\vert \nabla _{\mathbf{R}}\left\vert u^{\pm }\right\rangle $ where $%
\mathbf{R}=\left( r,\theta ,\phi \right) $. Since the gradient is $\mathbf{%
\nabla }f=\frac{\partial f}{\partial h}\widehat{\mathbf{h}}+\frac{1}{h}\frac{%
\partial f}{\partial \theta }\,\widehat{\mathbf{\theta }}+\frac{1}{h\sin
\theta }\frac{\partial f}{\partial \phi }\,\widehat{\mathbf{\phi }}$,
\begin{align}
A_{\theta }^{-}& =i\left\langle u^{-}\right\vert \frac{\partial }{h\partial
\theta }\left\vert u^{-}\right\rangle =0,\ \ A_{\theta }^{+}=i\left\langle
u^{+}\right\vert \frac{\partial }{h\partial \theta }\left\vert
u^{+}\right\rangle =0, \\
A_{\phi }^{-}& =i\left\langle u^{-}\right\vert \frac{1}{h\sin \theta }\frac{%
\partial }{\partial \phi }\left\vert u^{-}\right\rangle =\frac{\sin ^{2}%
\frac{1}{2}\theta }{h\sin \theta },\ \ A_{\phi }^{+}=i\left\langle
u^{+}\right\vert \frac{1}{h\sin \theta }\frac{\partial }{\partial \phi }%
\left\vert u^{+}\right\rangle =\frac{\cos ^{2}\frac{1}{2}\theta }{h\sin
\theta }.
\end{align}   
The Berry curvature is 
\begin{equation}
\mathbf{F}^{\pm }=\frac{1}{h\sin \theta }\left( \frac{\partial }{\partial
\theta }\left( A_{\phi }^{-}\sin \theta \right) \right) \widehat{\mathbf{h}}%
=\pm \frac{1}{2}\frac{\mathbf{h}}{h^{3}},
\end{equation}%
which is the field generated by a monopole (in parameter space) at the
origin. Obviously, the Berry curvature has a singularity at $h=0$ (i.e., $B_0=0$). This singularity
is due to a degeneracy between $\lambda ^{+}=h$ and $\lambda ^{-}=-h$ at the
origin of parameter space ($h=0$); these degeneracy points serve as
\textquotedblleft sources\textquotedblright\ (for $\lambda ^{-}$, producing
monopole strength $1/2$) and \textquotedblleft sinks\textquotedblright\ (for 
$\lambda ^{+}$, producing monopole strength $-1/2$) of Berry curvature (like
any monopole). Similar to Gauss's law, when we integrate around a closed
surface containing the monopole we get an integer (here we call it the
Chern number). 
The Chern number is  
\begin{equation}
C=\oint\nolimits_{S}\mathbf{F}\cdot d\mathbf{S}=\pm \oint\nolimits_{S}\frac{1%
}{2}\frac{\mathbf{h}}{h^{3}}\cdot \widehat{\mathbf{h}}h^{2}\sin \theta \
d\theta d\phi =\pm 2\pi =\pm \frac{1}{2}\Omega .
\end{equation}
Given that 
\begin{equation}
\gamma _{n}=\int_{S}d\mathbf{S}\cdot \mathbf{F}_{n}(\mathbf{R}),
\end{equation}%
the Berry phase can be viewed as $1/2$ the solid angle
subtended by the closed path,%
\begin{equation}
\gamma =\pm \int_{S}\frac{1}{2}\frac{\mathbf{h}}{h^{3}}\cdot \widehat{%
\mathbf{h}}h^{2}\sin \theta \ d\theta d\phi =\pm \frac{1}{2}\int_{S}\sin
\theta \ d\theta d\phi =\pm \frac{1}{2}\Omega .
\end{equation}%
In fact, in general the answer is $s\Omega $, where $\Omega $ is the solid
angle and $s$ is the spin. 

As a related example, but considering momentum space as the parameter space,
consider a two-dimensional material with a hexagonal lattice and two
inequivalent Dirac points, such as graphene. The two in-equivalent Dirac points lead to two different valleys sufficiently separated in momentum space so that inter-valley transitions can usually be ignored. In the absence of a magnetic field, graphene respects both TR and I symmetry (and, hence, has zero Berry curvature, but posses a Berry phase). The tight-binding Hamiltonian
near the $K$ and $K^{\prime }$ valleys has the same form as the magnetic field problem considered above, 
\begin{equation}
H=\tau \hslash v_{F}\mathbf{q}_{\tau }\cdot \boldsymbol{\sigma }\text{,}
\end{equation}%
where $\mathbf{q}_{\tau }=\left( q_{x},\tau q_{y}\right) $ is momentum
relative to the degeneracy point, $\tau =\pm 1$ is the valley index (for the $%
K$ and $K^{\prime }$ points, respectively), and $s=\pm 1$ is the conduction
and valance band index. In this case the two inequivalent valleys play the role of spin,
and so here $\mathbf{\sigma }$ represents pseudospin, not actual spin. In the $K$ valley
conduction band, the projection of pseudospin onto momentum is parallel to
momentum, whereas in the valance band it is antiparallel to momentum. In the 
$K^{\prime }$ valley these are reversed. 

The eigenvalues and eigenvectors are \footnote{The often used eigenfunctions
\begin{equation}
\left\vert u^{s,\tau }\right\rangle =\frac{1}{\sqrt{2}}\left( 
\begin{array}{c}
e^{-i\phi _{\mathbf{q}}/2} \\ 
s\tau e^{i\phi _{\mathbf{q}}/2}%
\end{array}%
\right) e^{i\mathbf{q}_{\tau }\cdot \mathbf{r}},
\end{equation}
are not appropriate since they are not single-valued \cite{BPG}     }
\begin{equation}
\lambda ^{s,\tau }=s\hslash v_{F}\left\vert \mathbf{q}_{\tau }\right\vert 
\text{, \ }\left\vert u^{s,\tau }\right\rangle =\frac{1}{\sqrt{2}}\left( 
\begin{array}{c}
1 \\ 
s\tau e^{i\phi _{\mathbf{q}}}%
\end{array}%
\right) e^{i\mathbf{q}_{\tau }\cdot \mathbf{r}},
\end{equation}%
where $\left( q_{x}+iq_{y}\right) =\left\vert \mathbf{q}\right\vert e^{i\phi
_{\mathbf{q}}}$, $\phi _{\mathbf{q}}$ being the angle between $\mathbf{q}$
and the $x$-axis, $\phi _{\mathbf{q}}=\tan ^{-1}\left( q_{y}/q_{x}\right) $.
Then, since the gradient is $\mathbf{\nabla }f=\frac{\partial f}{\partial q}%
\widehat{\mathbf{q}}+\frac{1}{q}\frac{\partial f}{\partial \phi _{\mathbf{q}}%
}\,\widehat{\mathbf{\phi _{\mathbf{q}} }}+\frac{\partial f}{\partial z}\,\widehat{\mathbf{z%
}}$, 
\begin{equation}
A_{\phi _{\mathbf{q}}}^{s,\tau }=i\left\langle u^{s,\tau }\right\vert \frac{1%
}{q}\frac{\partial }{\partial \phi _{\mathbf{q}}}\left\vert u^{s,\tau
}\right\rangle =-\frac{1}{2}\frac{1}{q},
\end{equation}%
and \cite{SA}  
\begin{equation}
\gamma _{n}=\oint\nolimits_{C}d\mathbf{q}\cdot \mathbf{A}(\mathbf{q}%
)=\int_{0}^{2\pi }-\frac{1}{2}\frac{1}{q}qd\phi _{\mathbf{q}}=-\pi .
\end{equation}%
The Berry phase of $\pi$ manifests itself in various ways, including in the suppression of backscattering and a phase shift in Shubinikov de-Haas
(SdH) oscillations \cite{YZK}. Note, however, that
\begin{equation}
\mathbf{F}=\frac{1}{q}\frac{\partial }{\partial q}\left( qA_{\phi _{\mathbf{q%
}}}\right) \widehat{\mathbf{z}}=-\frac{1}{q}\frac{\partial }{\partial q}%
\left( \frac{1}{2}\right) \widehat{\mathbf{z}}=0,
\end{equation}%
so that in (non-gapped) graphene, which has both time-reversal and inversion symmetry, the Berry
curvature vanishes. This is considered further in Section \ref{VHE}. 

Finally, let us consider the optical fiber wound into a helix as mentioned previously. For a linearly polarized optical field we have $\left( \mathbf{\sigma }\cdot \mathbf{%
k}\right) \left\vert u\right\rangle =\lambda \left\vert u\right\rangle $, which is the projection of spin (e.g., polarization) onto the direction of
momentum. This has the same general form as the magnetic field problem.
However, for photons (spin 1), the spin matrices are different and Berry
phase is equal to the solid angle, $\gamma =\Omega $.

\subsubsection{Generalized equations of motion in a crystal }

For electronic applications, an important aspect of the Berry curvature is
that it plays a role in the equations of motion \cite{Xiao}. In a crystal, the usual expression for the velocity, $\mathbf{v} =\overset{\cdot }{\mathbf{r}}={\partial \varepsilon }/{%
\hslash \partial \mathbf{k}}$, is modified by a non-zero Berry curvature,
\begin{align}
\mathbf{v}& =\overset{\cdot }{\mathbf{r}}=\frac{\partial \varepsilon }{%
\hslash \partial \mathbf{k}}-\overset{\cdot }{\mathbf{k}}\times \mathbf{F} \\
\hslash \overset{\cdot }{\mathbf{k}}& =-e\left( \mathbf{E}+\overset{\cdot }{%
\mathbf{r}}\times \mathbf{B}\right) ,
\end{align}%
where $\mathbf{k}$ is crystal momentum. The term $\overset{\cdot }{\mathbf{k}%
}\times \mathbf{F}$ is the anomalous (Hall) velocity due to Berry curvature, and is
transverse to the momentum. If we ignore the magnetic field contribution,
then $\hslash \overset{\cdot }{\mathbf{k}}=-e\mathbf{E}$ and 
\begin{equation}
\mathbf{v}=\overset{\cdot }{\mathbf{r}}=\frac{\partial \varepsilon }{\hslash
\partial \mathbf{k}}+\frac{e}{\hslash }\mathbf{E}\times \mathbf{F.}
\end{equation}%
In the photonic case, analogous equations of motion for the geometrical optics field are presented in \cite{Hall}.  

\subsubsection{The Effect of the Hall Velocity: Quantum Hall and Valley Hall Effects}

Obviously, the anomalous Hall velocity will give rise to an anomalous Hall
current, 
\begin{align}
\mathbf{J}& =q_{e}\sum_{n,\tau}\int \mathbf{v}_{e}\left( \mathbf{k}\right)
f\left( E\left( \mathbf{k}\right) \right) \left[ d\mathbf{k}\right]  \\
& \rightarrow \mathbf{J}_{\text{Hall}}=\frac{e^{2}}{\hslash }\sum_{n,\tau}\int
\left( \mathbf{E}\times \mathbf{F}\right) f\left( E\left( \mathbf{k}\right)
\right) \left[ d\mathbf{k}\right]  \\
& =\left( \mathbf{E}\times \mathbf{I}\right) \cdot \frac{e^{2}}{\hslash }%
\sum_{n,\tau}\int \mathbf{F}f\left( E\left( \mathbf{k}\right) \right) \left[ d%
\mathbf{k}\right]  \\
& =\mathbf{E}\cdot \left( \mathbf{I}\times \frac{e^{2}}{\hslash }%
\sum_{n,\tau}\int \mathbf{F}f\left( E\left( \mathbf{k}\right) \right) \left[ d%
\mathbf{k}\right] \right) 
\end{align}%
where $\tau$ is spin and $\left[ d\mathbf{k}\right] =\frac{d^{d}k}{\left( 2\pi \right) ^{d}}$
in $d$ dimensions (and we have ignored any magnetic field effect). Therefore, the Hall conductivity tensor is 
\begin{equation}
\underline{\mathbf{\sigma }}=\mathbf{I}\times \frac{e^{2}}{\hslash }%
\sum_{n,\tau}\int \mathbf{F}f\left( E\left( \mathbf{k}\right) \right) \left[ d%
\mathbf{k}\right] 
\end{equation}%
whenever the Berry curvature is nonzero. In fact, unlike the usual current,
the Hall current will be non-zero even when $f=f_{0}$, the equilibrium Fermi
distribution \footnote{Under typical perturbation conditions where a small electric field causes $f=f_0 +\delta f$, $\delta f <<f$, Hall current will be associated with both terms, and the term associated with $\delta f$ is second-order small since $\delta f$ itself is proportional to the electric field}. In the following we restrict attention to insulators, with the Fermi-level in the band gap. We further only consider 2D materials, where $\mathbf{F}=\widehat{\mathbf{z}}F_{z}$ ($z$ out-of-plane). Then, e.g.,
\begin{equation}
\sigma_{x,y}= \frac{e^{2}}{\hslash }%
\sum_{n,\tau}\int F _{z}f\left( E\left( \mathbf{k}\right) \right) \left[ d%
\mathbf{k}\right] 
\end{equation}%
where we sum over filled bands below the bandgap. Let $\mathbf{E}=E_{0}\widehat{\mathbf{y}}$; from the tensor 
\begin{equation}
F=-\mathbf{F}\times \mathbf{I}_{3\times 3}=\left[ 
\begin{array}{ccc}
0 & F _{z} & -F _{y} \\ 
-F _{z} & 0 & F _{x} \\ 
F _{y} & -F _{x} & 0%
\end{array}%
\right] 
\end{equation}%
we note that $F_{xy}=F_{z}$. Assuming a single valance band and spin component,
\begin{equation}
\mathbf{J}_{\text{Hall}}=-\widehat{\mathbf{x}}\left( \frac{q_{e}^{2}}{%
\hslash }\int F _{xy}\left[ d\mathbf{k}\right] \right)
E_{0}=-\widehat{\mathbf{x}}\sigma _{xy}E_{y}
\end{equation}%
where the Hall conductivity is%
\begin{equation}
\sigma _{xy}=\left( \frac{q_{e}^{2}}{\hslash }\int F _{xy}\left[ d\mathbf{k}\right] \right) , 
\end{equation}%
such that \cite{TKNN} 
\begin{equation}
\sigma _{xy}=\frac{q_{e}^{2}}{\hslash }C_{n},
\end{equation}%
where the Chern number is%
\begin{equation}
C_{n}=\int F _{xy}\left[ d\mathbf{k}\right] .
\end{equation}%

\subsubsection{Valley Hall Effect} \label{VHE}

Although graphene respects both TR and I symmetry (and, hence, has zero Berry curvature), if we consider materials like gapped graphene (gap opening by, say, an applied strain) or MoS2, inversion symmetry will be broken and Berry curvature will be nonzero. In this case, the effective Hamiltonian is

\begin{equation}
H=at\mathbf{q}_{\tau }\cdot \boldsymbol{\sigma }+\frac{\Delta }{2}\sigma
_{z}-\nu \tau \frac{\sigma _{z}-1}{2}\hat{s}_{z}\text{,}
\end{equation}%
where $\mathbf{q}_{\tau }=\left( \tau q_{x},q_{y}\right) $, $\tau =\pm $ describes the
two valleys, $\Delta $ is the energy gap (effectively, a mass term), the last term accounts for spin-orbit
coupling (negligible in graphene systems) where $2\nu $ is the spin-orbit splitting. The Berry
curvature is \cite{Xiao1}
\begin{equation}
F_{z,c}=\tau \left( \frac{2a^{2}t^{2}\Delta ^{\prime }}{\left( \Delta
^{\prime 2}+4a^{2}q^{2}t^{2}\right) ^{3/2}}\right) 
\end{equation}%
in the conduction band, where $\Delta ^{\prime }=\Delta -\nu \tau s$ with $%
s=\pm 1$ a spin index. For
the valance band, $F_{v}\left( \mathbf{k}\right) =-F_{c}\left( \mathbf{k}%
\right) $.  
Upon the application of an electric field, electrons in different valleys will flow to opposite directions transverse to the
electric field, giving rise to a valley Hall current. As $\Delta \rightarrow 0$ the system exhibits Dirac cones, and the Berry curvature vanishes everywhere except at the
Dirac points where it diverges. For $\Delta>0$, the presence of Berry curvature leads to an anomalous velocity transverse to momentum, and to a Hall conductivity for each valley. However, it is important to note that the Berry curvatures in the two valleys have opposite signs, so that, upon summing over the two valleys, the net Hall conductivity (as seen by an electromagnetic field) will vanish. This can be overcome by pumping the material with circularly-polarized light tuned to the bandgap \cite{Xiao1}, which will preferentially populate one of the valley conduction bands, leading to a net optical Hall conductivity and chiral edge SPPs\cite{KN2016}.

\subsubsection{Bulk-Edge Correspondence}
\begin{figure}[ht]
\begin{center}
\noindent  \includegraphics[width=3in]{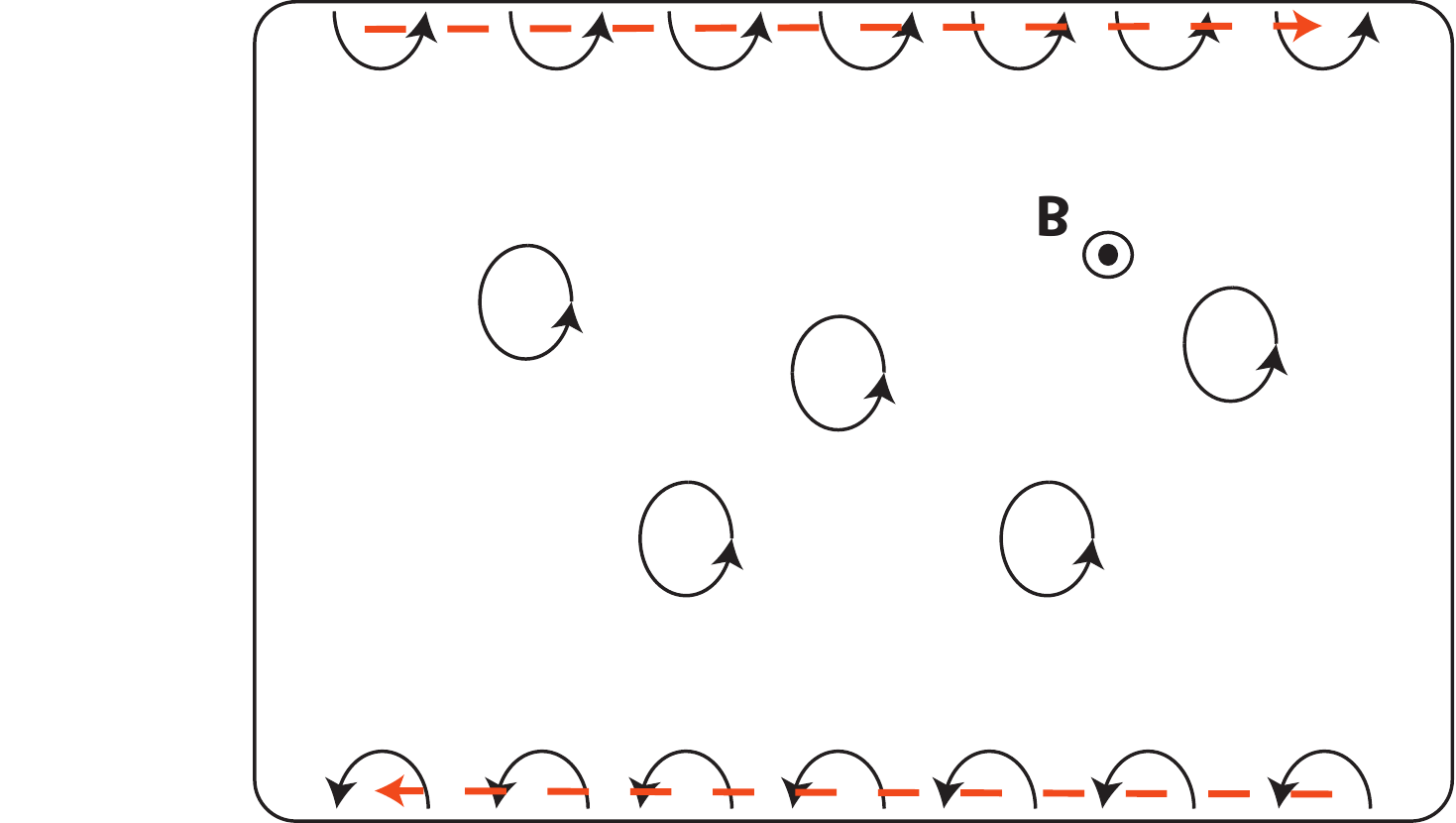}  
\end{center}
\caption{Depiction of electron orbits in an insulator in the presence of a magnetic field, and interrupted orbits at the edge.}
\label{hall}
\end{figure}

An aspect of Berry curvature that is of immense interest in both electronic and photonic applications is the presence of one-way edge modes that are topologically protected from backscattering. The idea of Hall conductivity in an insulator gives some intuition about the one-way nature of these modes. Consider a finite-sized rectangle of thin material, immersed in a perpendicular magnetic field as depicted in Fig. \ref{hall}. Bound electrons will circulate in response to the applied field, and those near the edge will have their orbits terminated by the edge\cite{BT}. The net effect is to have a uni-directional movement of electrons at the edge (orange arrows). 
The presence of a Hall conductivity (whether due to a magnetic field in the ordinary manner, or due to non-zero Berry curvature associated with broken TR or I symmetry) will elicit a similar response, although the response is quantized as described above. Thus, the bulk properties of the insulating material will result in a conducting edge state. This happes in both the electronic case, and the photonic case to be described below.

\begin{figure}[ht]
\begin{center}
\noindent  \includegraphics[width=4in]{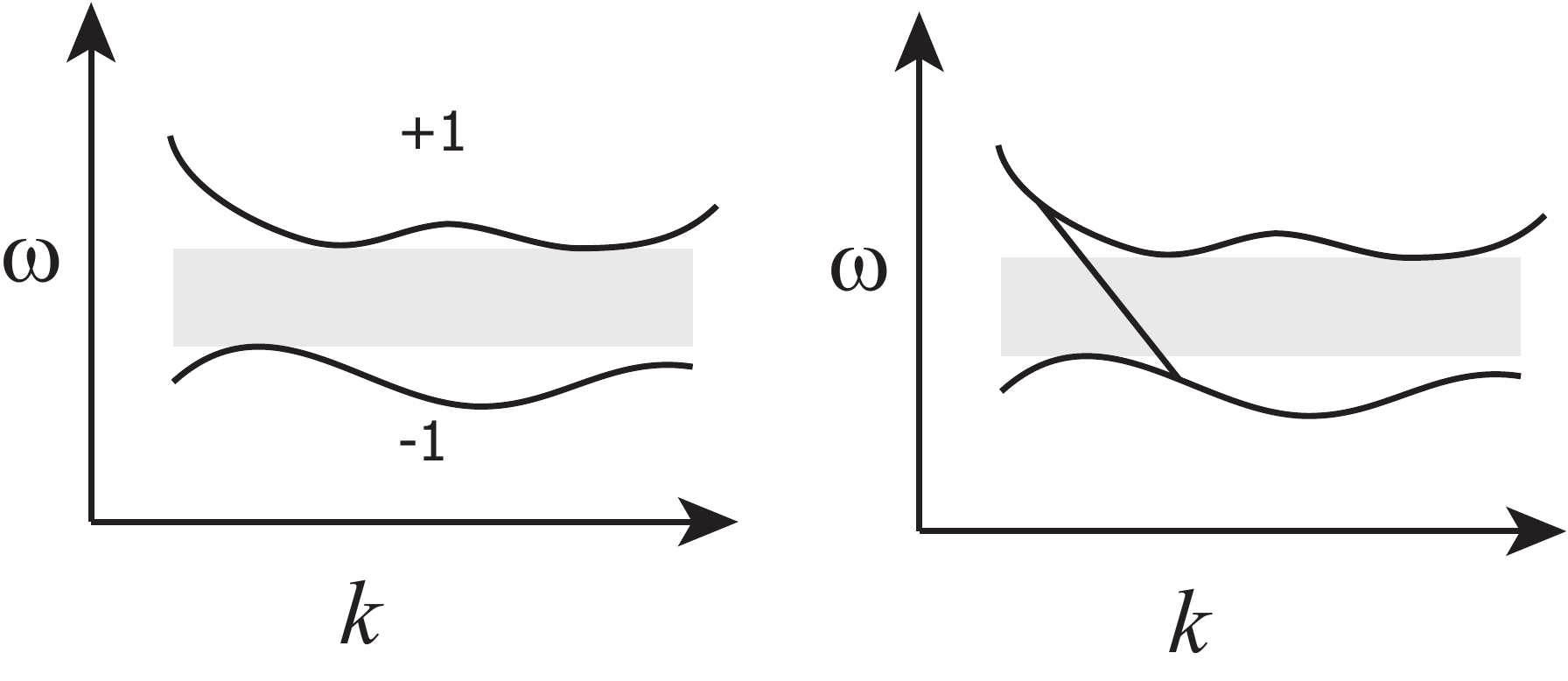}  
\end{center}
\caption{Bulk-edge correspondence. Materials with common bandgap and different Chern numbers share an interface where a uni-directional edge state closes the gap.}
\label{BE}
\end{figure}

Furthermore, consider that the Chern number and all
Berry properties are related to an infinite bulk material, which generates bandstructure. However,
in any practical application the material is finite, and has an interface
with another medium. Let's consider a planar interface between medium $1$
and medium $2$. Far from the interface, in each region, particles
(electrons, photons) will be governed by the respective Hamiltonians $%
H_{1,2} $. Let's assume that both materials share a common bandgap, and that 
$C_{\text{gap},1}=\sum_{n<n_{g}}C_{n}^{\left( 1\right) }$, the Chern number
sum over bands below the gap for material 1, and $C_{\text{gap}%
,2}=\sum_{n<n_{g}}C_{n}^{\left( 2\right) }$, the corresponding sum for
material $2$, differ, $C_{\text{gap},\Delta }=C_{2}-C_{1}\neq 0$. For some parameter value the
shared bandgap between the two mediums can close and then reopen. At the
closing point there is a degeneracy, and as the gap reopens it can be crossed by
a surface mode, as depicted in Fig. \ref{BE}. The edge-modes are
circularly-polarized (spin-polarized), and in periodic media are localized
to a few lattice constants from the material boundary.

The existence of the surface/edge state is necessitated by the bulk
material characteristics, and so is independent of interface details. Therefore, the
interface can possess discontinuities, defects, etc., which will not affect
the surface wave. It can also be seen that the fact that the surface/edge
states connect different energy levels indicates that they will have a group
velocity that has definite sign (i.e., one-directional propagation). Therefore, in
summary, the surface states are unidirectional and topologically protected
from backscattering.

\section{Electromagnetic Description - Berry quantities for photons} \label{SECEM}

Although the concept of Berry phase is general for any cyclic variation
through some parameter space $\mathbf{R}$, a primary application is to
periodic solid solid state systems (e.g., electrons in a crystal lattice),
although here we are primarily interested in the photonic analogous of those
systems, photonic topological insulators (PTIs) for both photonic crystals
and for continuum media.

For simplicity in observing the correspondence between Maxwell's
equations and the evolution equation (\ref{evol}), we first assume lossless and
dispersionless materials characterized by dimensionless real-valued
parameter $\overline{\epsilon },~\overline{\mu },~\overline{\xi },~\overline{%
\varsigma }$, representing permittivity, permeability and magneto-electric
coupling tensors. Although any real material must have frequency dispersion,
this simple model allows a straightforward conversion of various Berry
quantities from the electronic to the electromagnetic case. The inclusion of
both frequency and spatial dispersion will be discussed later.

In this case, Maxwell's equations are 
\begin{align}
& \nabla \times \mathbf{E}=-\mu _{0}\overline{\mu }\cdot \frac{\partial 
\mathbf{H}}{\partial t}-\frac{\overline{\varsigma }}{c}\cdot \frac{\partial 
\mathbf{E}}{\partial t}-\mathbf{J}_{m}  \notag  \label{Eq:17} \\
& \nabla \times \mathbf{H}=\epsilon _{0}\overline{\epsilon }\cdot \frac{%
\partial \mathbf{E}}{\partial t}+\frac{\overline{\xi }}{c}\cdot \frac{%
\partial \mathbf{H}}{\partial t}+\mathbf{J}_{e}.
\end{align}%
By defining the matrices%
\begin{align}
M& =\left( 
\begin{array}{cc}
\epsilon _{0}\overline{\epsilon } & \frac{1}{c}\overline{\xi } \\ 
\frac{1}{c}\overline{\varsigma } & \mu _{0}\overline{\mu }%
\end{array}%
\right) ,~N=\left( 
\begin{array}{cc}
0 & i\nabla \times \mathbf{I}_{3\times 3} \\ 
-i\nabla \times \mathbf{I}_{3\times 3} & 0%
\end{array}%
\right) ,~  \label{mat} \\[5pt]
f& =\left( 
\begin{array}{c}
\mathbf{E} \\ 
\mathbf{H}%
\end{array}%
\right) ,~g=\left( 
\begin{array}{c}
\mathbf{D} \\ 
\mathbf{B}%
\end{array}%
\right) =Mf,~J=\left( 
\begin{array}{c}
\mathbf{J}_{e} \\ 
\mathbf{J}_{m} \notag
\end{array}%
\right) \ 
\end{align}%
where M is Hermitian and real-valued, we can write Maxwell's equations in a
compact form \cite{Mario1},%
\begin{equation}
N\cdot f=i\left[ \frac{\partial g}{\partial t}+J\right] =i\left[ M\frac{%
\partial f}{\partial t}+\frac{\partial M}{\partial t}f+J\right] . \label{ME2}
\end{equation}%
Note that the units of the sub-blocks of $M$ differ (as do the dimensions of the 6-vectors $f$ and $g$) , and that $\epsilon, \xi, \varsigma$, and $\mu $ are dimensionless. In the absence of an external excitation ($J=0$) and assumption of
non-dispersive (instantaneous) materials, Maxwell's equations become%
\begin{equation}
i\frac{\partial f}{\partial t}=H_{cl}\cdot f  \label{ME1}
\end{equation}%
where $H_{cl}=M^{-1}\cdot N$, which has the same form as the evolution
equation (\ref{evol}) (e.g., the Schr\"{o}dinger equation) with $\hbar =1$, where the
operator $H_{cl}$ plays the role of a classical Hamiltonian. Because of this
similarity between Maxwell's equations and the evolution equation (\ref{evol}) it is
straightforward to extend the Berry potential concept to electromagnetic
energy (photons); rather then, say, electrons acquiring a Berry phase while
transversing a path in parameters space, photons will do the same (the
polarization of the photon plays the role of particle spin). In this case,
we define $f_{n}$ as a six-component eigenmode satisfying\footnote{%
In (\ref{mat})\ and (\ref{ME1}) $f$ is real-valued, unlike in the Schr\"{o}%
dinger equation where the wavefunction is complex-valued. The eigenfunctions in (\ref{PEE1})
$f_{n}$ are complex-valued.} 
\begin{equation}
H_{cl}\cdot f_{n}=E_{n}f_{n}  \label{PEE1}
\end{equation}%
where $E_{n}=\omega _{n}$. Under a suitable inner product (discussed below) $%
H_{cl}$ is Hermitian, and assuming the normalization condition $\left\langle
f_{n}|f_{m}\right\rangle =\delta _{nm}$, the Berry vector potential has the
same form as (\ref{AEM})%
\begin{equation}
\mathbf{A}_{n}=i\left\langle f_{n}|\nabla _{\mathbf{R}%
}f_{n}\right\rangle .  \label{PEE}
\end{equation}

If we assume a photonic crystal (periodic structure), $f_{n}$ has the Bloch
form $f_{n}\left( \mathbf{r}\right) =u_{n}\left( \mathbf{r}\right) e^{i%
\mathbf{kr}}$, where $u_{n}\left( \mathbf{r}\right) $ is the periodic Bloch
function and $\mathbf{k}$ is the Block wavevector. In this case, $\nabla _{%
\mathbf{R}}=\nabla _{\mathbf{k}}$ operates over parameter space $\mathbf{k}%
=(k_{x},~k_{y},~k_{z})$ and 
\begin{equation}\label{PEE3}
\mathbf{A}_{n}=i\left\langle u_{n}|\nabla _{\mathbf{k}}u_{n}\right\rangle
\end{equation} 
where the inner product is%
\begin{equation}
\left\langle u_{n}\left\vert u_{m}\right. \right\rangle =\frac{1}{2}\int_{%
\text{BZ}}u_{n}^{\ast }\left( \mathbf{r}\right) M\left( \mathbf{r}\right)
u_{m}\left( \mathbf{r}\right) d\mathbf{r.}  \label{IP1}
\end{equation}%
This is the dispersionless special case of the result presented in \cite{Haldane} (see (41) in that reference), generalized to account for magnetoelectric
coupling parameters in $M$.

The extension to dispersive media (i.e., real materials) would seem
difficult since the simple product $g=Mf$ in (\ref{mat}) becomes a
convolution in time. However, it is shown in \cite{Haldane} (omitting magnetoelectric parameters, although this can also be included) that the
only necessary modification to allow for dispersive materials $M=M\left(
\omega \right) $ is to replace $M$ in (\ref{IP1}) with $\partial \left(
\omega M\left( \omega \right) \right) /\partial \omega $, so that 
\begin{equation}
\left\langle u_{n}\left\vert u_{m}\right. \right\rangle =\frac{1}{2}\int_{%
\text{BZ}}u_{n}^{\ast }\left( \mathbf{r}\right) \frac{\partial \left( \omega
M\left( \omega \right) \right) }{\partial \omega }u_{m}\left( \mathbf{r}%
\right) d\mathbf{r.}  \label{IP2}
\end{equation}%
The material continuum model will be considered below.

\subsection{Some electromagnetic material classes that posses non-trivial
Chern numbers}

Although the field of topological photonic insulators is still being
developed, there are several classes of materials and structures which
posses topological protection and non-trivial Chern numbers. The approaches
to design PTIs can be roughly divided into two categories. The first one
relies on breaking of time-reversal symmetry by applying a static magnetic
field to a gyromagnetic material \cite{Solja} or time-harmonic modulation of
coupled resonators \cite{Kejie}, \cite{Hafezi}. Another approach involves
time-reversal-invariant metamaterials, where photon states are separated in
two `spin' sub-spaces (usually through geometry such as via a hexagonal
lattice), and `spin-orbit' coupling is introduced through symmetry-breaking
exploring such non-trivial characteristics of metamaterials as chirality,
bianisotropy and hyperbolicity \cite{Alexander}, \cite{PTI}. For an
electromagnetic standpoint, the most important aspect of a PTI is the
presence of surface/edge states that are topologically protected from
backscattering (having non-trivial Chern number).

In classical electromagnetics, the fields $\mathbf{E}$, $\mathbf{D}$, and $%
\mathbf{P}$ are even under time reversal (do not change with time-reversal),
whereas $\mathbf{A}_{\text{mag}}$, $\mathbf{B}$, $\mathbf{H}$, $\mathbf{J}$,
and $\mathbf{S}$ (Poynting vector) are odd under time reversal (negated
under time reversal). For systems with time-reversal (TR) symmetry, 
\begin{equation}
F_{\alpha \beta }^{n}\left( \mathbf{k}\right) =-F_{\alpha \beta }^{n}\left( -%
\mathbf{k}\right) .
\end{equation}%
Furthermore, the fields $\mathbf{B}$ and $\mathbf{H}$ are even under space
inversion, whereas $\mathbf{E}$, $\mathbf{D}$, $\mathbf{J}$, $\mathbf{P}$, $%
\mathbf{A}_{\text{mag}}$, and $\mathbf{S}$ are odd under spacial inversion.
For systems with parity/inversion (I) symmetry, 
\begin{equation}
F_{\alpha \beta }^{n}\left( \mathbf{k}\right) =F_{\alpha \beta }^{n}\left( -%
\mathbf{k}\right) ,
\end{equation}%
so that if both symmetries are present, 
\begin{equation}
F_{\alpha \beta }^{n}\left( \mathbf{k}\right) =0.  \label{2S}
\end{equation}%
Systems having both spatial-inversion and time-reversal symmetry will
exhibit trivial topology in momentum space, so that no one-way edge mode
will exist (all bands have $C_{n}=0$). 

Regarding periodic materials, Dirac (linear) degeneracies will occur for
hexagonal lattices, and other lattice types may exhibit other degeneracies
(e.g., quadratic degeneracies consisting of double Dirac degeneracies for a
cubic lattice \cite{Solja2}, \cite{Solja3}, but, regardless of degeneracy
type, for, e.g., a simple lattice of material cylinders in a host medium, if
the cylinders are made of simple isotropic materials have scalar material
properties $\varepsilon $ and $\mu $, the system will be both
space-inversion and time-reversal symmetric, and all bands will have trivial
Chern number.

In the periodic case the degeneracies can be broken in several ways. One way
that has been widely studied is to use rods with materials that themselves
break TR symmetry \cite{Solja3}, or to embed, say, a hexagonal array of
nonreciprocal rods into another array of simple rods \cite{Kejie2} so that
both arrays share a common bandgap. The resulting nonreciprocal structure
will generally have bands of non-trivial Chern number, leading to a non-zero
gap Chern number. A detailed example is provided below. Large Chern numbers
can be achieved by increasing spatial symmetry to result in point
degeneracies of higher order (e.g., several co-located Dirac points), and
then to, say, introduce TR breaking \cite{Solja4}.

Another method to create a nontrivial Chern number is to use simple
materials (simple dielectrics and metals), but to break inversion symmetry
by deforming the lattice. For example, in \cite{Hua} simple dielectric rods
are used in a hexagonal pattern, resulting in a Dirac degeneracy and trivial
Chern number. Slightly deforming the lattice can result in
inversion-symmetry breaking, and $C_{\text{gap}}\neq 0$. Various other
schemes have also been proposed \cite{Alexander}, \cite{Dong}.

\subsection{Berry quantities for continuum media}

Although the electronic case, and, by analogy, the photonic case, were
developed for periodic systems (for which the relations provided in sections \ref{sec1} and \ref{sec2} hold), it turns out that continuum material models
can also lead to nontrivial Chern numbers. The simplest example is of a
biased plasma (magneto-plasma) as considered in Fig. 1, with permittivity
tensor 
\begin{equation}
\overline{\epsilon }=\left( 
\begin{array}{ccc}
\epsilon _{11} & \epsilon _{12} & 0 \\ 
\epsilon _{21} & \epsilon _{22} & 0 \\ 
0 & 0 & \epsilon _{33}%
\end{array}%
\right)
\end{equation}%
where typically $\varepsilon _{21}=\varepsilon _{12}^{\ast }$ and $%
\varepsilon _{11}=\varepsilon _{22}=\varepsilon _{33}$ (in the absence of a
bias field $\varepsilon _{12}=0$, the material is reciprocal, and $\overline{%
\mathbf{\varepsilon }}$ reduces to a scalar). An example involving this type
of material is provided below.  At the interface between the magneto-plasma
and an ordinary (unbiased) plasma, a topologically protected edge mode can
exist \cite{Mario2}, \cite{Arthur}, \cite{Biao}. In addition, more
complicated materials combining hyperbolic and chiral response have been
shown to be topologically nontrivial \cite{PTI}.

A continuum material presents a difficulty in that, rather than have a
periodic Brillouin zone that is, effectively, a closed surface (equivalent
to a torus), providing the surface over which the Chern number can be
computed, the momentum-space of an infinite homogeneous material continuum
model does not form a closed surface. However, in \cite{Mario2} it is shown
that 2D momentum space can be mapped to the Riemann sphere, forming the
necessary surface (north and south poles being exceptional points, as
discussed below).

Another issue, for both periodic and continuum models, is to account for
material dispersion. Following the result in \cite{Haldane} for lossless
dispersive local periodic media, in \cite{Mario2} continuum models of
dispersive lossless, and possibly wavevector-dependent (nonlocal) materials
are considered. The Berry potential is again given by (\ref{AEM}), with the
inner product\footnote{%
The most general result in \cite{Mario2} is more complicated, but for a wide
range of material classes the simpler result shown here holds.} 
\begin{equation}
\left\langle f_{n}|f_{m}\right\rangle =\frac{1}{2}f_{n}^{\ast }\frac{%
\partial \left( \omega M\left( \omega \right) \right) }{\partial \omega }%
f_{m}.
\end{equation}

\section{Photonic Examples}
\subsection{Periodic photonic system example}\label{Ex1}

This example is taken directly from \cite{Kejie2}.

\begin{figure}[ht]
\begin{center}
\noindent  \includegraphics[width=3in]{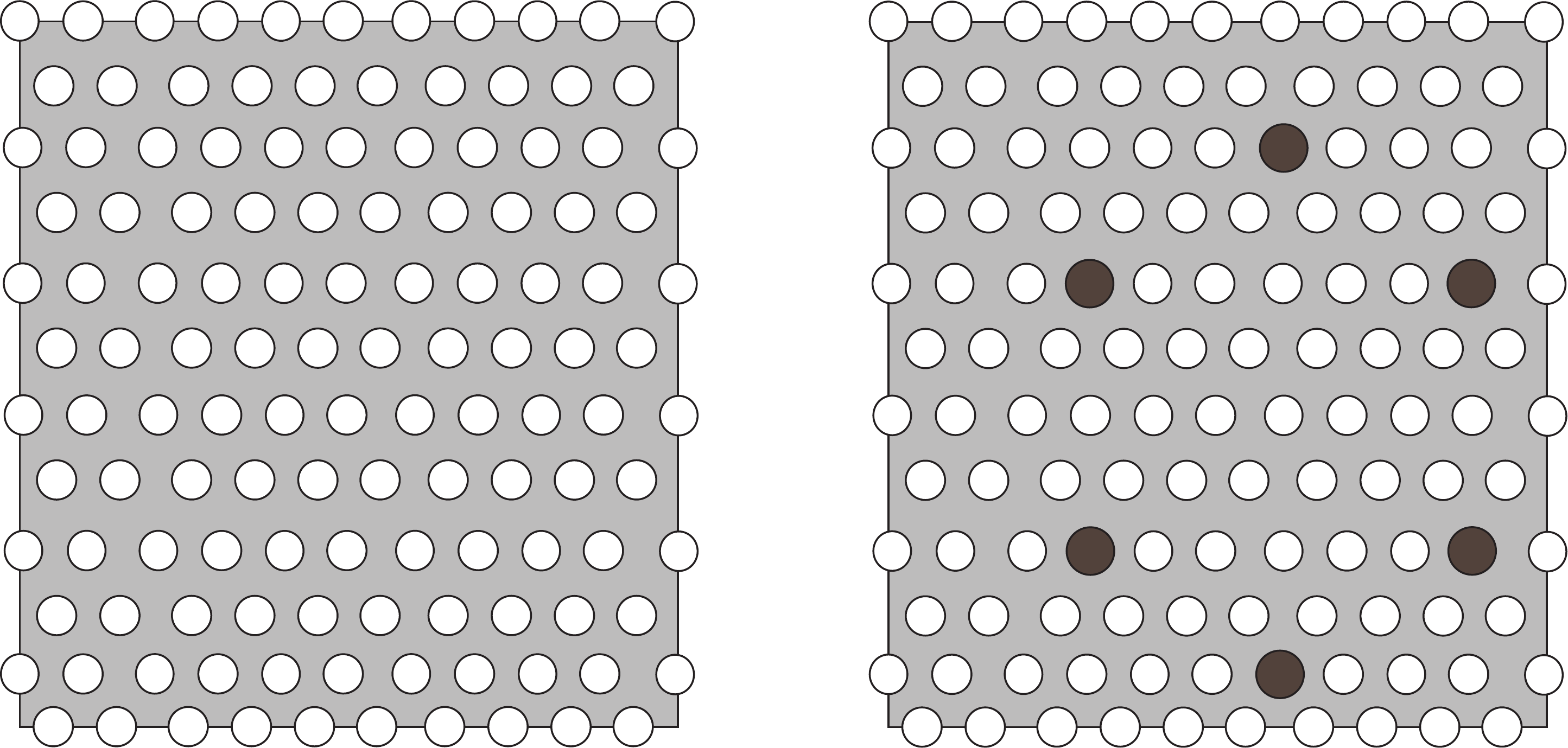}  
\end{center}
\caption{(a) Top view of photonic crystal made from air holes (cylinders) in a host insulating medium. Lattice constant is $a$, cylinders have radius $r_1=\alpha_1 a$, and the host medium is characterized by $\varepsilon$ (equivalently, one could have dielectric cylinders in a host material), (b) Defected structure; here, magneto-optic cylinders forming a hexagonal array of ``defects" in the air-hole medium. Magneto-optic cylinders have radius $r_2=\alpha_2 a$ and lattice constant $a'=\alpha_3 a$.}
\label{P1}
\end{figure}

One way to create a PTI is via a hexagonal array of infinite cylinders. Consider a simple dielectric of $%
\varepsilon $ with a periodic array of air holes (cylinders of radius $%
r_1=\alpha_1 a$, with $a$ the lattice constant) in the form of a triangular
lattice, as shown in Fig. \ref{P1}a. The periodicity is chosen to create a bandgap
in the allowed modes of the system \cite{Kejie2}. A single defect, such as
making one hole a different radius, or filling the hole with some material,
can establish a resonator having frequency in the bandgap. Making a periodic
array of defects can create bandstructure within the original bandgap, in
this case creating four modes in the bandgap. Here the array of defects is
created using cylinders of radius $r_2=\alpha_2 a$ of magneto-optic material, 
\begin{equation}
\varepsilon _{\text{rod}}=\left[ 
\begin{array}{ccc}
\varepsilon _{r} & -i\varepsilon _{i} & 0 \\ 
i\varepsilon _{i} & \varepsilon _{r} & 0 \\ 
0 & 0 & \varepsilon _{r}%
\end{array}%
\right] 
\end{equation}%
and arranging them in a hexagonal lattice with lattice constant $a^{\prime
}=\alpha_3 a$, as shown in Fig. \ref{P1}b. Due to the hexagonal symmetry, for the unbiased defect array ($%
\varepsilon _{i}=0$) there are degeneracies in the modes at the $\Gamma $
and $K$ points. Time-reversal symmetry can be broken by applying a bias
parallel to the cylinders ($\varepsilon _{i}\neq 0$), lifting the degeneracy (see Fig. \ref{BD1} discussed later). 
 
To determine the bandstructure, we could solve the eigenvalue equation $H_{\mathrm{cl}}f_{n%
\mathbf{k}}\left( \mathbf{r}\right) =E_{n\mathbf{k}}f_{n\mathbf{k}}\left( 
\mathbf{r}\right) $, where $H_{cl}$ is the $6\mathrm{x}6$ electromagnetic Hamiltonian and 
$f_{n\mathbf{k}}$ is the six-vector of fields, both defined previously. In
general, this is a quite complicated electromagnetic problem, which,
however, can be solved using commercial simulators. We can simplify the problem from the $6\mathrm{x}6$ formulation by noting that from Maxwell's
equations for a material characterized by this magneto-optic permittivity, TE
modes have a single magnetic field component $H_{z}$ parallel to the
infinite cylinders, and an in-plane electric field. The magnetic field
satisfies the eigenvalue equation \cite{AD}--\cite{FC} 
\begin{equation}
H(\mathbf{z}H_z)=\left( \frac{\omega }{c}\right) ^{2}(\mathbf{z}H_z)
\end{equation}%
where the operator $H=H_{\mathrm{cl}}=\nabla \times \varepsilon ^{-1}\nabla \times$ is Hermitian for lossless media, under the usual inner product $\left\langle 
\mathbf{f},\mathbf{g}\right\rangle =\int \mathbf{f}^{\ast }\cdot \mathbf{g}d%
\mathbf{r}$. Thus, we can solve a scalar equation for $H_{z}$. Furthermore,
an approximate solution can be obtained that gives considerable insight into
the problem; the typical-binding method of condensed matter physics can
be used to obtain an effective four-band Hamiltonian in the electromagnetic case \cite{TB} (and, as a special case
we recover the graphene result). 

The individual resonators support two $p$%
-type (dipole-like) modes at the same frequency $\omega _{0}$. Considering
that the honeycomb lattice has two inequivalent sites $A$ and $B$, each
having two different states $p_{x,y}$, then considering Bloch's theorem the wavefunction $f_{n\mathbf{k}}\left( 
\mathbf{r}\right) =H_{z}\left( \mathbf{r}\right) $ is expanded as%
\begin{equation}
f_{n\mathbf{k}}\left( \mathbf{r}\right) =\frac{1}{\sqrt{N}}\sum_{\mathbf{R}%
}e^{i\mathbf{k}\cdot \mathbf{R}}\sum_{\beta =A,B}\sum_{\alpha =x,y}c_{\alpha
}^{\beta }\phi _{p_{\alpha }}\left( \mathbf{r}-\mathbf{d}_{\beta }-\mathbf{R}%
\right) 
\end{equation}%
where $\mathbf{R}$ is the lattice vector and $\phi $ the mode function.
Plugging into the energy eigenvalue equation $Hf_{n\mathbf{k}}\left( \mathbf{%
r}\right) =E_{n\mathbf{k}}f_{n\mathbf{k}}\left( \mathbf{r}\right) $,
multiplying through by $\int d\mathbf{r\ }\phi _{p\alpha }\left( \mathbf{r}-%
\mathbf{d}_{\gamma }\right) $ and exploiting the assumed orthogonality of
the modes, it is easy to obtain the Hermitian (effective, 4-band) Hamiltonian matrix%
\begin{equation}
H=\left[ 
\begin{array}{cccc}
\omega _{0} & g_{xy}^{AA}\left( \mathbf{k}\right)  & g_{xx}^{AB}\left( 
\mathbf{k}\right)  & g_{xy}^{AB}\left( \mathbf{k}\right)  \\ 
& \omega _{0} & g_{yx}^{AB}\left( \mathbf{k}\right)  & g_{yy}^{AB}\left( 
\mathbf{k}\right)  \\ 
&  & \omega _{0} & g_{xy}^{BB}\left( \mathbf{k}\right)  \\ 
&  &  & \omega _{0}%
\end{array}%
\right] 
\end{equation}%
where 
\begin{equation}
g_{\gamma \delta }^{\alpha \beta }\left( \mathbf{k}\right) =\sum_{\mathbf{R}%
}e^{i\mathbf{k}\cdot \mathbf{R}}H_{\gamma \delta }^{\alpha \beta }\left( 
\mathbf{R}\right) ,
\end{equation}%
with $H_{\gamma \delta }^{\alpha \beta }\left( \mathbf{\tau }\right) $ being
overlap/hopping integrals having the form%
\begin{equation}
H_{\gamma \delta }^{\alpha \beta }\left( \mathbf{R}\right) =\int d\mathbf{r\ 
}\phi _{p\alpha }\left( \mathbf{r}-\mathbf{d}_{\gamma }\right) H\phi
_{p\beta }\left( \mathbf{r}-\mathbf{d}_{\delta }-\mathbf{R}\right) .
\end{equation}%
We assume that $H_{\gamma \gamma }^{\alpha \alpha }$ is dominated by the
self energy\footnote{%
\begin{align}
H_{\gamma \gamma }^{\alpha \alpha }& =\int d\mathbf{r\ }\phi _{p\alpha
}\left( \mathbf{r}-\mathbf{d}_{\gamma }\right) H\phi _{p\alpha }\left( 
\mathbf{r}-\mathbf{d}_{\gamma }-\mathbf{R}\right) \simeq \int d\mathbf{r\ }%
\phi _{p\alpha }\left( \mathbf{r}-\mathbf{d}_{\gamma }\right) H_{\gamma
}\phi _{p\alpha }\left( \mathbf{r}-\mathbf{d}_{\gamma }-\mathbf{R}\right)  \\
& =\omega _{0}\int d\mathbf{r\ }\phi _{p\alpha }\left( \mathbf{r}-\mathbf{d}%
_{\gamma }\right) \phi _{p\alpha }\left( \mathbf{r}-\mathbf{d}_{\gamma }-%
\mathbf{R}\right) =\omega_0 \left\{ 
\begin{array}{c}
1\text{, \ }\mathbf{R}=\mathbf{0} \\ 
0\text{, \ }\mathbf{R}\neq \mathbf{0}%
\end{array}%
,\right.   \nonumber
\end{align}%
where $H_{\gamma }$ is the Hamiltonian of an isolated resonator}, which
leads to the diagonal components.

To evaluate the off-diagonal components, we consider only nearest neighbors.
However, let us first digress and consider graphene, which is arranged in a hexagonal lattice and has two carbon
atoms per unit cell. 

\subsubsection{Graphene interlude -- the hexagonal lattice}
\begin{figure}[ht]
\begin{center}
\noindent  \includegraphics[width=2in]{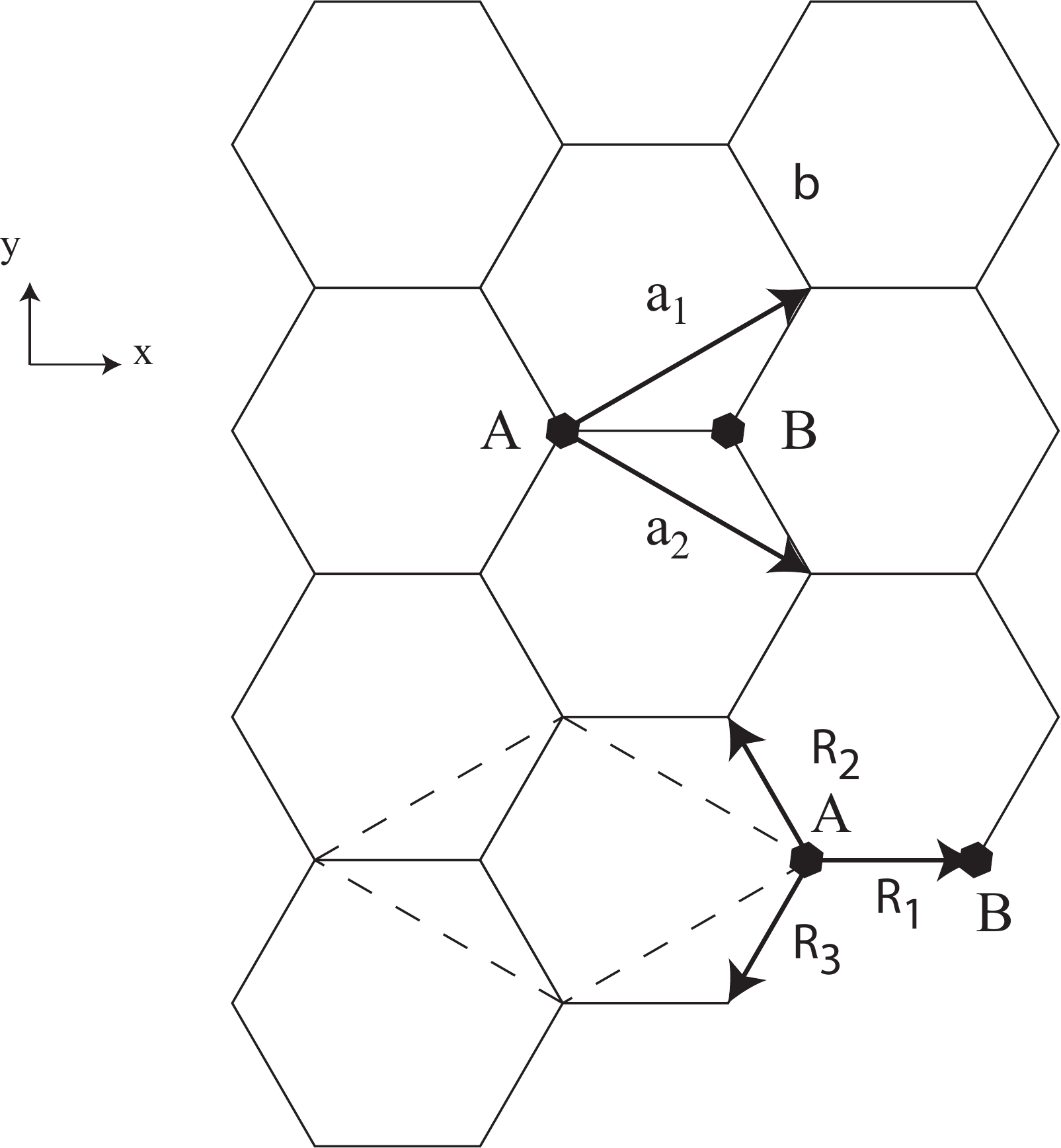}  
\end{center}
\caption{Direct space for a hexagonal lattice, two-atoms (A and B) per unit cell. Nearest-neighbor vectors are $\mathbf{R}_{1,2,3}$}
\label{G1}
\end{figure}

\begin{figure}[ht]
\begin{center}
\noindent  \includegraphics[width=2in]{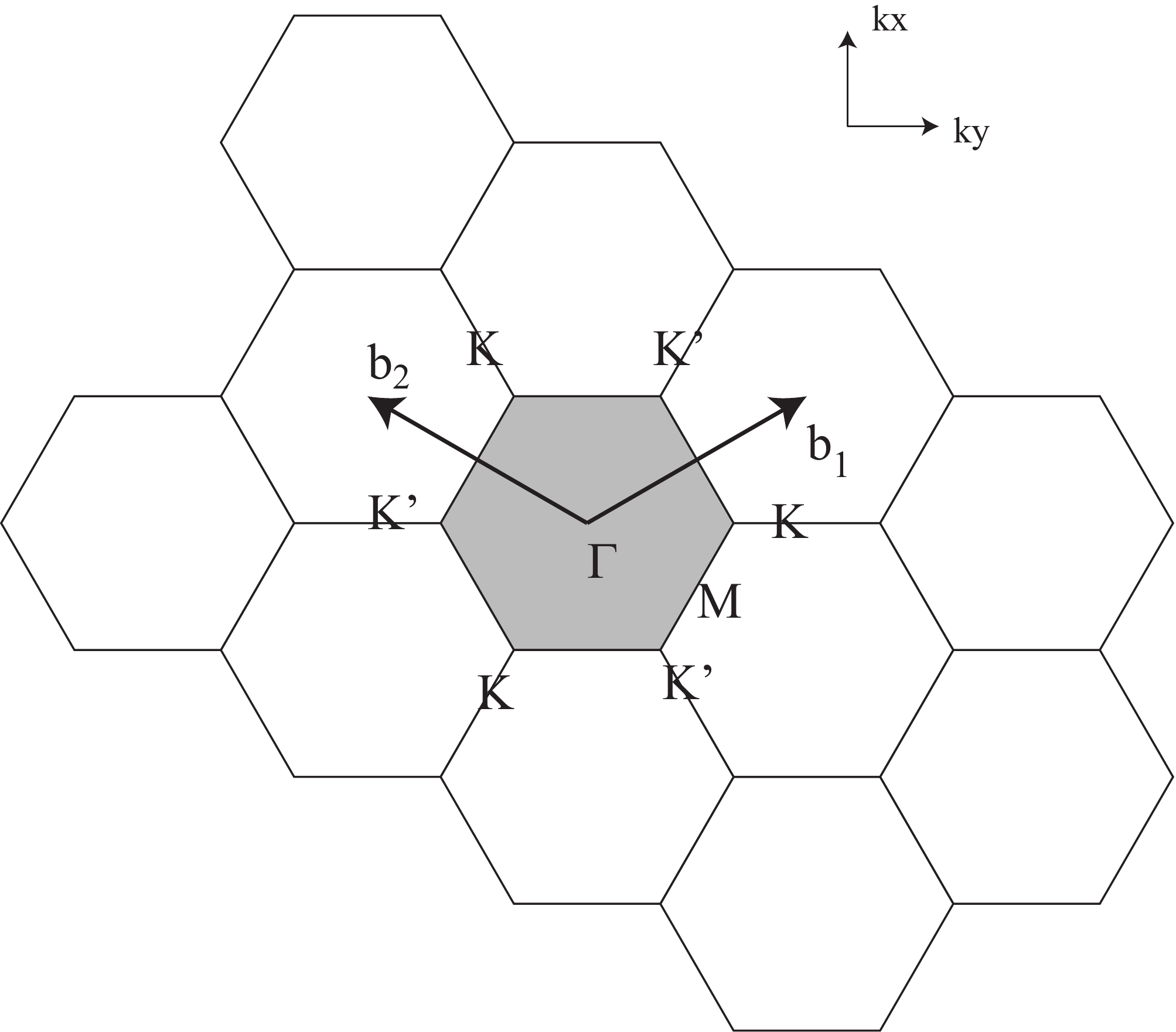}  
\end{center}
\caption{Reciprocal space for a hexagonal lattice. $\Gamma$, K, and M are the high-symmetry points.}
\label{G2a}
\end{figure}

\begin{figure}[ht]
\begin{center}
\noindent  \includegraphics[width=2in]{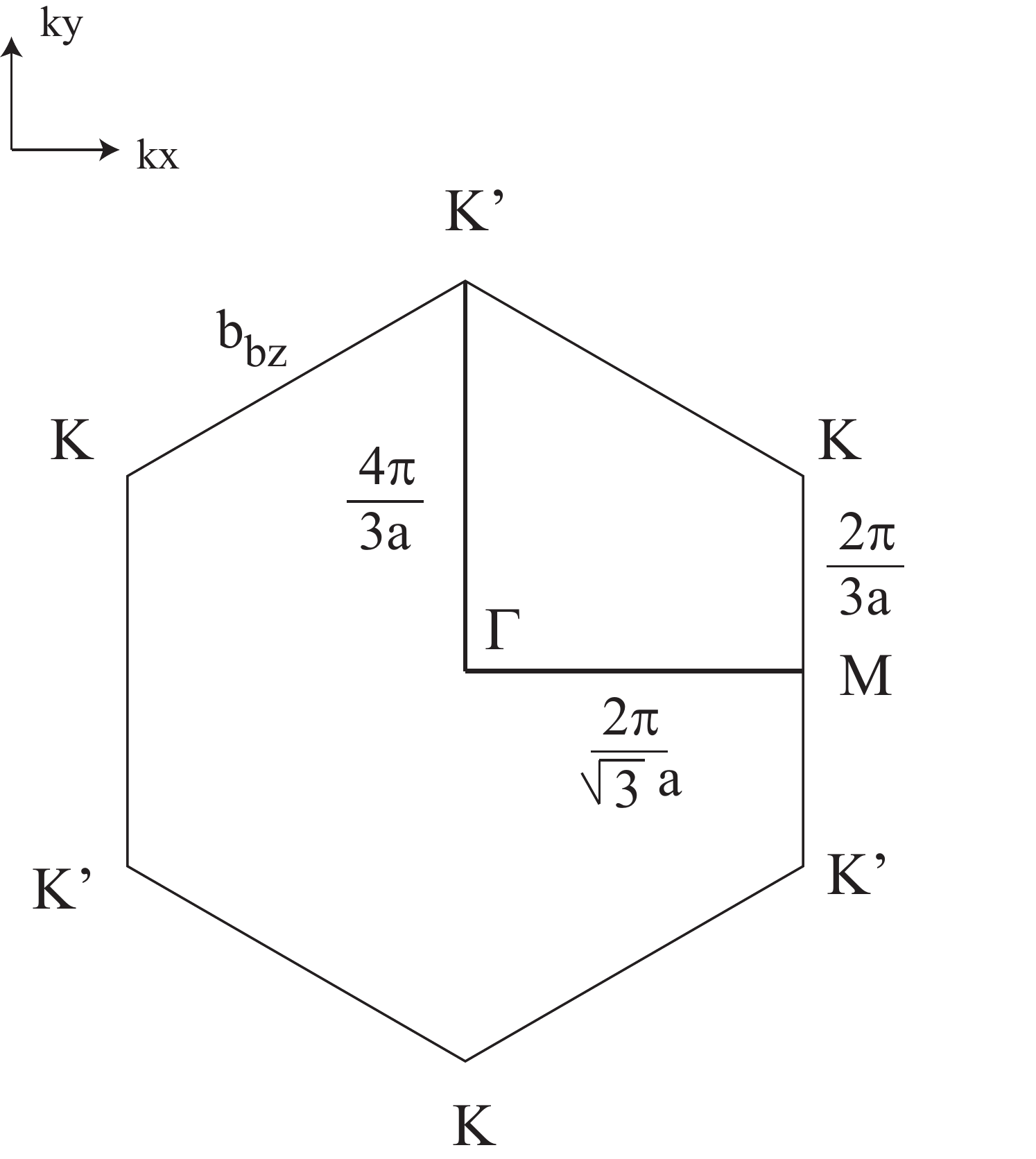}  
\end{center}
\caption{Close-up, reciprocal space for a hexagonal lattice.}
\label{G2}
\end{figure}

The direct and reciprocal lattices for a hexagon lattice are shown in Figs. \ref{G1}-\ref{G2}.

The direct-space vectors are 
\begin{equation}
\mathbf{a}_{1}=\frac{a}{2}\left( \sqrt{3},1\right) ,\ \mathbf{a}_{2}=\frac{a%
}{2}\left( \sqrt{3},-1\right) ,
\end{equation}%
where $a=\left\vert \mathbf{a}_{1}\right\vert =\left\vert \mathbf{a}%
_{2}\right\vert =\sqrt{3}b$ is the lattice constant. For graphene, $%
b=0.142$ nm is the interatomic distance between carbon atoms. The nearest
neighbor vectors are (see Fig. \ref{G1})
\begin{equation}
\mathbf{R}_{1}=\left( \frac{a}{\sqrt{3}},0\right) ,\ \ \mathbf{R}_{2}=-%
\mathbf{a}_{2}+\mathbf{R}_{1}=\left( -\frac{a}{2\sqrt{3}},\frac{a}{2}\right)
,\ \ \mathbf{R}_{3}=-\mathbf{a}_{1}+\mathbf{R}_{1}=\left( -\frac{a}{2\sqrt{3}%
},-\frac{a}{2}\right) ,
\end{equation}
with $\left\vert \mathbf{R}_{1}\right\vert =\left\vert \mathbf{R}%
_{2}\right\vert =\left\vert \mathbf{R}_{3}\right\vert =b=a/\sqrt{3}$.

The reciprocal lattice vectors are (see Fig. \ref{G2a})
\begin{equation}
\mathbf{b}_{1}=\frac{2\pi }{a}\left( \frac{1}{\sqrt{3}},1\right) ,\ \ 
\mathbf{b}_{2}=\frac{2\pi }{a}\left( \frac{1}{\sqrt{3}},-1\right) ,
\end{equation}%
where $\left\vert \mathbf{b}_{1}\right\vert =\left\vert \mathbf{b}%
_{2}\right\vert =4\pi /\sqrt{3}a$, (the side length of the reciprocal
lattice hexagon is $b_{bz}=\left\vert \mathbf{b}_{1}\right\vert /\sqrt{3}%
=4\pi /3a$).

The high-symmetry points in the Brillouin zone are

\begin{equation}
\mathbf{\Gamma }=\left( 0,0\right) ,\ \ \ \mathbf{K}=\left( \frac{2\pi }{%
\sqrt{3}a},\frac{2\pi }{3a}\right) ,\ \ \ \mathbf{M}=\left( \frac{2\pi }{%
\sqrt{3}a},0\right) ,
\end{equation}%
and $\left\vert \mathbf{\Gamma -M}\right\vert =2\pi /\sqrt{3}a$, $\left\vert 
\mathbf{\Gamma -K}\right\vert =4\pi /3a$, and $\left\vert \mathbf{M-K}%
\right\vert =2\pi /3a$.

For $\pi -$bonding in graphene (the usual low-energy case), each
carbon atom contributes one $2p_{z}$-orbital, so we have half as many
unknowns compared to the two-orbital photonic case described above. Then, 
\begin{equation}
f_{n\mathbf{k}}\left( \mathbf{r}\right) =\frac{1}{\sqrt{N}}\sum_{\mathbf{R}%
}e^{i\mathbf{k}\cdot \mathbf{R}}\sum_{\beta =A,B}c^{\beta }\phi _{p}\left( 
\mathbf{r}-\mathbf{d}_{\beta }-\mathbf{R}\right) .
\end{equation}%
Since there is only one symmetric orbital, confining the summation to the
three nearest neighbors $\mathbf{\tau }$,  
\begin{equation}
g^{\alpha \beta }\left( \mathbf{k}\right) =\sum_{\mathbf{\tau }}e^{i\mathbf{k%
}\cdot \mathbf{\tau }}H^{\alpha \beta }\left( \mathbf{\tau }\right) \simeq
H^{\alpha \beta }\sum_{\mathbf{\tau }}e^{i\mathbf{k}\cdot \mathbf{\tau }}
\end{equation}%
For atom $A$, the three nearest neighbors are the three nearby $B\,$\ atoms,
located at $\mathbf{\tau }_{1,2,3}=\mathbf{R}_{1,2,3}$,
so that%
\begin{equation}
h\left( \mathbf{k}\right) =\sum_{\mathbf{\tau }}e^{i\mathbf{k}\cdot \mathbf{%
\tau }}=e^{ik_{x}\frac{a}{\sqrt{3}}}+e^{i\left( -k_{x}\frac{a}{2\sqrt{3}}%
+k_{y}\frac{a}{2}\right) }+e^{-i\left( k_{x}\frac{a}{2\sqrt{3}}+k_{y}\frac{a%
}{2}\right) }=e^{ik_{x}\frac{a}{\sqrt{3}}}+2\cos \left( k_{y}\frac{a}{2}%
\right) e^{-ik_{x}\frac{a}{2\sqrt{3}}}.
\end{equation}%
The Hamiltonian matrix is then 
\begin{equation}
\left[ 
\begin{array}{cc}
\omega _{0} & \gamma h\left( \mathbf{k}\right)  \\ 
\gamma h^{\ast }\left( \mathbf{k}\right)  & \omega _{0}%
\end{array}%
\right] \left[ 
\begin{array}{c}
c_{\mathbf{k}}^{A} \\ 
c_{\mathbf{k}}^{B}%
\end{array}%
\right] =E_{n\mathbf{k}}\left[ 
\begin{array}{c}
c_{\mathbf{k}}^{A} \\ 
c_{\mathbf{k}}^{B}%
\end{array}%
\right] 
\end{equation}%
where $\gamma =H^{\alpha \beta }$ is the overlap integral (with typical values
of several eV). Then, the energy dispersion is $\left( \omega _{0}-E_{n\mathbf{k%
}}\right) ^{2}-\left\vert \gamma h\left( \mathbf{k}\right) \right\vert ^{2}=0
$, so that, since $\omega _{0}=0$, we have the celebrated graphene result 
\begin{equation}
E_{n\mathbf{k}}=\pm \gamma \sqrt{1+4 \cos
\left( k_{x}\frac{\sqrt{3}a}{2}\right)\cos \left( k_{y}\frac{a}{2}\right) +4\cos ^{2}\left( k_{y}\frac{a}{2}%
\right) }.
\end{equation}
There is no bandgap because the two atoms are identical. The above is the
simplest formulation; we obtain several corrections to this result if we do
not drop the $\mathbf{R}\neq 0$ terms in the diagonal components, and also
from the fact that the $A$ and $B$ orbitals have some overlap. 

Returning to the electromagnetic cylinder case considered in \cite{Kejie2},
since there are two orbitals per mode $\phi _{p_{x,y}}$, evaluation of the $%
g_{\gamma \delta }^{\alpha \beta }\left( \mathbf{k}\right) $ functions is
more difficult because we cannot factorize $\sum_{\mathbf{\tau }}e^{i\mathbf{%
k}\cdot \mathbf{\tau }}H^{\alpha \beta }\left( \mathbf{\tau }\right) \simeq
H^{\alpha \beta }\sum_{\mathbf{\tau }}e^{i\mathbf{k}\cdot \mathbf{\tau }}$
as above for the graphene case, due to the complexity of the orbitals.
However, this is a common occurrence in condensed matter physics, and the
following two-center interaction integrals are widely used,%
\begin{equation}
H_{\gamma \delta }^{\alpha \neq \beta }\left( \mathbf{R}\right) =\int d%
\mathbf{r\ }\phi _{p\alpha }\left( \mathbf{r}-\mathbf{d}_{\gamma }\right)
H\phi _{p\beta }\left( \mathbf{r}-\mathbf{d}_{\delta }-\mathbf{R}\right)
=\left\{ 
\begin{array}{c}
l_{\alpha }^{2}V_{\sigma }+\left( 1-l_{\beta }^{2}\right) V_{\pi }\ \ \text{%
for }\alpha =\beta =x,y \\ 
l_{\alpha }l_{\beta }\left( V_{\sigma }-V_{\pi }\right) \ \ \text{for }%
\alpha \neq \beta ,%
\end{array}%
\right. 
\end{equation}%
where $l_{x,y}$ are the direction cosines $l_{\alpha }=\mathbf{\alpha }\cdot 
\mathbf{R}/\left\vert \mathbf{R}\right\vert $. For $\mathbf{R}=\mathbf{R}_{1}$, $l_{x}=1$, $l_{y}=0$, for $\mathbf{%
R}=\mathbf{R}_{2}$, $l_{x}=-1/2$ and $l_{y}=\sqrt{3}/2$, and for $\mathbf{R}=%
\mathbf{R}_{3}$, $l_{x}=-1/2$ and $l_{y}=-\sqrt{3}/2$. Therefore, 
\begin{align}
\sum_{\mathbf{R}}e^{i\mathbf{k}\cdot \mathbf{R}}H_{ABxx}\left( \mathbf{R}%
\right) & =\sum_{\mathbf{R}}e^{i\mathbf{k}\cdot \mathbf{R}}\int d\mathbf{r\ }%
\phi _{px}\left( \mathbf{r}-\mathbf{d}_{A}\right) H\phi _{p_{x}}\left( 
\mathbf{r}-\mathbf{d}_{B}-\mathbf{R}\right)  \\
& =e^{ik_{x}\frac{a}{\sqrt{3}}}V_{\sigma }+\cos \left( k_{y}\frac{a}{2}%
\right) e^{-ik_{x}\frac{a}{2\sqrt{3}}}\left( \frac{1}{2}V_{\sigma }+\frac{3}{%
2}V_{\pi }\right) .
\end{align}%
For the next element, 
\begin{eqnarray}
\sum_{\mathbf{R}}e^{i\mathbf{k}\cdot \mathbf{R}}H_{ABxy}\left( \mathbf{R}%
\right)  &=&\sum_{\mathbf{R}}e^{i\mathbf{k}\cdot \mathbf{R}}\int d\mathbf{r\ 
}\phi _{px}\left( \mathbf{r}-\mathbf{d}_{A}\right) H\phi _{p_{y}}\left( 
\mathbf{r}-\mathbf{d}_{B}-\mathbf{R}\right)  \\
&=&-2i\sin \left( k_{y}\frac{a}{2}\right) e^{-ik_{x}\frac{a}{2\sqrt{3}}}%
\frac{\sqrt{3}}{4}\left( V_{\sigma }-V_{\pi }\right) .
\end{eqnarray}%
The other elements are evaluated in a similar fashion.

In the following, we assume \cite{Kejie2} $\varepsilon =16$, $r_{1}=0.35a$, $%
r_{2}=0.5a$, and $a^{\prime }=6a$. Making the substitution $a\rightarrow 6a$
since in the derivation $a=a^{\prime }$ is the hexagon lattice constant, but
it is convenient to express the matrix entries in terms of the original
(undeformed) lattice constant $a$, the final matrix is \cite{Kejie2}%
\footnote{%
Note that \cite{Kejie2} uses a coordinate system where $x$ and $y$ are
interchanged from those used here.} 
\begin{equation}
H=\left[ 
\begin{array}{cccc}
\omega _{0} & -iV_{p} & \left( \frac{3}{2}V_{\pi }+\frac{1}{2}V_{\sigma
}\right) \cos \left( 3k_{y}a\right) e^{-i\sqrt{3}k_{x}a}+V_{\alpha }e^{i2%
\sqrt{3}k_{x}a} & -i\frac{\sqrt{3}}{2}\left( V_{\sigma }-V_{\pi }\right)
\sin \left( 3k_{y}a\right) e^{-i\sqrt{3}k_{x}a} \\ 
\  & \omega _{0} & -i\frac{\sqrt{3}}{2}\left( V_{\sigma }-V_{\pi }\right)
\sin \left( 3k_{y}a\right) e^{-i\sqrt{3}k_{x}a} & \left( \frac{3}{2}%
V_{\sigma }+\frac{1}{2}V_{\pi }\right) \cos \left( 3k_{y}a\right) e^{-i\sqrt{%
3}k_{x}a}+V_{\pi }e^{i2\sqrt{3}k_{x}a} \\ 
\  & \  & \omega _{0} & -iV_{p} \\ 
\  & \  & \  & \omega _{0}%
\end{array}%
\right] .
\end{equation}%
The bond integrals $V_{\sigma ,\pi }$ are evaluated by matching the
resulting bandstructure to the commercial simulation. This results in the
absence of applied magnetization (time-reversal invariant case) \cite{Kejie2}
$V_{\sigma }=-0.001185$, $V_{\pi }=0.000085$, and $V_{p}=0$, the overlap
between $x$ and $y$ orbitals for the same atom. In the presence of
magnetization (time-reversal symmetry broken, assuming $\varepsilon _{i}=1$%
), $V_{\sigma }=-0.001192$, $V_{\pi }=0.000092$, and $V_{p}=0.0007$, where
all terms have units of radian frequency $2\pi c/a$.

\begin{figure}[ht]
\begin{center}
\noindent  \includegraphics[width=5in]{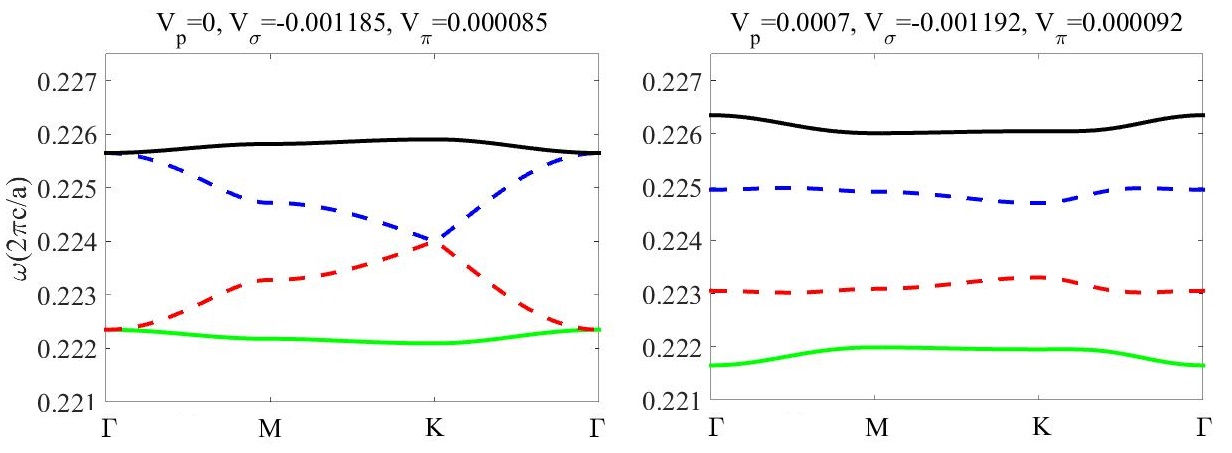}  
\end{center}
\caption{Bandstructure for the lattice depicted in Fig. \ref{P1}b. Left side shows the reciprocal case, $V_p=0$, where there are modal degeneracies that close the bandgap, and $C_n=0$, and the right side shows the nonreciprocal case, $V_p\neq 0$, for which the degeneracies are lifted and two bands have $C_n=\pm 1$.}
\label{BD1}
\end{figure}

From the projected Hamiltonian matrix $H_{n,m}$ it is easy to solve the
eigenvalue problem $Hf_{n}=\omega _{n}f_{n}$. Results are shown in Fig. \ref%
{BD1}, showing (a) the case for no magnetic bias ($V_{p}=0$ and (b) with
bias applied, breaking TR symmetry and lifting the degeneracies. Note that
the high-symmetry points in Fig. \ref{BD1} are with respect to the $%
a^{\prime }$ lattice, 
\begin{equation}
\mathbf{\Gamma }=\left( 0,0\right) ,\ \ \ \mathbf{K}=\left( \frac{2\pi }{%
\sqrt{3}a^{\prime }},\frac{2\pi }{3a^{\prime }}\right) ,\ \ \ \mathbf{M}=\left( \frac{2\pi }{%
\sqrt{3}a^{\prime }},0\right) .
\end{equation}

To compute the Chern number, the form of the Berry curvature (\ref{Eq:15})
is convenient to use since derivatives of the effective 4-band Hamiltonian
can be taken analytically. Eigenvalues $\omega_n=E_{n}$ of the matrix $H$ were found
analytically using a symbolic solver, and eigenfunctions $\left\vert
n\right\rangle $ were found numerically. Integration over the first
Brillouin zone depicted in Fig. \ref{G2} results in the Chern number by (\ref{CNF}), 
\begin{equation}
C_{n}=\frac{1}{2\pi }\int_{\text{BZ}}F_{xy}^{n}dk_{x}dk_{y}
\end{equation}%
leading to bands 1 and 4 having Chern number $C_{1,4}=\mp 1$ and the middle
two bands having $C_{2,3}=0$ (note that the sum of Chern numbers is zero, as expected). For the reciprocal case $V_{p}=0$, $C_{n}=0$ for
all bands. 

\begin{figure}[ht]
\begin{center}
\noindent \includegraphics[width=4in]{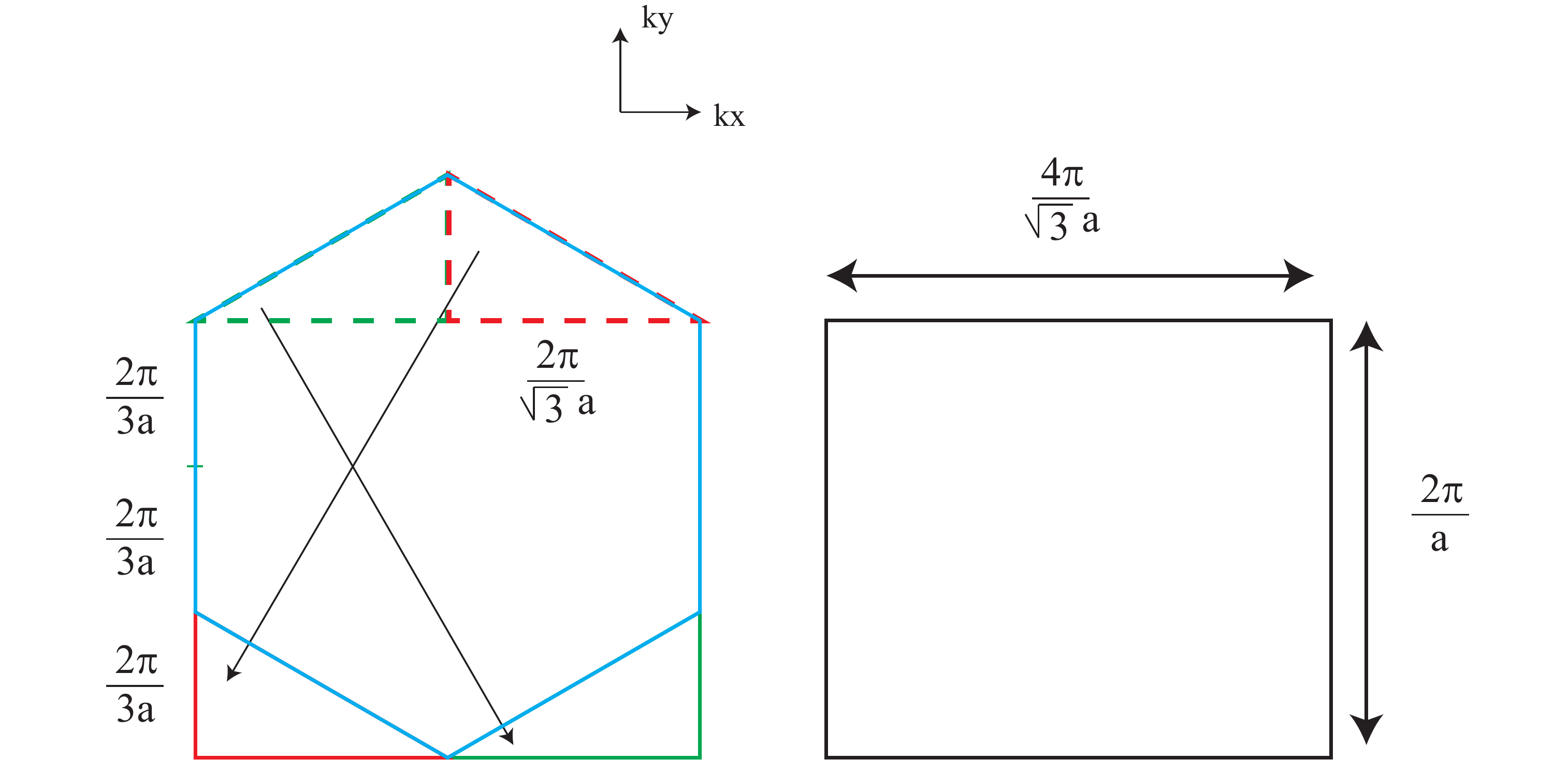}
\end{center}
\caption{Left: Original hexagonal Brillouin zone (blue), and Right: equivalent rectangular Brillouin zone.}
\label{Rect}
\end{figure}

For the numerical integration, it is convenient to use the fact that any two points in adjacent Brillouin zones (or any points connect by multiples of a basis vector) are equivalent. Figure \ref{Rect} shows that the upper two right triangles (red and green) that form the top of the hexagon (blue) can be mapped to the bottom of the hexagon, so that the integration reduces to being over the simple rectangle shown at the right of the figure.

\begin{figure}[ht!]
\begin{center}
\noindent \includegraphics[width=6in]{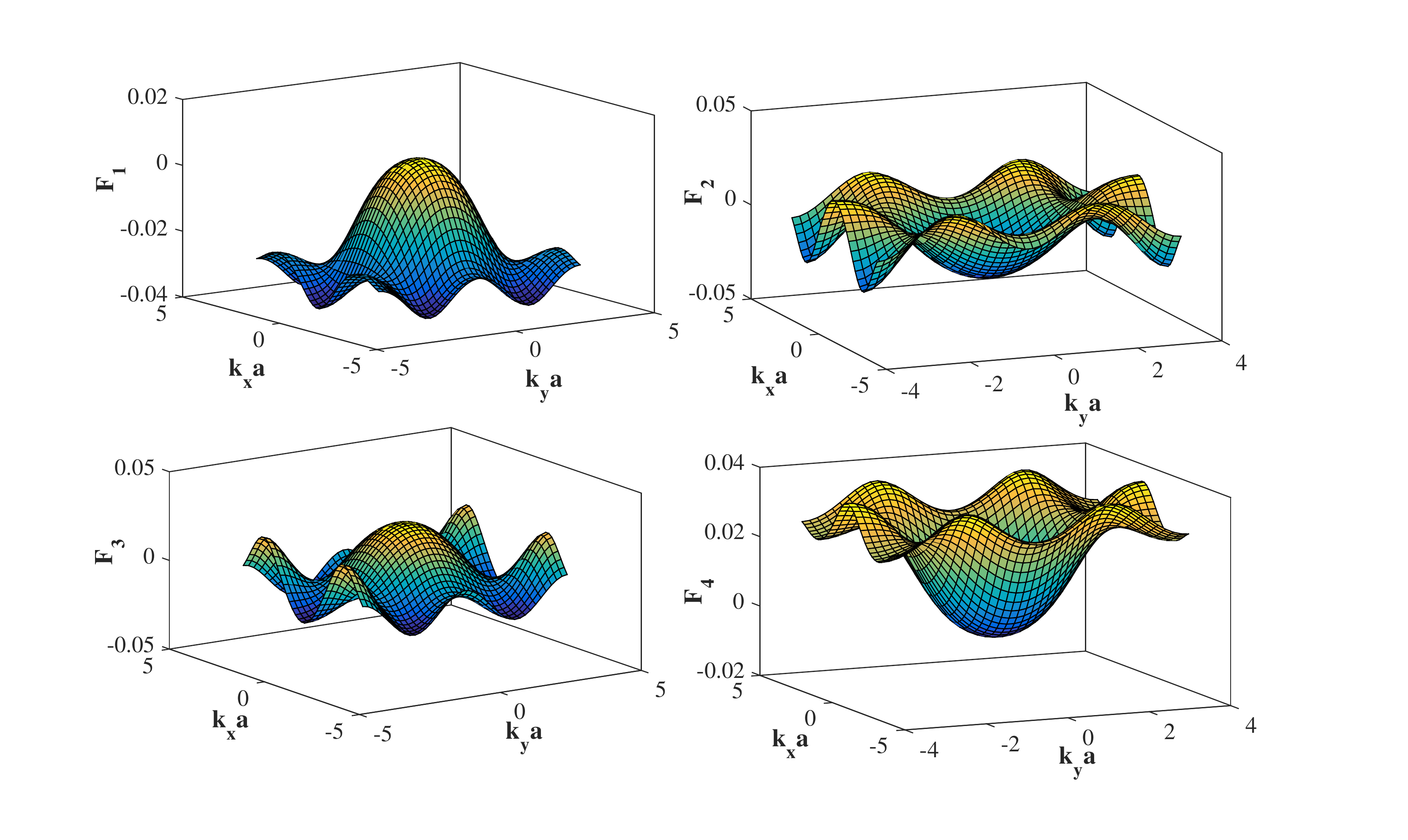}
\end{center}
\caption{Berry curvature for the four bands of Fig. \ref{BD1}; $C_{1,4}=\mp 1$ and $C_{2,3}=0$}
\label{BCP}
\end{figure}

The Berry curvature for each band in the non-reciprocal case is plotted as a function of $k_x-k_y$ in Fig. \ref{BCP}. The two bands $C_{2,3}=0$ have odd Berry curvature (so that they integrate to zero), and the two bands with $C_{1,4}=\pm 1$ have even, sinusoidal Berry curvature (so that they integrate to an integer). In the reciprocal case ($V_p=0$), the Berry curvature is identically zero, by (\ref{2S}), and which is also easy to confirm numerically.

\subsection{Continuum photonic example}

This example is related to \cite{Arthur} (see also \cite{Biao}), with Berry quantities and Chern number analysis directly taken from the seminal work \cite{Mario2}. 

\begin{wrapfigure}{R}{0.35\textwidth}
\begin{centering}
\noindent  \includegraphics[width=4.5in]{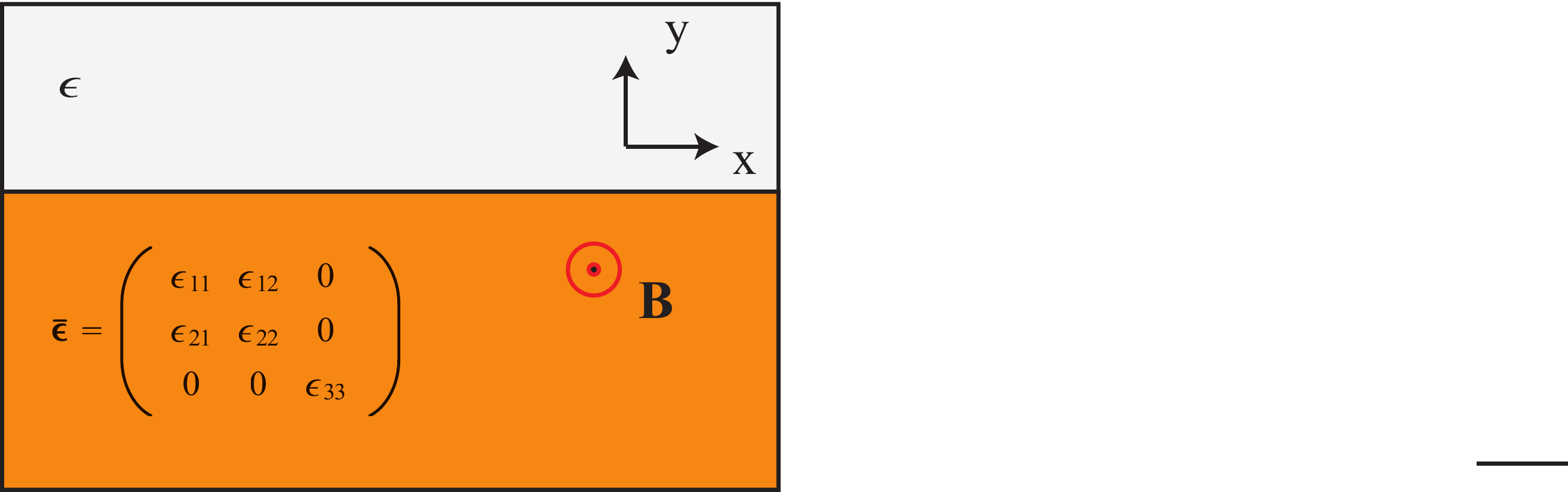}  
\end{centering}
\caption{Interface between a magnetic-field biased plasma (bottom) and a simple material (top). }
\label{BE3}
\end{wrapfigure}

As an example of a nonreciprocal continuous medium, we consider a magnetized plasma in the Voigt configuration (propagation perpendicular to the bias magnetic field $\mathbf{B}$), as depicted in Fig. \ref{BE3}.
For a single-component plasma biased with a static magnetic field $\mathbf{B}%
=\mathbf{z}B_{z}$, the permeability is $\mu=\mu_0$ and the relative permittivity has the form of a Hermitian antisymmetric tensor,
\begin{equation}\label{moe}
\overline{\epsilon }=\left( 
\begin{array}{ccc}
\epsilon _{11} & \epsilon _{12} & 0 \\ 
\epsilon _{21} & \epsilon _{22} & 0 \\ 
0 & 0 & \epsilon _{33}%
\end{array}%
\right) 
\end{equation}%
where
\begin{align}
\varepsilon _{11}& =\varepsilon _{22}=1-\frac{\omega _{p}^{2}}{\omega
^{2}-\omega _{c}^{2}}\text{,  \: }\varepsilon _{33}=1-\frac{\omega _{p}^{2}}{%
\omega ^{2}},\ \   \nonumber \\
\varepsilon _{12}& =-\varepsilon _{21}=i\frac{-\omega _{c}\omega _{p}^{2}}{%
\omega \left( \omega ^{2}-\omega _{c}^{2}\right) } \label{BMPM}
\end{align}%
where the cyclotron frequency is $\omega _{c}=\left( q_{e}/m_{e}\right)
B_{z}\ $and the plasma frequency is $\omega
_{p}^{2}=N_{e}q_{e}^{2}/\varepsilon _{0}m_{e}$. In the above, $N_{e}$ is the
free electron density, and $q_{e}$ and $m_{e}$ are the electron charge and
mass, respectively.

We will also consider the material model examined in \cite{Mario2}, where $\epsilon _{33}=1$ and
\begin{equation}
\epsilon _{11}=\epsilon _{22}=1-\frac{\omega _{0}\omega _{e}}{\omega^{2}-\omega_{0} ^{2}},~~\epsilon _{12}=\epsilon _{21}=i\frac{\omega \omega
_{e}}{\omega^{2}-\omega_{0}^{2}},\label{MBMPM}
\end{equation}
where $|\omega _{0}|$ a the resonance frequency and $\omega _{e}$
determines the resonance strength, with $\omega _{0}\omega _{e}>0$.

\begin{figure}[ht]
\begin{center}
\noindent \includegraphics[width=6in]{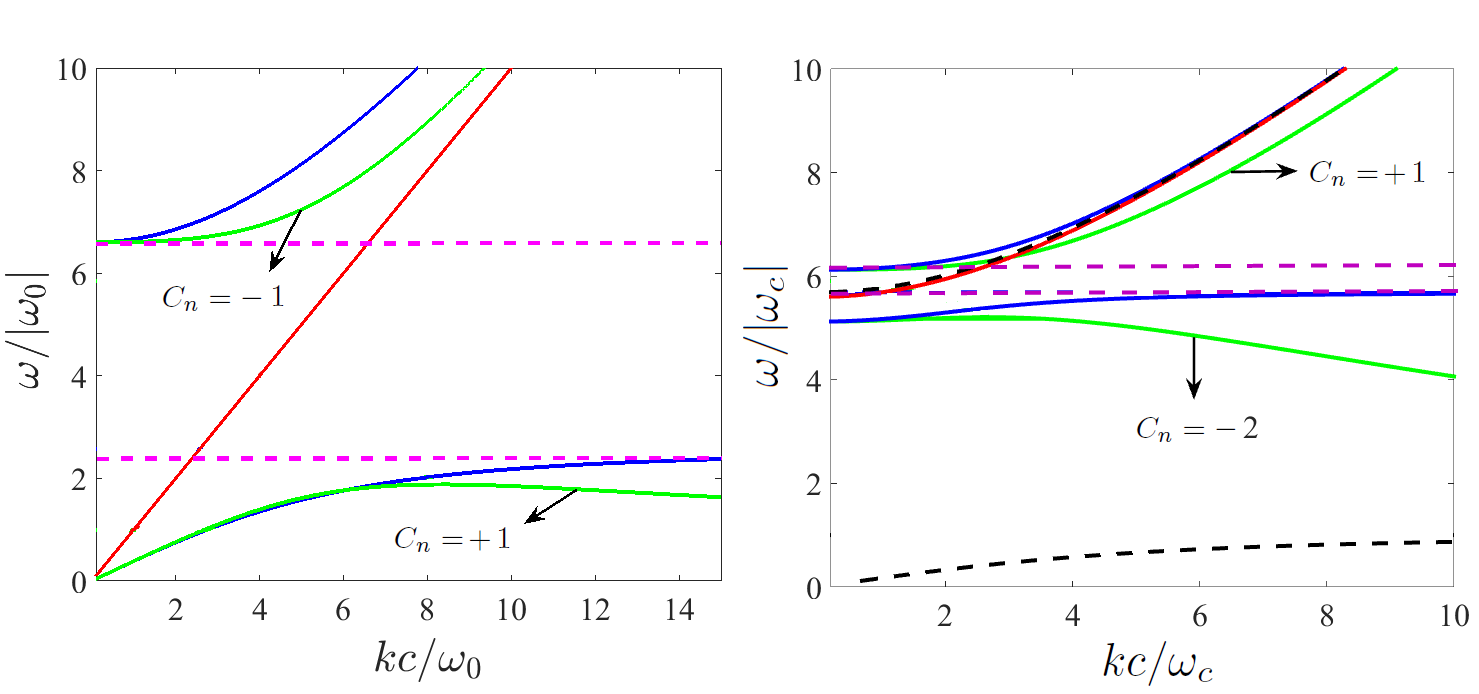}
\end{center}
\caption{Band diagram and Chern numbers (TM modes) for a magneto-optic material; blue: TM mode, no spatial cut-off, green: TM mode, with spatial cut-off, red: TE mode, purple: gap. Left: magneto-optic material (\ref{MBMPM}) with $\protect\omega_e/ \protect\omega_0=5.6$ and $k_{\text{max}}=10$ ($\omega_0/c$), right: magneto-optic material (\ref{BMPM}) with $ \omega_p / 2\pi=9.7 $ THz, $ \omega_c/ 2 \pi= 1.73 $ THz ($ \omega_p / \omega_c=5.6$), and $k_{\text{max}}=10$ ($\omega_c/c$), black: SPP dispersion.}
\label{DC1}
\end{figure}

For propagation in the $x-y$ plane, $\mathbf{k}=(k_{x},k_{y},0)$, the
plane wave supported by this medium can be decoupled into TE ($E_{z}\neq
0,~H_{z}=0$) and TM ($E_{z}=0,~H_{z}\neq 0$) waves. Since there is no
magneto-electric coupling $\overline{\xi }=\overline{\varsigma }=0$, the
dispersion of these modes is%
\begin{align}
& k^{2}=\frac{\epsilon _{11}^{2}+\epsilon _{12}^{2}}{\epsilon _{11}}\left( 
\frac{\omega _{n}}{c}\right) ^{2},~~\mathrm{TM~mode}  \label{dispC} \\
& k^{2}=\epsilon _{33}\left( \frac{\omega _{n}}{c}\right) ^{2},~~\mathrm{%
TE~mode}
\end{align}%
such that $\omega _{n}$ is the eigenfrequency of each mode. Despite the non-reciprocal nature of the medium itself, in the Voigt configuration the bulk dispersion behavior is reciprocal (an interface will break this reciprocity). The dispersion curves for these material are shown in Fig. \ref{DC1} (the spatial cutoff is described later).

The associated electromagnetic waves envelopes can be obtained by finding the solution $f=\left[\mathbf{E},\mathbf{H}\right] ^{T}$, of (\ref{PEE1}), $N \cdot f=\omega M \cdot f$, which is%
\begin{equation}
\left( 
\begin{array}{cc}
0 & -\mathbf{k}\times \mathbf{I}_{3\times 3} \\ 
\mathbf{k}\times \mathbf{I}_{3\times 3} & 0%
\end{array}%
\right) \cdot \left( 
\begin{array}{c}
\mathbf{E} \\ 
\mathbf{H}%
\end{array}%
\right) =\left( 
\begin{array}{cc}
\omega \epsilon _{0}\overline{\epsilon } & 0 \\ 
0 & \omega \mu _{0}\mathbf{I}_{3\times 3}%
\end{array}%
\right) \cdot \left( 
\begin{array}{c}
\mathbf{E} \\ 
\mathbf{H}%
\end{array}%
\right)   \label{EE4}
\end{equation}
so that 
\begin{equation}
\left( 
\begin{array}{cc}
-\mathbf{I}_{3\times 3} & -\frac{\overline{\epsilon }^{-1}}{\omega
\epsilon _{0}}\cdot \mathbf{k}\times \mathbf{I}_{3\times 3} \\ 
\frac{1}{\omega \mu _{0}}\cdot \mathbf{k}\times \mathbf{I}_{3\times 3} & 
-\mathbf{I}_{3\times 3}%
\end{array}%
\right) \cdot \left( 
\begin{array}{c}
\mathbf{E} \\ 
\mathbf{H}%
\end{array}%
\right) =0.
\end{equation}%
With $\mathbf{H}=\widehat{\mathbf{z}}\rightarrow \mathbf{E}=\overline{%
\mathbf{\epsilon }}^{-1}\cdot \frac{\widehat{\mathbf{z}}\times \mathbf{k}}{%
\omega \epsilon _{0}}~~(\mathrm{TM}),~~\mathbf{E}=\widehat{\mathbf{z}}%
\rightarrow \mathbf{H}=\frac{\mathbf{k}}{\omega \mu _{0}}\times \widehat{%
\mathbf{z}}~~(\mathrm{TE})$, we have the $6\times 1$ vectors%
\begin{align}
& f_{nk}^{\text{TM}}=\left( 
\begin{array}{c}
\overline{\mathbf{\epsilon }}^{-1}\cdot \widehat{\mathbf{z}}\times \frac{%
\mathbf{k}}{\epsilon _{0}\omega _{nk}} \\ 
\widehat{\mathbf{z}}%
\end{array}%
\right) ,  \nonumber \\
& f_{nk}^{\text{TE}}=\left( 
\begin{array}{c}
\widehat{\mathbf{z}} \\ 
\frac{\mathbf{k}}{\mu _{0}\omega _{nk}}\times \widehat{\mathbf{z}}%
\end{array}%
\right) .~
\end{align}%
Because the envelopes of the electromagnetic waves in the above equations are
not normalized, the Berry potential is computed using

\begin{equation}
\mathbf{A}_{nk}=\frac{\text{Re}\{if_{nk}^{\ast }\cdot \frac{\partial }{%
\partial \omega }(\omega M_{(\omega )})\partial _{k}f_{n,k}\}}{f_{nk}^{\ast
}\cdot \frac{\partial }{\partial \omega }(\omega M_{(\omega )})f_{n,k}}.
\label{Eq:60}
\end{equation}

Considering the Riemann sphere mapping of the $k_{x}-k_{y}$ plane as
detailed in \cite{Mario2}, it is possible to write the Chern number
associated with $n$th eigenmode branch as%
\begin{equation}
C_{n}=\frac{1}{2\pi }\int \mathbf{A}_{n,k=\infty }\cdot d\mathbf{l}-\frac{1}{2\pi 
}\int \mathbf{A}_{n,k=0^{+}}\cdot d\mathbf{l}
\end{equation}%
where the two line integrals are over infinite and infinitesimal radii
(north and south poles of the Riemann sphere), respectively. If we define $%
A_{nk}=\mathbf{A}_{nk}\cdot \hat{\mathbf{\phi}}$ then we have 
\begin{equation}
C_{n}=\lim\limits_{k\rightarrow \infty }(A_{n,\phi
=0}k)-\lim\limits_{k\rightarrow 0^{+}}(A_{n,\phi =0}k).  \label{Eq:7}
\end{equation}

For a lossless TM-mode in propagating in the $x-y$ plane we have $k=k_{x}%
\hat{\mathbf{x}}+k_{y}\hat{\mathbf{y}}=k \cos(\phi )\hat{\mathbf{x}}+k \sin(\phi )\hat{\mathbf{y}}$. Writing%
\begin{equation}
\overline{\epsilon }^{-1}=\left( 
\begin{array}{ccc}
\alpha _{11} & \alpha _{12} & 0 \\ 
\alpha _{21} & \alpha _{22} & 0 \\ 
0 & 0 & \alpha _{33}%
\end{array}%
\right) 
\end{equation}%
we have%
\begin{equation}
f_{nk}=\left( 
\begin{array}{c}
\overline{\mathbf{\epsilon }}^{-1}\cdot \widehat{\mathbf{z}}\times \frac{%
\mathbf{k}}{\epsilon _{0}\omega _{nk}} \\ 
\widehat{\mathbf{z}}%
\end{array}%
\right) =\left( 
\begin{array}{c}
\frac{-\alpha _{11}k_{y}+\alpha _{12}k_{x}}{\epsilon _{0}\omega _{n}} \\ 
\frac{-\alpha _{21}k_{y}+\alpha _{22}k_{x}}{\epsilon _{0}\omega _{n}} \\ 
0 \\ 
0 \\ 
0 \\ 
1%
\end{array}%
\right) ,~~\partial _{k}f_{nk}=\left( 
\begin{array}{c}
\frac{-\alpha _{11}\hat{y}+\alpha _{12}\hat{x}}{\epsilon _{0}\omega _{n}} \\ 
\frac{-\alpha _{21}\hat{y}+\alpha _{22}\hat{x}}{\epsilon _{0}\omega _{n}} \\ 
0 \\ 
0 \\ 
0 \\ 
0%
\end{array}%
\right) 
\end{equation}%
where%
\begin{equation}
\alpha _{11}=\frac{\epsilon _{22}}{\epsilon _{11}\epsilon _{22}-\epsilon
_{12}\epsilon _{21}},~~\alpha _{22}=\frac{\epsilon _{11}}{\epsilon
_{11}\epsilon _{22}-\epsilon _{12}\epsilon _{21}},~~\alpha _{12}=\frac{%
-\epsilon _{12}}{\epsilon _{11}\epsilon _{22}-\epsilon _{12}\epsilon _{21}}%
,~~\alpha _{21}=\frac{-\epsilon _{21}}{\epsilon _{11}\epsilon _{22}-\epsilon
_{12}\epsilon _{21}},
\end{equation}%
such that%
\begin{equation}
f_{nk}^{\ast }=\frac{1}{\epsilon _{0}\omega _{n}}\left( 
\begin{array}{cccccc}
(-\alpha _{11}k_{y}+\alpha _{12}k_{x})^{\ast } & (-\alpha _{21}k_{y}+\alpha
_{22}k_{x})^{\ast } & 0 & 0 & 0 & 1%
\end{array}%
\right) .
\end{equation}%
From the frequency derivative of the material response matrix, $\partial
_{\omega }(\omega M)$, we have $\beta _{ij}=\partial _{\omega }(\omega
\epsilon _{0}\epsilon _{ij})$. So, for the Berry potential we have%
\begin{equation}
\mathbf{A}_{nk}=\frac{Re\{if_{nk}^{\ast }\cdot \frac{1}{2} \frac{\partial }{\partial
\omega }(\omega M_{(\omega )})\partial _{k}f_{n,k}\}}{f_{nk}^{\ast }\cdot \frac{1}{2} 
\frac{\partial }{\partial \omega }(\omega M_{(\omega )})f_{n,k}}=\frac{%
Re\{N_{x}+N_{y}\}}{D}  \label{AB1}
\end{equation}%
where%
\begin{align}
& N_{x}=\frac{i}{2(\epsilon _{0}\omega _{n})^{2}}\{-2\alpha _{11}\alpha
_{12}[k_{x}\beta _{12}+k_{y}\beta _{11}]+(|\alpha _{11}|^{2}+|\alpha
_{12}|^{2})[k_{x}\beta _{11}-k_{y}\beta _{12}]\}\hat{x}  \label{AB2} \\
& N_{y}=\frac{i}{2(\epsilon _{0}\omega _{n})^{2}}\{2\alpha _{11}\alpha
_{12}[k_{x}\beta _{11}-k_{y}\beta _{12}]+(|\alpha _{11}|^{2}+|\alpha
_{12}|^{2})[k_{x}\beta _{12}+k_{y}\beta _{11}]\}\hat{y}  \nonumber \\
& D=\frac{|k|^{2}}{2(\epsilon _{0}\omega _{n})^{2}}[(|\alpha
_{11}|^{2}+|\alpha _{12}|^{2})\beta _{11}-2\alpha _{11}\alpha _{12}\beta
_{12}]+\mu_0 .
\end{align}%
Therefore, for the Chern number calculation we obtain%
\begin{align}
& A_{n}=\mathbf{A}_{n}\cdot \hat{\phi}=\frac{Re\{N_{y}cos(\phi
)-N_{x}sin(\phi )\}}{D}  \label{BCa} \\
& A_{n}(\phi =0)=\frac{Re\{N_{y}(\phi =0)\}}{D},~~N_{y}(\phi =0)=\frac{ik}{%
(\epsilon _{0}\omega _{n})^{2}}\{2\alpha _{11}\alpha _{12}\beta
_{11}+(|\alpha _{11}|^{2}+|\alpha _{12}|^{2})\beta _{12}\}  \nonumber \\
& A_{n}(\phi =0)k=\frac{Re(\frac{i|k|^{2}}{(\epsilon _{0}\omega _{n})^{2}}%
\{2\alpha _{11}\alpha _{12}\beta _{11}+(|\alpha _{11}|^{2}+|\alpha
_{12}|^{2})\beta _{12}\})}{\frac{|k|^{2}}{(\epsilon _{0}\omega _{n})^{2}}%
[(|\alpha _{11}|^{2}+|\alpha _{12}|^{2})\beta _{11}-2\alpha _{11}\alpha
_{12}\beta _{12}]+\mu_0 } \label{Eq:84}.
\end{align}%
These expressions are used below in calculating the Chern number from (\ref%
{Eq:7}).

\subsubsection{Chern number calculation as a surface integral over the $k_x-k_y$
plane}

From (\ref{AB2}) the Berry curvature is%
\begin{align}
& \mathbf{F}_{k}=\mathrm{Re}\{\frac{\partial A_{x}(k_{x},k_{y})}{\partial k_{y}}-\frac{%
\partial A_{y}(k_{x},k_{y})}{\partial k_{x}}\}(-\hat{\mathbf{z}})  \nonumber \\
& \mathbf{F}_{k}=\mathrm{Re}\{\frac{i\hat{\mathbf{z}}}{D(\epsilon _{0}\omega _{n})^{2}}\{2\alpha
_{11}\alpha _{12}\beta _{11}+(\rvert \alpha _{11}\rvert ^{2}+\rvert \alpha
_{12}\rvert ^{2})\beta _{12}\}\}.
\end{align}%
If we consider propagation in $\mathbf{k}-$space such that $k_{z}=0$
then the Chern number computed over the infinite surface is%
\begin{align}
& C=\frac{1}{2\pi }=\int_{k_{x}=-\infty }^{k_{x}=+\infty
}\int_{k_{y}=-\infty }^{k_{y}=+\infty }dk_{x}dk_{y}\cdot \mathbf{F}_{k}=\frac{1}{2\pi 
}\int_{\phi =0}^{\phi =2\pi }\int_{k=0}^{k=\infty }kdkd\phi \cdot \mathbf{F}_{k} 
\nonumber \\
& C=\left( \delta (k)\right) _{k=0}^{k=\infty }=\left( \mathrm{Re}\{\frac{i}{%
D(\epsilon _{0}\omega _{n})^{2}} \frac{1}{2} \{2\alpha _{11}\alpha _{12}\beta
_{11}+(\rvert \alpha _{11}\rvert ^{2}+\rvert \alpha _{12}\rvert ^{2})\beta
_{12}\}\}\right) _{k=0}^{k=\infty }
\end{align}%
which leads to (\ref{Eq:7}) with $\delta (k)=A_{n}(\phi =0)k$ in (\ref{Eq:84}). Therefore the Chern number computed as an infinite surface
integral is the same as computed via the line integral near the north and
south poles of the Riemann sphere, as shown in \cite{Mario2}. 

\subsubsection{Low frequency band of the TM-mode, material model (\ref{MBMPM})}

For material model (\ref{MBMPM}), we will denote the lower curve in Fig. \ref{DC1} as the low frequency band of the TM-mode. When $k\rightarrow \infty $ from the TM-dispersion relation (\ref{dispC}), then $\omega _{n}$ should tend
to the zero of $\epsilon _{11}$, which is $\omega _{n}=\sqrt{\omega
_{0}^{2}+\omega _{0}\omega _{e}}$. Since%
\begin{equation}
k^{2}=\frac{\epsilon _{11}^{2}+\epsilon _{12}^{2}}{\epsilon _{11}}\left( 
\frac{\omega _{n}}{c}\right) ^{2}=\frac{\omega _{0}^{2}-\omega ^{2}+2\omega
_{0}\omega _{e}+\omega _{e}^{2}}{\omega _{0}^{2}-\omega ^{2}+\omega
_{0}\omega _{e}}\left( \frac{\omega _{n}}{c}\right) ^{2},~~\mathrm{TM~mode,}
\end{equation}%
in the limit $k\rightarrow \infty $ we obtain $\epsilon _{11}=0$, $\alpha
_{11}=0$ and $(k/\omega _{n})^{2}\rightarrow \infty $. So, for $A_{n}(\phi
=0)$ we get%
\begin{equation}
A_{n}(\phi =0)k=\frac{Re\frac{i|k|^{2}}{(\epsilon _{0}\omega _{n})^{2}}%
\{|\alpha _{12}|^{2}\beta _{12}\}}{\frac{|k|^{2}}{(\epsilon _{0}\omega
_{n})^{2}}\{|\alpha _{12}|^{2}\beta _{11}\}+\mu_0 }=\frac{Re\frac{i|k|^{2}}{%
(\epsilon _{0}\omega _{n})^{2}}\{|\frac{1}{\epsilon _{12}}|^{2}\beta _{12}\}%
}{\frac{|k|^{2}}{(\epsilon _{0}\omega _{n})^{2}}\{|\frac{1}{\epsilon _{12}}%
|^{2}\beta _{11}\}+\mu_0 }  \label{Eq.13}
\end{equation}%
such that%
\begin{equation}
\beta _{11}=\epsilon _{0}(1+\omega _{0}\omega _{e}\frac{\omega
_{0}^{2}+\omega ^{2}}{(\omega _{0}^{2}-\omega ^{2})^{2}}),~~\beta
_{12}=-i\epsilon _{0}\omega _{e}\frac{2\omega \omega _{0}^{2}}{(\omega
_{0}^{2}-\omega ^{2})^{2}}.
\end{equation}%
For $\omega _{n}=\sqrt{\omega _{0}^{2}+\omega _{0}\omega _{e}}$ we have $%
\beta _{11}=2\epsilon _{0}(1+\frac{\omega _{0}}{\omega _{e}})$ and $\beta
_{12}=-\left( 2\epsilon _{0}i/\omega _{e}\right) \sqrt{\omega
_{0}^{2}+\omega _{0}\omega _{e}}$. Finally, taking the limit $k\rightarrow
\infty $ ,%
\begin{equation}
\lim\limits_{k\rightarrow \infty }A_{n}(\phi =0)k=\frac{\sqrt{\omega
_{0}^{2}+\omega _{0}\omega _{e}}/\omega _{e}}{(1+\frac{\omega _{0}}{\omega
_{e}})}=\frac{|\omega _{0}|\sqrt{1+\frac{\omega _{e}}{\omega _{0}}}}{\omega
_{0}(1+\frac{\omega _{e}}{\omega _{0}})}=\frac{\text{sgn}(\omega _{0})}{%
\sqrt{1+\frac{\omega _{e}}{\omega _{0}}}}=\frac{\text{sgn}(\omega _{e})}{%
\sqrt{1+|\frac{\omega _{e}}{\omega _{0}}}|}
\end{equation}

When $k\rightarrow 0$, the lower band of TM-mode tends to the light line
(which means $\omega _{n}\rightarrow 0$). Therefore $\epsilon _{12}=0$ and $%
\beta _{12}=0$, which leads to $\lim\limits_{k\rightarrow 0}A_{n}(\phi =0)k=0$%
. Eventually, for Chern number we obtain%
\begin{equation}
C_{n=1}=\lim\limits_{k\rightarrow \infty }(A_{n,\phi
=0}k)-\lim\limits_{k\rightarrow 0^{+}}(A_{n,\phi =0}k)=\frac{\text{sgn}%
(\omega _{e})}{\sqrt{1+|\frac{\omega _{e}}{\omega _{0}}}|}  .\label{Eq:32}
\end{equation}
The fact that this Chern number is not an integer will be addressed below, and the solution of this issue is a fundamental contribution of \cite{Mario2}.

\subsubsection{High frequency band of the TM-mode, material model (\ref{MBMPM})}

For material model (\ref{MBMPM}), denoting the upper curve in Fig. \ref{DC1}) as the high-frequency band, from the dispersion relation (\ref{dispC}),%
\begin{equation}
k^{2}=\frac{\omega _{0}^{2}-\omega ^{2}+2\omega _{0}\omega _{e}+\omega
_{e}^{2}}{\omega _{0}^{2}-\omega ^{2}+\omega _{0}\omega _{e}}\left( \frac{%
\omega _{n}}{c}\right) ^{2}\rightarrow 0~~\mathrm{if}~~\omega
_{n}=0,~~\omega _{n}=\left|\omega _{0}+\omega _{e} \right |.
\end{equation}%
If $\omega _{n}=0$, then%
\begin{equation}
\epsilon _{11}=1+\frac{\omega _{e}}{\omega _{0}},~~\epsilon _{12}=0,~~\alpha
_{11}=\frac{1}{1+\omega _{e}/\omega _{0}},~~\alpha _{12}=0,~~\beta
_{11}=1+\omega _{e}/\omega _{0},~~\beta _{12}=0,~~\left( \frac{k}{\omega _{n}%
}\right) ^{2}=\frac{1}{c^{2}}\left( 1+\frac{\omega _{e}}{\omega _{0}}\right)
^{2}
\end{equation}%
so for the numerator of $A_{n}(\phi =0)k$ we have $2\alpha _{11}\alpha
_{12}\beta _{11}+(|\alpha _{11}|^{2}+|\alpha _{12}|^{2})\beta _{12}=0$.
Therefore there is no contribution for the eigenfrequency $\omega _{n}=0$ as 
$k\rightarrow 0$. In fact it is obvious that $ \omega_n=0 $ has no contribution because this eigenfrequency as $ k \rightarrow 0 $ belongs to the TM low frequency band and it has no effect on the high frequency TM band.

If $\omega _{n}= \left | \omega _{0}+\omega _{e} \right | $, it can be shown that%
\begin{equation}
\epsilon _{12}=i\epsilon_{11} \mathrm{sgn}(\omega_{e}),~~\alpha _{11}=\frac{\epsilon _{11}}{\epsilon
_{11}^{2}+\epsilon _{12}^{2}}\rightarrow \infty ,~~\alpha _{12}=\frac{%
-\epsilon _{12}}{\epsilon _{11}^{2}+\epsilon _{12}^{2}}\rightarrow \infty
,~~\beta _{12}=i\beta _{11} \mathrm{sgn}(\omega_{e}),~~\left( \frac{k}{\omega _{n}}\right) ^{2}=\frac{%
1}{c^{2}}\frac{\epsilon _{11}^{2}+\epsilon _{12}^{2}}{\epsilon _{11}}%
\rightarrow 0
\end{equation}%
and by carefully treating the limit $ \lim\limits_{k\rightarrow 0}A_{n}(\phi =0)k$ and considering the fact that for $ \omega _{n}= \left | \omega _{0}+\omega _{e} \right | $ we have $ \epsilon _{11}=\left( \omega _{e}^{2}+\omega _{0}\omega _{e}\right) /\left(
\omega _{e}^{2}+2\omega _{0}\omega _{e}\right) >0$ then it can be shown that 

\begin{equation}
\lim\limits_{k\rightarrow 0}A_{n}(\phi =0)k=\mathrm{sgn}(\omega_e).
\end{equation}%
When $k\rightarrow \infty $ then the high-frequency mode tends to the light
line, so that $\lim\limits_{k\rightarrow \infty }A_{n}(\phi =0)k=0$.
Eventually for the high frequency TM-band we have%
\begin{equation}
C_{n}=\lim\limits_{k\rightarrow \infty }(A_{n,\phi
=0}k)-\lim\limits_{k\rightarrow 0^{+}}(A_{n,\phi =0}k)=-\mathrm{sgn}(\omega_e).  \label{Eq:37}
\end{equation}

\subsubsection{TE-Mode}

Using same procedure as above, it is straightforward to show that for the TE-Mode we have%
\begin{equation}
C_{n}=\lim\limits_{k\rightarrow \infty }(A_{n,\phi
=0}k)-\lim\limits_{k\rightarrow 0^{+}}(A_{n,\phi =0}k)=0.  \label{Eq:37a}
\end{equation}

\subsubsection{Material model (\ref{BMPM})}

Considering material model (\ref{BMPM}), from the dispersion equation (\ref{dispC}),%
\begin{equation}
k^{2}=\frac{\omega ^{2}(\omega ^{2}-\omega _{c}^{2})-2\omega ^{2}\omega
_{p}^{2}+\omega _{p}^{4}}{\omega ^{2}-\omega _{c}^{2}-\omega _{p}^{2}}\frac{1%
}{c^{2}}.\label{keq}
\end{equation}

As $k\rightarrow \infty $ and regarding Fig. (\ref{DC1}) we have $\omega
_{n}\rightarrow \infty $ for the high frequency band, and, for the low
frequency band, $\omega ^{2}-\omega _{c}^{2}-\omega _{p}^{2}=0$, such that $%
\omega _{n}=\sqrt{\omega _{c}^{2}+\omega _{p}^{2}}$.

For the TM mode if $k\rightarrow 0$ we have%
\[
\omega ^{2}(\omega ^{2}-\omega _{c}^{2})-2\omega ^{2}\omega _{p}^{2}+\omega
_{p}^{4}=0~\rightarrow ~%
\begin{cases}
\omega _{n}^{2}=\frac{\omega _{h}^{2}}{2}\left\{ 1+\sqrt{1-4(\frac{\omega
_{p}}{\omega _{h}})^{4}}\right\} , & \text{for high frequency TM } \\ 
\omega _{n}^{2}=\frac{\omega _{h}^{2}}{2}\left\{ 1-\sqrt{1-4(\frac{\omega
_{p}}{\omega _{h}})^{4}}\right\} , & \text{for low frequency TM}%
\end{cases}%
\]%
where $\omega _{h}^{2}=\omega _{c}^{2}+2\omega _{p}^{2}$.

The Chern number is (\ref{Eq:7}) with (\ref{Eq:84}), and 
\begin{equation}
\beta _{11}=1+\omega _{p}^{2}\frac{\omega ^{2}+\omega _{c}^{2}}{(\omega
^{2}-\omega _{c}^{2})^{2}},~~\beta _{12}=2i\omega _{c}\omega _{p}^{2}\frac{%
\omega }{(\omega ^{2}-\omega _{c}^{2})^{2}}.
\end{equation}

For the low frequency TM band when $k\rightarrow \infty $ ($\omega _{n}=%
\sqrt{\omega _{c}^{2}+\omega _{p}^{2}}$), $\epsilon _{11}=0$ and $\alpha
_{11}=0$. Therefore,%
\begin{equation}
\lim_{k\rightarrow \infty }(A_{n,\phi =0}k)=\mathrm{Re}\left\{ \frac{i\beta
_{12}}{\beta _{11}}\right\} _{\omega _{n}=\sqrt{\omega _{c}^{2}+\omega
_{p}^{2}}}=-\frac{\mathrm{sgn}(\omega _{c})}{\sqrt{1+(\frac{\omega _{p}}{%
\omega _{c}})^{2}}}
\end{equation}

For the case of $k\rightarrow 0$, we have $\omega _{n}^{2}=\frac{\omega
_{h}^{2}}{2}\left\{ 1-\sqrt{1-4\left( \frac{\omega _{p}}{\omega _{h}}\right)
^{4}}\right\} $ which is the pole of $\alpha _{11}$ and $\alpha _{12}$, so $%
\alpha _{11}\rightarrow \infty $, $\alpha _{12}\rightarrow \infty $. Then, 
\begin{align}
\lim_{k\rightarrow 0}(A_{n,\phi =0}k)& =\lim_{k\rightarrow 0}\frac{\mathrm{Re%
}(\frac{i}{(\epsilon _{0}c)^{2}\alpha _{11}}\{2\alpha _{11}\alpha _{12}\beta
_{11}+(|\alpha _{11}|^{2}+|\alpha _{12}|^{2})\beta _{12}\})}{\frac{1}{%
(\epsilon _{0}c)^{2}\alpha _{11}}\{(|\alpha _{11}|^{2}+|\alpha
_{12}|^{2})\beta _{11}-2\alpha _{11}\alpha _{12}\beta _{12}\}+\mu _{0}} 
\nonumber \\
& =\left\{ \frac{\mathrm{Re}(\frac{i}{(\epsilon _{0}c)^{2}}\{2\frac{\alpha
_{12}}{\alpha _{11}}\beta _{11}+(1+\frac{|\alpha _{12}|^{2}}{\alpha _{11}^{2}%
})\beta _{12}\})}{\frac{1}{(\epsilon _{0}c)^{2}}\{(1+\frac{|\alpha _{12}|^{2}%
}{\alpha _{11}^{2}})\beta _{11}-2\frac{\alpha _{12}}{\alpha _{11}}\beta
_{12}\}}\right\} _{\omega _{n}^{2}=\frac{\omega _{h}^{2}}{2}\left\{ 1-\sqrt{%
1-4\left( \frac{\omega _{p}}{\omega _{h}}\right) ^{4}}\right\} }=1,
\end{align}%
and the Chern number of the low frequency band is%
\begin{equation}
C_{n}=-\frac{\mathrm{sgn}(\omega _{c})}{\sqrt{1+(\frac{\omega _{p}}{\omega
_{c}})^{2}}}-1. \label{CNLF}
\end{equation}%
$\allowbreak $

For the high frequency band when $k\rightarrow \infty $ we have $\omega
_{n}\rightarrow \infty $, $\epsilon _{11}=1$, $\epsilon _{12}=0$, $\alpha
_{11}=1$, $\alpha _{12}=0$ and $\beta _{12}=0$, so $\lim_{k\rightarrow
\infty }(A_{n,\phi =0}k)=0$ and for the case of $k\rightarrow 0$ we have $%
\omega _{n}^{2}=\frac{\omega _{h}^{2}}{2}\left\{ 1+\sqrt{1-4(\frac{\omega
_{p}}{\omega _{h}})^{4}}\right\} $, which is a pole of $\alpha _{11}$ and $%
\alpha _{12}$ so $\alpha _{11}\rightarrow \infty ,~\alpha _{12}\rightarrow
\infty $. Then, 
\[
\lim_{k\rightarrow 0}(A_{n,\phi =0}k)=\left\{ \frac{\mathrm{Re}(\frac{i}{%
(\epsilon _{0}c)^{2}}\{2\frac{\alpha _{12}}{\alpha _{11}}\beta _{11}+(1+%
\frac{|\alpha _{12}|^{2}}{\alpha _{11}^{2}})\beta _{12}\})}{\frac{1}{%
(\epsilon _{0}c)^{2}}\{(1+\frac{|\alpha _{12}|^{2}}{\alpha _{11}^{2}})\beta
_{11}-2\frac{\alpha _{12}}{\alpha _{11}}\beta _{12}\}}\right\} _{\omega
_{n}^{2}=\frac{\omega _{h}^{2}}{2}\left\{ 1+\sqrt{1-4(\frac{\omega _{p}}{%
\omega _{h}})^{4}}\right\} }=-1,
\]%
and so for the high frequency band the Chern number is%
\[
C_{n}=0-(-1)=1.
\]

It can be seen that generally the Chern number of the high frequency band is an integer, but that of the low frequency TM band is not (as was found for the material model (\ref{BMPM})). In both material models, when off-diagonal permittivity elements in (\ref{moe}) are set to zero, all Chern numbers are $C_n =0$.

\subsubsection{Integer Chern numbers and wave vector cutoff for magneto-optic material response}

The non-integer Chern numbers for the low TM band for both material models, (\ref{Eq:32}) and (\ref{CNLF}), arise from the continuum nature of the material \cite{Mario2}, associated with the Hamiltonian not being
sufficiently well-behaved at infinity (mapped to the north pole of the
Riemann sphere). The problem is thoroughly discussed in \cite{Mario2}, and
here we merely repeat the solution therein. The issue can be solved by
introducing a high-frequency spatial cutoff by defining a nonlocal material 
\begin{equation}
M_{\text{reg}}(\omega ,k)=M_{\infty }+\frac{1}{1+k^{2}/k_{\text{max}}^{2}}%
[M(\omega )-M_{\infty }]
\end{equation}%
where $M_{\infty }=\lim\limits_{\omega \rightarrow \infty }M(\omega )$. This material response tends to the local response as $k_{\text{%
max}}\rightarrow \infty \footnote{%
Noting the Fourier transform pair 
\begin{equation}
\frac{1}{1+k^{2}/k_{\text{max}}^{2}}\leftrightarrow \frac{k_{\text{max}}}{2}%
e^{-k_{\text{max}}r}
\end{equation}%
and as $k_{\text{max}}\rightarrow \infty $ we have $\lim_{k_{\text{max}%
}\rightarrow \infty }\frac{k_{\text{max}}}{2}e^{-k_{\text{max}}r}=\delta (r)$%
, which indicates locality.}$  By considering a wave vector cutoff for, e.g., the material model (\ref{MBMPM}), the
non-local parameters of the material are 
\begin{equation}
\epsilon _{11}=\epsilon _{22}=1+\frac{1}{1+k^{2}/k_{\text{max}}^{2}}\frac{%
\omega _{0}\omega _{e}}{\omega _{0}^{2}-\omega ^{2}},~~\epsilon
_{12}=-\epsilon _{21}=\frac{-i}{1+k^{2}/k_{\text{max}}^{2}}\frac{\omega
_{e}\omega }{\omega _{0}^{2}-\omega ^{2}},~~\epsilon _{33}=1,~~\mathbf{%
\mu }=\mathrm{diag}\{\mu ,\mu ,\mu \}
\end{equation}

For the low frequency TM-band we have%
\begin{equation}
k^{2}=\frac{\omega _{0}^{2}-\omega ^{2}+2\gamma \omega _{0}\omega
_{e}+\gamma ^{2}\omega _{e}^{2}}{\omega _{0}^{2}-\omega ^{2}+\gamma \omega
_{0}\omega _{e}}(\frac{\omega }{c})^{2}  \label{Eq:57}
\end{equation}%
such that $\gamma =\frac{1}{1+k^{2}/k_{\text{max}}^{2}}$ describes the
non-locality. If $k\rightarrow \infty $ then $\gamma \rightarrow 0$ and  $%
\omega _{n}\rightarrow |\omega _{0}|$, which is a zero of $\epsilon _{11}$
in the limit of $k\rightarrow \infty $. Therefore for the case of $%
k\rightarrow \infty $ we have%
\begin{equation}
\epsilon _{11}=0,~~\alpha _{11}=0,~~\beta _{11}=\gamma \omega _{0}\omega _{e}%
\frac{2\omega ^{2}}{(\omega _{0}^{2}-\omega ^{2})^{2}},~~\beta
_{12}=-i\gamma \omega _{e}\frac{2\omega \omega _{0}^{2}}{(\omega
_{0}^{2}-\omega ^{2})^{2}}.
\end{equation}
It can be shown that%
\begin{equation}
\lim\limits_{k\rightarrow \infty }A_{n}(\phi =0)k=\frac{i|\alpha
_{12}|^{2}\beta _{12}}{|\alpha _{12}|^{2}\beta _{11}}=\frac{i\beta _{12}}{%
\beta _{11}}=\frac{\omega _{0}}{|\omega _{0}|}=\text{sgn}(\omega _{0})=\text{%
sgn}(\omega _{e}),
\end{equation}%
and for the case that $k\rightarrow 0$ the low frequency band of the TM-mode
tends to the light line and so $\lim\limits_{k\rightarrow 0}A_{n}(\phi =0)k=0
$. Eventually for the Chern number of the low frequency TM band we obtain%
\begin{equation}
C_{n}=\lim\limits_{k\rightarrow \infty }(A_{n,\phi
=0}k)-\lim\limits_{k\rightarrow 0^{+}}(A_{n,\phi =0}k)=\text{sgn}(\omega
_{e}),  \label{Eq:43}
\end{equation}%
the desired integer.

For the high frequency TM-band nothing changes from the previous development
because the contribution to Chern number comes from $k\rightarrow 0$, and in
this limit the non-local response turn into local response and the Chern
number is the same as previously obtained. So, for high frequency TM-band we
have%
\begin{equation}
C_{n}=\lim\limits_{k\rightarrow \infty }(A_{n,\phi
=0}k)-\lim\limits_{k\rightarrow 0^{+}}(A_{n,\phi =0}k)=-\mathrm{sgn}(\omega_e)  \label{Eq:44}
\end{equation}%

Introducing the wave number cutoff has no effect on the TE-mode because $%
\epsilon _{33}$ does not change, and so the Chern number of this mode
remains the same as in the previous section ($C_{n}=0$). 

Therefore, we have Chern numbers $C_\mathrm{high}=\text{sgn}(\omega_{e})$ and $C_\mathrm{low}=\text{sgn}(\omega_{e})$ for the higher and lower band, respectively, so that the sum of the Chern numbers is zero. The band dispersion and integer Chern numbers are shown in Fig. \ref{DC1}.

\noindent \textbf{Biased Plasma Case}

For the material model (\ref{MBMPM}), the permittivity tensor components become%
\begin{equation}
\epsilon _{11}=\epsilon _{22}=1-\gamma \frac{\omega _{p}^{2}}{\omega
^{2}-\omega _{c}^{2}},~~\epsilon _{12}=-\epsilon _{21}=-i\gamma \frac{\omega
_{c}\omega _{p}^{2}}{\omega (\omega ^{2}-\omega _{c}^{2})}
\end{equation}%
such that $\gamma =1/({1+k^{2}/k_{\mathrm{max}}^{2}})$.

For this case, the dispersion equation is%
\begin{equation}
k^{2}=\frac{\epsilon _{11}(k)^{2}+\epsilon _{12}(k)^{2}}{\epsilon _{11}(k)}(%
\frac{\omega }{c})^{2}.
\end{equation}%
We have $k\rightarrow \infty $ if $\epsilon _{11}(k)=0$ and $\epsilon
_{12}(k)\neq 0$, or $\omega _{n}\rightarrow \infty $, or $\omega _{n}=\omega
_{c}$. The eigen frequency of the higher TM band is $\omega _{n}\rightarrow
\infty $ and that of lower frequency band comes from the zero of $\epsilon
_{11}(k)$,

\[
\epsilon _{11}(k)=1-\gamma \frac{\omega _{p}^{2}}{\omega ^{2}-\omega _{c}^{2}%
}=0\rightarrow \omega _{n}=\sqrt{\omega _{c}^{2}+\gamma \omega _{p}^{2}}.
\]%
When $k\rightarrow \infty $ then $\gamma \rightarrow 0$ so for the low
frequency band the eigenfrequency is $\omega _{n}=\lim\limits_{\gamma
\rightarrow 0}\sqrt{\omega _{c}^{2}+\gamma \omega _{p}^{2}}=|\omega _{c}|$. 

For $k\rightarrow \infty $, $\omega _{n}=|\omega _{c}|$, we have $\epsilon
_{11}(k)=0$, $\alpha _{11}(k)=0$ and so%
\begin{equation}
\lim_{k\rightarrow \infty }(A_{n,\phi =0}k)=\mathrm{Re}\left\{ \frac{i\beta
_{12}(k)}{\beta _{11}(k)}\right\} _{\omega _{n}=|\omega _{c}|}
\end{equation}%
such that $\beta _{11}(k)=1+\gamma \omega _{p}^{2}\frac{\omega ^{2}+\omega
_{c}^{2}}{(\omega ^{2}-\omega _{c}^{2})^{2}},~~\beta _{12}=2i\gamma \omega
_{c}\omega _{p}^{2}\frac{\omega }{(\omega ^{2}-\omega _{c}^{2})^{2}}$ so the
contribution form $k\rightarrow \infty $ in low frequency TM band is%
\begin{equation}
	\lim_{k\rightarrow \infty ,~\gamma \rightarrow 0}(A_{n,\phi =0}k)=\mathrm{Re}%
	\left\{ \frac{i\beta _{12}(k)}{\beta _{11}(k)}\right\} _{\omega _{n}=|\omega
		_{c}|}=-1.
	\end{equation}
For the case of $k\rightarrow 0$ ( $\gamma \rightarrow 1$) and we have same
dispersion equation as when there is no wave vector cut-off, and so that
limit remains the same as before, $\lim_{k\rightarrow 0}(A_{n,\phi =0}k)=1$.
Therefore, for the low frequency band we obtain%
\[
	C_{n}=-1-1=-2.
	\]

For the high frequency TM band as $k\rightarrow \infty ~(\gamma \rightarrow
0)$, $\lim_{k\rightarrow \infty }(A_{n,\phi =0}k)=0$ and when $k\rightarrow
0~(\gamma \rightarrow 1)$, $\lim_{k\rightarrow 0}(A_{n,\phi =0}k)=-1$ (as
before), and so for the high frequency band the Chern number is%
\[
C_{n}=0-(-1)=+1.
\]%

The sum of the Chern numbers is -1. However, in addition to needing a wavenumber cutoff to obtain integer Chern numbers, the continuum model presents another complication. As detailed in \cite{M5}, to predict edge states in general for continuum media, one should compute Chern numbers for an "`interpolated material response"'. This means, for example, that to see bulk-edge correspondence for the magnetized plasma and a Drude metal interface, we should define a function $\epsilon(\tau)$ where $\tau$ varies from $0$ to $1$, such that when $\tau=0$ we obtain the permittivity of the magnetized plasma, and when $\tau=1$ we obtain the Drude metal. Then, one needs to compute the topological numbers for $\tau=1^-$ and $\tau=0^+$. With this model, we obtain one additional low frequency band for the magnetized plasma, very near $\omega=0$, having Chern number $1$. In this case, all band Chern numbers are integers and sum to zero.

\subsubsection{Full-wave simulation of one-way propagation}
\begin{figure}[ht]
\begin{center}
\noindent \includegraphics[width=6in]{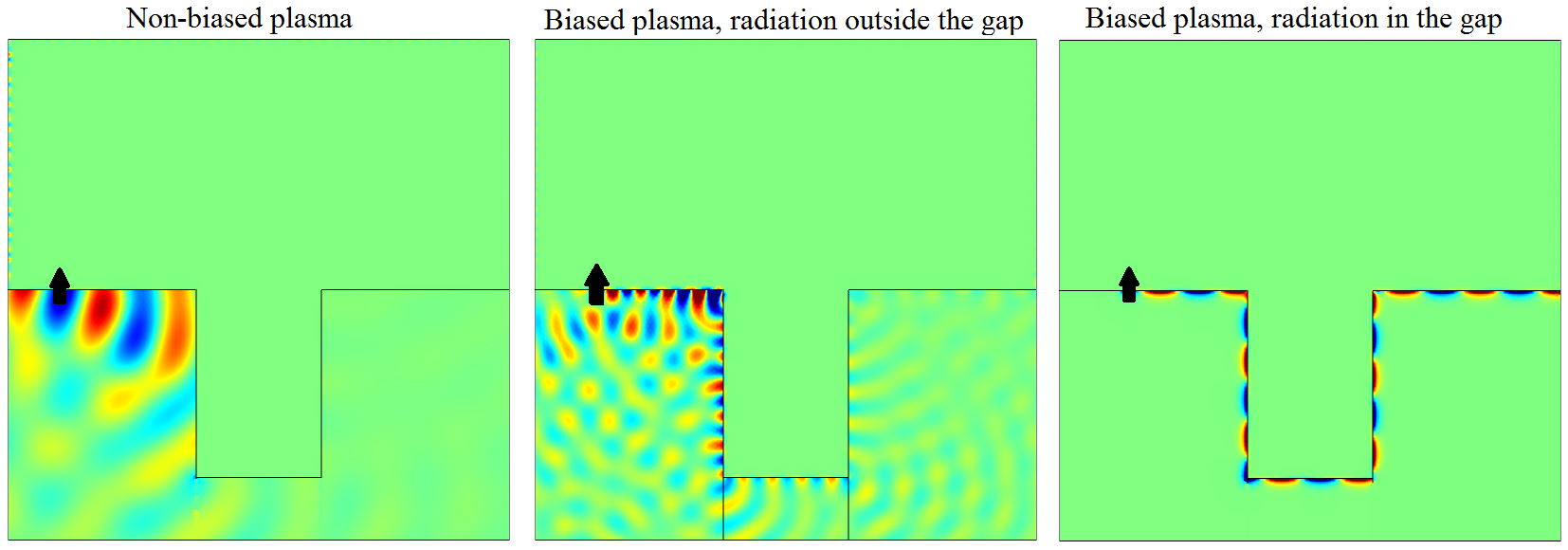}
\end{center}
\caption{Electric field due to a 2D vertical dipole and $ \omega_p / 2\pi=9.7 $ THz for three cases: left: unbiased (reciprocal) case that respects TR symmetry, $ \omega_c= 0 $, at 10 THz ($\lambda=30 \mu$ m), center: biased with $ \omega_c/ 2 \pi= 1.73 $ THz  at 12 THz, outside of the band gap ($\omega/\omega_c=6.93$), and right: biased with $ \omega_c/ 2 \pi= 1.73 $ THz inside the bandgap at 10 THz, ($\omega/\omega_c=5.78$). }
\label{2Dmp}
\end{figure}

We first consider a 2D structure. A 2D dipole (i.e., a line source) is at the interface between a simple plasma (upper region) having $\varepsilon =-5$ (this specific value relatively is unimportant; we simply need a negative-permittivity material such as a metal) and a magnetoplasma (lower region) having permittivity (\ref{BMPM}). Fig. \ref{2Dmp} shows the electric field profile for three cases, unbiased, biased but operating outside the band gap, and biased operating within the bandgap. It can be seen that in the unbiased (reciprocal) case energy flows in both directions, in the biased (non-reciprocal) case operating outside the gap we have one-way propagation but energy can leak into the lower region, and in the biased case operating within the bandgap energy just flows in one direction, is well-contained at the interface, and goes around discontinuities.

\begin{figure}[h!]
\begin{center}
\noindent \includegraphics[width=3in]{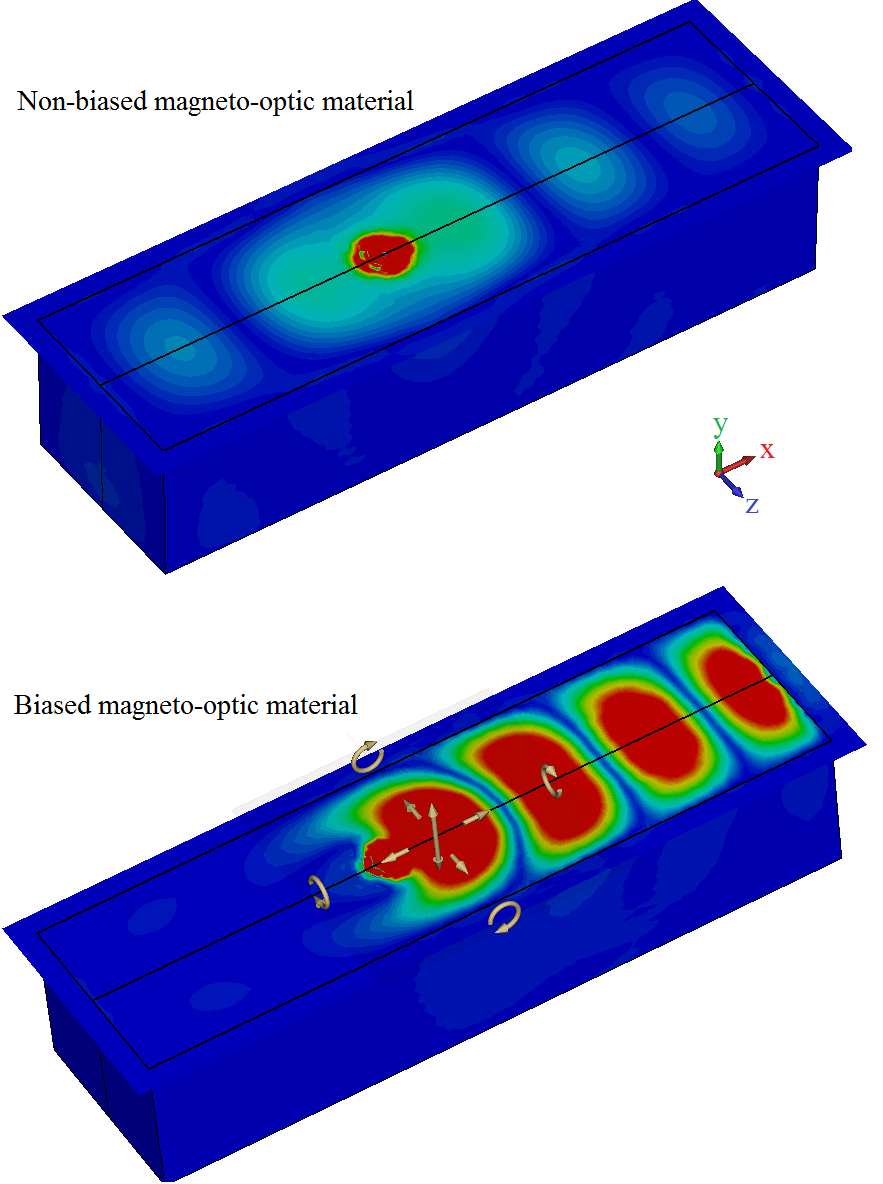}
\end{center}
\caption{Electric field at 10 THz for a 3D vertical dipole at a magnetoplasma--plasma interface (top interface is between the magnetoplasma and the $\varepsilon =-5$ simple plasma, all other interfaces are between the magnetoplasma and vacuum). Top:  unbiased ($ \omega_c=0 $, reciprocal) case. Bottom: non-reciprocal case when $ \omega_c/ 2 \pi= 1.73 $ THz inside the bandgap ($\omega/\omega_c=5.78$).}
\label{FFI}
\end{figure}

Fig. \ref{FFI} shows a 3D simulation for a 420x120x90 um rectangular block of magnetoplasma with an $\varepsilon =-5$ plasma on the top surface and vacuum on all other sides: a (top) shows the electric field profile in the reciprocal case, $ \omega_c=0 $, at 10 THz. Fig. \ref{FFI}b (lower) shows the non-reciprocal case at 10 THz (in the bandgap). It can be seen that in the reciprocal case energy flows in both directions, whereas in the non-reciprocal case energy just flows to the right.

\begin{figure}[h!]
\begin{center}
\noindent \includegraphics[width=3in]{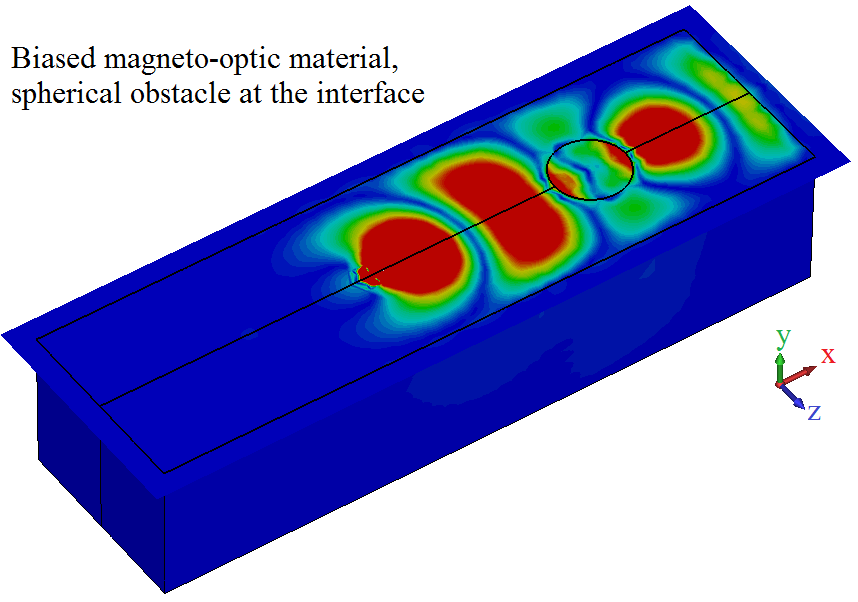}
\end{center}
\caption{Electric field near a magnetoplasma--plasma interface, as in Fig. \ref{FFI}, in the non-reciprocal case when a large ($1 \lambda$) spherical vacuum obstacle is placed in the SPP path.}
\label{OBS}
\end{figure}

Figure \ref{OBS} shows the non-reciprocal case when an obstacle (a half-sphere) is hollowed out of each material at the interface, forming a spherical vacuum obstacle having radius $30 \mu$m ($1 \lambda$) in the SPP path. It can be seen that the wave goes past the obstacle without backscattering. 

\begin{figure}[h!]
\begin{center}
\noindent \includegraphics[width=3.5in]{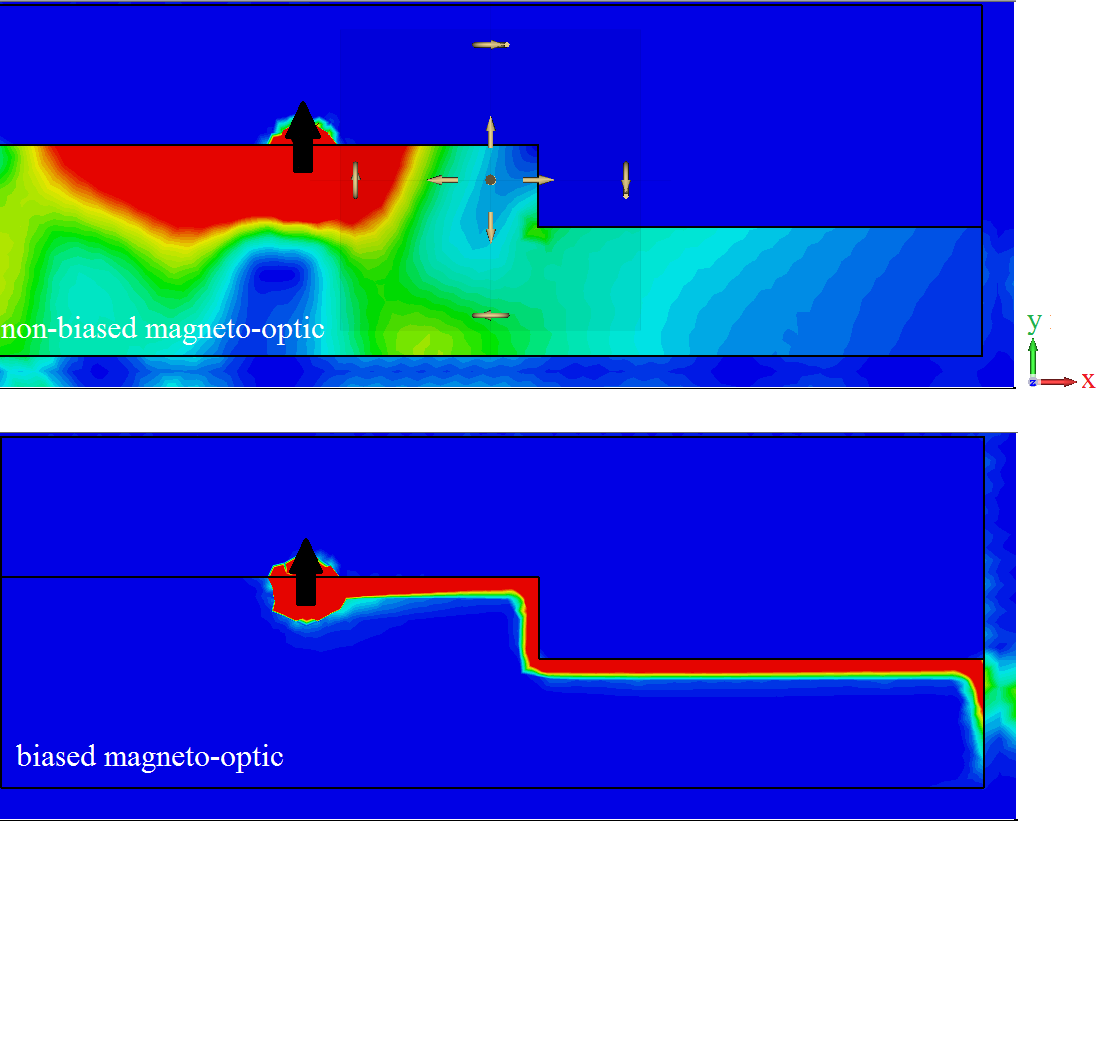}
\end{center}
\caption{Side view of power density due to a vertical point dipole source at the interface between a magnetoplasma--plasma interface (top surface, all other sides interface with vacuum). a. Power density in the reciprocal case. b. Non-reciprocal case.}
\label{Power}
\end{figure}

\begin{figure}[h!]
\begin{center}
\noindent \includegraphics[width=3.5in]{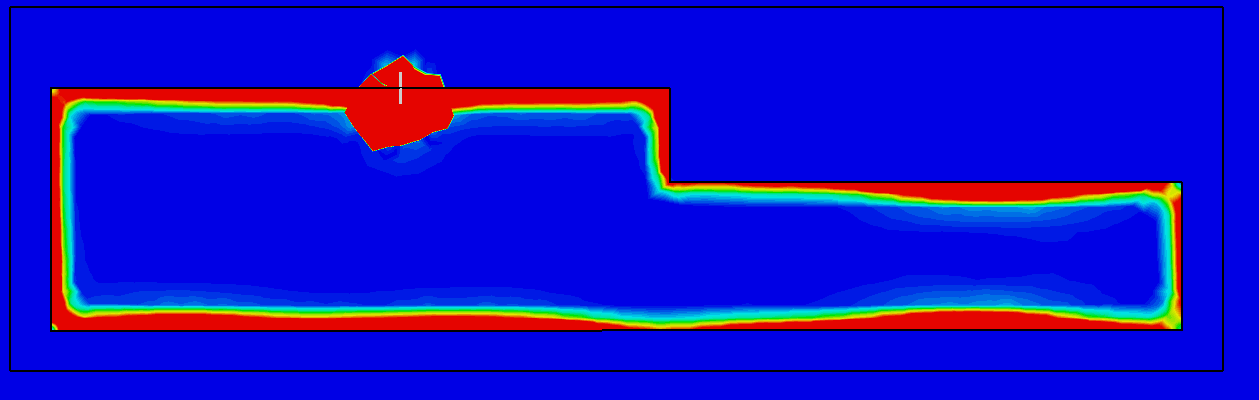}
\end{center}
\caption{Side view of power density due to a vertical point dipole source at the interface between a magnetoplasma--plasma interface. a. Power density in the reciprocal case. b. Non-reciprocal case.}
\label{rotating}
\end{figure}

Finally, Figs. \ref{Power} and \ref{rotating} show the power density for the case of an interface with a step discontinuity in height. The step height is 30 $\mu$m ($1 \lambda$). For Fig. \ref{Power}, as in Figs. \ref{FFI} and \ref{OBS}, the top interface is with the $\varepsilon =-5$ simple plasma, all other interfaces are between the magnetoplasma and vacuum. A vertical dipole source is located on the left side as indicated. Fig. \ref{Power}a shows the side view of the power density in the reciprocal case, and Fig. \ref{Power}b shows the non-reciprocal case. It can be seen that in the reciprocal case energy flows in both directions as well as interacting with and reflecting from the step, whereas in the non-reciprocal case energy just flows to the right, and doesn't scatter off of the step discontinuity. In Fig. \ref{rotating} we surround all sides of the magnetoplasma with $\varepsilon =-5$ plasma. In this case energy circulates around the entire structure.

The dispersion relation for the surface mode is \cite{Arthur} 
\begin{equation}
	\frac{\sqrt{k_{x}^{2}-k_{0}^{2}\varepsilon _{s}}}{\varepsilon _{s}}+\frac{%
		\sqrt{k_{x}^{2}-k_{0}^{2}\varepsilon _{eff}}}{\varepsilon _{eff}}=-\frac{%
		\varepsilon _{12}ik_{x}}{\varepsilon _{11}\varepsilon _{eff}}
	\end{equation}
where $\varepsilon _{s}$ is the top material permittivity and 
\begin{equation}
\varepsilon _{eff}=\frac{\varepsilon _{11}^{2}+\varepsilon _{12}^{2}}{%
\varepsilon _{11}}
\end{equation}
where $\varepsilon _{\alpha,\beta}$ are the magnetoplasma permittivity components.

\subsubsection{Numerical Computation of the Chern Number}

In these continuum examples the Chern number can be found analytically. However, often
this will not be the case, and numerical methods must be used (as in the
photonic crystal example in Section \ref{Ex1}). As discussed previously, the Berry
potential (\ref{AEM}) and associated Chern computation (\ref{CNNF}) may not be convenient for numerical computations since
it involves derivatives of the eigenfunctions, which generally need to be
taken numerically. The curvature form (\ref{Eq:15}) and associated Chern
number (\ref{CNF}) provide a convenient method, since only the Hamiltonian
matrix needs to be differentiated. 

In the non-dispersive case the formulation in Section \ref{SECEM} suffices, the classical Hamiltonian $H_{\text{cl}}=M^{-1}N$ is Hermitian under the indicated inner product, the eigenvalue problem 
$H_{cl} \cdot f_{n}=\omega _{n}f_{n}\,$ is a standard eigenvalue problem, and the 6-vector of natural modes $f_{n}=\left[ \mathbf{E\ \ H}\right] ^{T}$ from (\ref{PEE1}) from a complete set of eigenfunctions. In principle, either the formulation (\ref{CNNF}) or (\ref{CNF}) can be used to compute the Chern number.

In the dispersive case this does not hold, but, nevertheless, only the natural modes $f_n$ are need to compute the Berry curvature  (\ref{PEE3}) \cite{Mario2}. However, these modes are not appropriate for the form (\ref{CNF}), in particular, since the eigenmodes depend on frequency and (\ref{CNF}) involves terms with different eigenmodes. Furthermore, in the dispersive
case the Hamiltonian $H_{\text{cl}}=M^{-1}N$ does not admit a
complete set of eigenvectors, which, in principle, is needed in the
computation (\ref{CNF}). Moreover, the 6-vector of natural modes (if one is going to use (\ref{CNNF})) is not so
easily computed in practice, since $H_{cl}\left( \omega _{n}\right) \cdot
f_{n}=\omega _{n}f_{n}\,$ is a non-standard eigenvalue problem, and
eigenvalues would generally need to be found via a root search or similar
method.$\ $

In \cite{Haldane} (and other works, see, e.g., \cite{Fan}) the non-standard
eigenvalue problem in the dispersive case is avoided by introducing auxiliary variables (additional
degrees of freedom), and in \cite{Mario2} this approach is extended to allow for
both temporal and spatial dispersion of general linear media. The resulting standard Hermitian eigenvalue problem to be solved is 
\begin{equation}
\left(M_{g}^{-1} \cdot L \right) Q=\omega Q,
\end{equation}%
where, in block-matrix form (all elements in $M_{g}$ and $L$ are 6x6 blocks),%
\begin{equation}
M_{g}=\left( 
\begin{array}{cccc}
M_{\infty } & \mathbf{0} & \mathbf{0} & \cdots  \\ 
\mathbf{0} & \mathbf{I} & \mathbf{0} & \cdots  \\ 
\mathbf{0} & \mathbf{0} & \mathbf{I} & \cdots  \\ 
\vdots  & \vdots  & \vdots  & \ddots 
\end{array}%
\right) ,\ \ M_{\infty }=\lim_{\omega \rightarrow \infty }M\left( \omega ,%
\mathbf{k}\right) ,
\end{equation}
and%
\begin{equation}
L=\left( 
\begin{array}{cccc}
N+\sum_{\alpha }\mathrm{sgn}\left( \omega _{p,\alpha }\right) \mathbf{A}%
_{\alpha }^{2} & \left\vert \omega _{p,1}\right\vert ^{1/2}\mathbf{A}_{1}
& \left\vert \omega _{p,2}\right\vert ^{1/2}\mathbf{A}_{2} & \cdots  \\ 
\left\vert \omega _{p,1}\right\vert ^{1/2}\mathbf{A}_{1} & \omega _{p,1}%
\mathbf{I} & \mathbf{0} & \cdots  \\ 
\left\vert \omega _{p,2}\right\vert ^{1/2}\mathbf{A}_{2} & \mathbf{0} & 
\omega _{p,2}\mathbf{I} & \cdots  \\ 
\vdots  & \vdots  & \vdots  & \ddots 
\end{array}%
\right) ,
\end{equation}%
\begin{equation}
N=\left( 
\begin{array}{cc}
0 & -\mathbf{k}\times \mathbf{I}_{3\times 3} \\ 
\mathbf{k}\times \mathbf{I}_{3\times 3} & 0%
\end{array}%
\right) =\left( 
\begin{array}{cccccc}
0 & 0 & 0 & 0 & k_{z} & -k_{y} \\ 
0 & 0 & 0 & -k_{z} & 0 & k_{x} \\ 
0 & 0 & 0 & k_{y} & -k_{x} & 0 \\ 
0 & -k_{z} & k_{y} & 0 & 0 & 0 \\ 
k_{z} & 0 & -k_{x} & 0 & 0 & 0 \\ 
-k_{y} & k_{x} & 0 & 0 & 0 & 0%
\end{array}%
\right) ,  \label{N}
\end{equation}%
where $M_g$ and $L$ are independent of frequency, $Q=\left[ f\ \ Q_{1}\ \ Q_{2}\ \ ...\right] 
$ where each element in $Q$ is 6x1, and $f$ is defined as before. The elements $\mathbf{A}_{\alpha }$ are the 6x6 residues of the material
matrix, $\mathbf{A}_{\alpha }^{2}=-\mathrm{sgn}\left( \omega _{p,\alpha
}\right) \mathrm{Res}\left(M\right) _{\alpha }$, and $\omega
_{p,\alpha }$ is the $\alpha $th pole of $M$. More
details are available in \cite{Mario2}, and here we focus on the specific
material example (\ref{MBMPM})considered above.

Given the permittivity form (\ref{MBMPM}), the material matrix has two poles, at $%
\omega=\pm \omega _{0}$. Therefore, $L$ is an 18x18 matrix, $M_{g}$ is the diagonal matrix $\left( \varepsilon _{0},\varepsilon
_{0},\varepsilon _{0},\mu _{0},\mu _{0},\mu _{0},1,1,1...1\right) $  and%
\begin{equation}
\mathbf{A}_{1}^{2}=\left( 
\begin{array}{cccccc}
\varepsilon _{0}\frac{\omega _{e}}{2} & -i\varepsilon _{0}\frac{\omega _{e}}{%
2} & 0 & 0 & 0 & 0 \\ 
i\varepsilon _{0}\frac{\omega _{e}}{2} & \varepsilon _{0}\frac{\omega _{e}}{2%
} & 0 & 0 & 0 & 0 \\ 
0 & 0 & 0 & 0 & 0 & 0 \\ 
0 & 0 & 0 & 0 & 0 & 0 \\ 
0 & 0 & 0 & 0 & 0 & 0 \\ 
0 & 0 & 0 & 0 & 0 & 0%
\end{array}%
\right) ,\ \ \mathbf{A}_{2}^{2}=\left( 
\begin{array}{cccccc}
\varepsilon _{0}\frac{\omega _{e}}{2} & i\varepsilon _{0}\frac{\omega _{e}}{2%
} & 0 & 0 & 0 & 0 \\ 
-i\varepsilon _{0}\frac{\omega _{e}}{2} & \varepsilon _{0}\frac{\omega _{e}}{%
2} & 0 & 0 & 0 & 0 \\ 
0 & 0 & 0 & 0 & 0 & 0 \\ 
0 & 0 & 0 & 0 & 0 & 0 \\ 
0 & 0 & 0 & 0 & 0 & 0 \\ 
0 & 0 & 0 & 0 & 0 & 0%
\end{array}%
\right) .
\end{equation}%

\noindent Setting $k_{z}=0$, the final Hamiltonian matrix is 

\begin{equation}
H=M_{g}^{-1}L=\left( 
\begin{array}{cccccccccccccccccc}
0 & -i\omega _{e}^{\prime } & 0 & 0 & 0 & -\frac{k_{y}^{\prime }}{%
\varepsilon _{0}} & -\alpha  & i\alpha  & 0 & 0 & 0 & 0 & -\alpha  & 
-i\alpha  & 0 & 0 & 0 & 0 \\ 
i\omega _{e}^{\prime } & 0 & 0 & 0 & 0 & \frac{k_{x}^{\prime }}{\varepsilon
_{0}} & -i\alpha  & -\alpha  & 0 & 0 & 0 & 0 & i\alpha  & -\alpha  & 0 & 0 & 
0 & 0 \\ 
0 & 0 & 0 & \frac{k_{y}^{\prime }}{\varepsilon _{0}} & -\frac{k_{x}^{\prime }%
}{\varepsilon _{0}} & 0 & 0 & 0 & 0 & 0 & 0 & 0 & 0 & 0 & 0 & 0 & 0 & 0 \\ 
0 & 0 & \frac{k_{y}^{\prime }}{\mu _{0}} & 0 & 0 & 0 & 0 & 0 & 0 & 0 & 0 & 0
& 0 & 0 & 0 & 0 & 0 & 0 \\ 
0 & 0 & -\frac{k_{x}^{\prime }}{\mu _{0}} & 0 & 0 & 0 & 0 & 0 & 0 & 0 & 0 & 0
& 0 & 0 & 0 & 0 & 0 & 0 \\ 
-\frac{k_{y}^{\prime }}{\mu _{0}} & \frac{k_{x}}{\mu _{0}} & 0 & 0 & 0 & 0 & 
0 & 0 & 0 & 0 & 0 & 0 & 0 & 0 & 0 & 0 & 0 & 0 \\ 
-\beta  & i\beta  & 0 & 0 & 0 & 0 & 1 & 0 & 0 & 0 & 0 & 0 & 0 & 0 & 0 & 0 & 0
& 0 \\ 
-i\beta  & -\beta  & 0 & 0 & 0 & 0 & 0 & 1 & 0 & 0 & 0 & 0 & 0 & 0 & 0 & 0 & 
0 & 0 \\ 
0 & 0 & 0 & 0 & 0 & 0 & 0 & 0 & 1 & 0 & 0 & 0 & 0 & 0 & 0 & 0 & 0 & 0 \\ 
0 & 0 & 0 & 0 & 0 & 0 & 0 & 0 & 0 & 1 & 0 & 0 & 0 & 0 & 0 & 0 & 0 & 0 \\ 
0 & 0 & 0 & 0 & 0 & 0 & 0 & 0 & 0 & 0 & 1 & 0 & 0 & 0 & 0 & 0 & 0 & 0 \\ 
0 & 0 & 0 & 0 & 0 & 0 & 0 & 0 & 0 & 0 & 0 & 1 & 0 & 0 & 0 & 0 & 0 & 0 \\ 
-\beta  & -i\beta  & 0 & 0 & 0 & 0 & 0 & 0 & 0 & 0 & 0 & 0 & 1 & 0 & 0 & 0 & 
0 & 0 \\ 
i\beta  & -\beta  & 0 & 0 & 0 & 0 & 0 & 0 & 0 & 0 & 0 & 0 & 0 & 1 & 0 & 0 & 0
& 0 \\ 
0 & 0 & 0 & 0 & 0 & 0 & 0 & 0 & 0 & 0 & 0 & 0 & 0 & 0 & 1 & 0 & 0 & 0 \\ 
0 & 0 & 0 & 0 & 0 & 0 & 0 & 0 & 0 & 0 & 0 & 0 & 0 & 0 & 0 & 1 & 0 & 0 \\ 
0 & 0 & 0 & 0 & 0 & 0 & 0 & 0 & 0 & 0 & 0 & 0 & 0 & 0 & 0 & 0 & 1 & 0 \\ 
0 & 0 & 0 & 0 & 0 & 0 & 0 & 0 & 0 & 0 & 0 & 0 & 0 & 0 & 0 & 0 & 0 & 1%
\end{array}%
\right) \label{BigM}
\end{equation}
where $\alpha =\frac{1}{2}\sqrt{\omega _{e}^{\prime }/\varepsilon _{0}}$, 
$\beta =\frac{1}{2}\sqrt{\varepsilon _{0}\omega _{e}^{\prime }}$, $\omega
_{e}^{\prime }=\omega _{e}/\omega _{0}$, $k_{x,y}^{\prime }=k_{x,y}/\omega
_{0}$. 

From (\ref{BigM}) the eigenvalues and associated eigenvectors can easily be found numerically (or symbolically), and the Chern number computed from (\ref{CNF}). Of the 18 branches, two are the positive-frequency TM modes and one is the positive-frequency TE mode described previously. In addition to static-like (longitudinal) modes, there are dispersionless dark modes with $\mathbf{E}=\mathbf{H}=\mathbf{0}$ which don't contribute to the Chern number. For each TM band, the other TM band and, to a lesser extent, the TE band, provide the most important contributions to the Chern number calculation (\ref{CNF}).

\subsection{Acknowledgments}

We would like to thank M\'{a}rio G. Silveirinha, Shuang Zhang, and Kejie Fang for help
with this topic. Any mistakes or misconceptions are our own.


\begin{thebibliography}{10}
	\newcommand{\enquote}[1]{``#1''}


\bibitem{Arthur} Arthur R. Davoyan and Nader Engheta, \emph{Theory of wave propagation in magnetized near-zero-epsilon metamaterials: evidence for one-way photonic states and magnetically switched transparency and opacity}, Phys. Rev. Lett. 111, 257401, December 2013.

\bibitem{Biao} Biao Yang, Mark Lawrence, Wenlong Gao, Qinghua Guo, Shuang Zhang, \emph{One-way helical electromagnetic wave propagation supported by magnetized plasma}, arXiv:1410.4596 [physics.optics].

\bibitem{Mario2} M\'{a}rio G. Silveirinha, \emph{Chern invariants for continuous media}, Phys. Rev. B 92, 125153, 2015.	

\bibitem{Haldane2} F. D. M. Haldane and S. Raghu, \emph{Possible realization of directional optical waveguides in photonic crystals with broken time-reversal symmetry}, Phys. Rev. Lett. 100, 013904, Jan. 2008.

\bibitem{Haldane} S. Raghu and F. D. M. Haldane, \emph{Analogs of quantum-Hall-effect edge states in photonic crystals}, Phys. Rev. A 78, 033834, September 2008.

\bibitem{Rechtsman} M. C. Rechtsman, J. M. Zeuner, Y. Plotnik, Y. Lumer, D. Podolsky, F. Dreisow, S. Nolte,	M. Segev and A. Szameit, \emph{Photonic Floquet topological insulators}, Nature 496, 196–200, April 2013.

\bibitem{Rechtsman1} M. C. Rechtsman, Y. Plotnik, J. M. Zeuner, D. Song, Z. Chen, A. Szameit, and M. Segev \emph{Topological Creation and Destruction of Edge States in Photonic Graphene}, Phys. rev. Lett. 111, 103901, 2013.

\bibitem{Rechtsman2} M. C. Rechtsman, Y. Plotnik, J. M. Zeuner, D. Song, Z. Chen, A. Szameit, and M. Segev \emph{Topological Creation and Destruction of Edge States in Photonic Graphene}, Phys. rev. Lett. 111, 103901, 2013.

\bibitem{Poo} Y. Poo, R-X Wu, Z. Lin, Y. Yang, and C. T. Chan, \emph{Experimental realization of self-guiding unidirectional electromagnetic edge states}, Phy. Rev. lett. 106, 093903, 2011.

\bibitem{Dong} Wen-Jie Chen,	Shao-Ji Jiang, Xiao-Dong Chen, Baocheng Zhu, Lei Zhou, Jian-Wen Dong and C. T. Chan, \emph{Experimental realization of photonic topological insulator in a uniaxial metacrystal waveguide}, Nature Communications,5, 5782, December 2014.

\bibitem{Skirlo} S. A. Skirlo, L. Lu, Y. Igarashi, Q. Yan, J. D. Joannopoulos, and Marin Soljačić, \emph{Experimental observation of large Chern numbers in photonic crystals}, Phys. Rev. Lett. 115, 253901, Dec. 2015.

\bibitem{MGA} S. A. Hassani Gangaraj, M. G. Silveirinha, and G. W. Hanson, \emph{Berry phase, Berry Potential, and Chern Number for Continuum Bianisotropic Material from a Classical Electromagnetics Perspective}, IEEE Journal on multiscale and multiphysics computational techniques, to be published 2017.

\bibitem{Solja} Zheng Wang, Yidong Chong, J. D. Joannopoulos and Marin Soljačić, \emph{Observation of unidirectional backscattering-immune topological electromagnetic states}, Nature 461, 772-775, Oct. 2009.

\bibitem{ballentine} Leslie E. Ballentine, \emph{Quantum mechanics: A modern
development}, Prentice Hall, New Jersey, 1990.

\bibitem{griffiths} David J. Griffiths, \emph{Introduction to quantum
mechanics}, Prentice Hall, New Jersey, 1995.

\bibitem{TC1986} A. Tomita and R. Y. Chiao, \emph{Observation of Berry's Topological Phase by Use of an Optical Fiber}, Phys. Rev. Letts. 57, 937, 1986.

\bibitem{xu2014} Q. Xu, L. Chen, M. G. Wood, P. Sun, and R. M. Reano, \emph{Electrically tunable optical polarization rotation on a silicon chip using Berry’s phase}, Nat. Comm. DOI: 10.1038/ncomms6337, 2014.

\bibitem{Xiao} Di Xiao, Ming-Che Chang and Qian Niu, \emph{Berry phase effects on electronic properties}, Rev. Mod. Phys. 82, 1959–6 July 2010.

\bibitem{Berry} M. V. Berry, \emph{Quantal phase factors accompanying
adiabatic changes}, Proc. R. Soc. Lond. A 392, 45-57 (1984).

\bibitem{Anandan} Y. Aharonov and J. Anandan, \emph{Phase change during a
cyclic quantum evolution}, Phys. Rev. Lett. 58, 1593-20 April 1987.

\bibitem{RL1} J. Anandan, J. Christian, and K. Wanelik, \emph{Resource Letter GPP-1: Geometric Phases in Physics}, Am. J. Phys. 65, 180 (1997).

\bibitem{TKNN} D. J. Thouless, M. Kohmoto, M. P. Nightingale, and M. den Nijs, \emph{Quantized Hall Conductance in a Two-Dimensional Periodic Potential}, Phys. Rev. Lett. 49, 405, 1982.

\bibitem{Xiao1} D. Xiao, G.-B. Liu, W. Feng, X. Xu, and W. Yao, \emph{Coupled Spin and Valley Physics in Monolayers of MoS2 and Other Group-VI Dichalcogenides}, Phys. Rev. Lett. 108, 196802.

\bibitem{KN2016} A. Kumar, A. Nemilentsau, K.H. Fung, G.W. Hanson, N.X. Fang, and T. Low, \emph{Chiral plasmon in gapped Dirac systems}, Phys. Rev. B (Rapid Communications) 93, 041413(R), 2016. 

\bibitem{SA} T. Ando, T. Nakanishi, and R. Saito, \emph{Berry's Phase and Absence of Back Scattering in Carbon Nanotubes}, J. Phy. Soc. Japan, 67, 2857-2862, 1998.

\bibitem{Hall} M. Masaru, S. Murakami, and N. Nagaosa, \emph{Hall effect of light}, Phys. Rev. Lett. 93, 083901, 2004.

\bibitem{BPG} J. Xue, \emph{Berry phase and the unconventional quantum Hall effect in graphene}, arXiv:1309.6714, 2013.

\bibitem{YZK} A. Young, Y. Zhang, and P. Kim, \emph{Experimental Manifestation of Berry Phase in Graphene}, in Physics of Graphene, H. Aoki, and M S. Dresselhaus (Eds), Springer, 2014.

\bibitem{BT} G. Tkachov, \emph{Topological Insulators: The Physics of Spin Helicity in Quantum Transport}, CRC Press, 2015.

\bibitem{Mario1} M\'{a}rio G. Silveirinha and Stanislav I. Maslovski, \emph{Exchange of momentum between moving matter induced by the zero-point fluctuations of the electromagnetic field}, Phys. Rev. A 86, 042118, October 2012.


\bibitem{Kejie}  Kejie Fang, Zongfu Yu and Shanhui Fan, \emph{Realizing effective magnetic field for photons by controlling the phase of dynamic modulation}, Nature Photonics 6, 782–787 (2012).

\bibitem{Hafezi} M. Hafezi, S. Mittal, J. Fan, A. Migdall and J. M. Taylor, \emph{Imaging topological edge states in silicon photonics}, Nature Photonics 7, 1001–1005 (2013).

\bibitem{Alexander} Alexander B. Khanikaev, S. Hossein Mousavi, Wang-Kong Tse,	Mehdi Kargarian, Allan H. MacDonald and Gennady Shvets, \emph{Photonic topological insulators}, Nature Materials 12, 233–239 (2013).

\bibitem{PTI} Wenlong Gao, Mark Lawrence, Biao Yang, Fu Liu, Fengzhou Fang, Benjamin Béri, Jensen Li and Shuang Zhang, \emph{Topological photonic phase in chiral hyperbolic metamaterials}, Phys. Rev. Lett. 114, 037402-22 Jan. 2015.

\bibitem{Solja2} Y. D. Chong, Xiao-Gang Wen, and Marin Soljačić, \emph{Effective theory of quadratic degeneracies}, Phys. Rev. B 77, 235125, 30 June 2008.

\bibitem{Solja3} Zheng Wang, Y. D. Chong, John D. Joannopoulos, and Marin Soljačić, \emph{Reflection-Free one-way edge modes in a gyromagnetic photonic crystal}, Phys. Rev. Lett. 100, 013905, January 2008.

\bibitem{Kejie2} Kejie Fang, Zongfu Yu, and Shanhui Fan, \emph{Microscopic theory of photonic one-way edge mode}, Phys. Rev. B 84, 075477, August 2011.

\bibitem{Solja4} Scott A. Skirlo, Ling Lu, and Marin Soljačić, \emph{Multimode one-way waveguides of large Chern numbers}, Phys. Rev. Lett. 113, 113904, September 2014.

\bibitem{Hua} Long-Hua Wu and Xiao Hu, \emph{Scheme for achieving a topological photonic crystal by using dielectric material}, Phys. Rev. Lett. 114, 223901, June 2015.

\bibitem{AB2013} B. Andrei Bernevig, \emph{Topological Insulators and Topological
Superconductors}, Princeton Univ. Press, NJ: 2013.

\bibitem{FK} L. Fu, C. L. Kane, \textquotedblleft Time reversal polarization
and a Z2 adiabatic spin pump\textquotedblright , Phys. Rev. B, 74,195312,
2006.

\bibitem{M2} M\'{a}rio G. Silveirinha, \textquotedblleft $\mathbb{Z}_{2}$ Topological Index for Homogeneous Continuous Photonic Materials,\textquotedblright arXiv:1601.02823, 2016.

\bibitem{AD} S. Fan, M.F. Yanik, Z. Wang, S. Sandhu, and M.L. Povinelli, \textquotedblleft Advances in Theory of Photonic Crystals\textquotedblright , J. Lightwave Tech. 24, 4493-4501,
2006.

\bibitem{FC} J.D. Joannopoulos, S.G. Johnson, J.N. Winn, and R.D. Meade, \textquotedblleft Photonic Crystals\textquotedblright ,  Princeton University Press, 2008.

\bibitem{TB} E. Lidorikis, M. M. Sigalas, E. N. Economou, and C. M. Soukoulis, \textquotedblleft Tight-Binding Parametrization for Photonic Band Gap Materials\textquotedblright ,  Phys. Rev. Lett. 81, 1405, 1998.

\bibitem{Fan} A. Raman and S. Fan,  \textquotedblleft Photonic Band Structure of Dispersive Metamaterials Formulated as a Hermitian Eigenvalue Problem\textquotedblright ,  Phys. Rev. Lett. 104, 087401, 2010.

\bibitem{M5} M\'{a}rio G. Silveirinha, \textquotedblleft Bulk-edge correspondence for topological photonic continua,\textquotedblright Phys. Rev. B 94, 205105, 2016.

\end{thebibliography}
\end{document}